\documentclass[aps,pre,10pt,showpacs,floatfix,onecolumn,nofootinbib,a4paper]{revtex4-2}

\usepackage{amssymb, amsmath, amsthm, mathtools}
\usepackage{amsfonts}
\usepackage{stmaryrd}
\usepackage{mathrsfs}
\usepackage{hyperref,bookmark}
\usepackage{graphicx}
\usepackage{epsfig,color}
\usepackage{float}
\usepackage{caption}
\usepackage{subcaption}
\usepackage{verbatim}
\usepackage{dcolumn}
\usepackage{tikz}
\usepackage{url}
\usepackage{cases}
\usepackage[utf8x]{inputenc}
\usepackage[english]{babel}
\usepackage{epsfig}
\usepackage{listings}
\usepackage{paralist}
\usepackage{mdwlist}
\usepackage{bm}
\usepackage{dsfont}
\usepackage{oldgerm}
\usepackage{geometry}
\usepackage{relsize}
\usepackage{afterpage}
\usepackage{lmodern}
\usepackage{tabularx}
\usepackage{cancel}
\usepackage{dutchcal} 
\usepackage{natbib} 
\usepackage[overload]{empheq}

\newcommand{\dd}{\mathrm{d}}
\newcommand{\ex}{\mathrm{e}}

\newcommand{\vt}{\vartheta}

\newcommand{\trans}{^\mathsf{T}}

\newcommand{\n}{\bm{n}}
\newcommand{\nA}{\n_\mathrm{A}}
\newcommand{\nS}{\n_\mathrm{S}}
\newcommand{\nU}{\n_\mathrm{U}}
\newcommand{\cv}{\bm{c}}

\newcommand{\dv}{\bm{d}}
\newcommand{\e}{\bm{e}}
\newcommand{\x}{\bm{x}}
\newcommand{\zero}{\bm{0}}
\newcommand{\y}{\bm{y}}

\newcommand{\normal}{\bm{\nu}}
\newcommand{\normalB}{\normal_{\partial\body}}
\newcommand{\normalW}{\normal_{\wall}}

\newcommand{\normalL}{\normal_{\Line}}
\newcommand{\surface}{\mathscr{S}}
\newcommand{\tplane}{\mathscr{T}}
\newcommand{\free}{\mathcal{F}}
\newcommand{\freepol}{\free_\mathrm{pol}}
\newcommand{\freela}{\free_\mathrm{ela}}
\newcommand{\freeint}{\free_\mathrm{int}}
\newcommand{\freewall}{\free_\mathrm{wall}}
\newcommand{\freebulk}{\free_\mathrm{bulk}}

\newcommand{\nablas}{\nabla\!_\mathrm{s}}
\newcommand{\nablac}{\nabla\!_\mathrm{c}}

\newcommand{\rv}{\bm{r}}

\newcommand{\curl}{\operatorname{curl}}

\newcommand{\curlc}{\curl_\mathrm{c}\!}
\newcommand{\divs}{\diver_\mathrm{s}\!}
\newcommand{\divc}{\diver_\mathrm{c}\!}
\newcommand{\tr}{\operatorname{tr}}

\newcommand{\jump}[1]{\left\llbracket{#1}\right\rrbracket}

\newcommand{\body}{\mathscr{B}}

\newcommand{\jumpset}{\mathscr{J}}

\newcommand{\frameC}{(\e_x,\e_y,\e_z)}
\newcommand{\framen}{(\n,\n_\perp,\normal)}

\newcommand{\tangent}{\bm{t}}

\newcommand{\tperp}{\bm{t}_\perp}
\newcommand{\nperp}{\n_\perp}

\newcommand{\Curve}{\mathscr{C}}
\newcommand{\pol}{\bm{P}}
\newcommand{\Ele}{\bm{E}}

\newcommand{\wall}{\mathscr{W}}
\newcommand{\Line}{\mathscr{L}}

\newcommand{\real}{{\mathbb{R}}}

\newcommand{\proj}{\mathbf{P}(\normal)}

\newcommand{\sphere}{\mathbb{S}^2}

\newcommand{\diver}{\operatorname{div}} 

\newcommand{\NF}{\mathsf{N}_\mathsf{F}}
\newcommand{\NX}{\mathsf{N}_\mathsf{X}}
\newcommand{\rob}{\varrho_\mathrm{b}}
\newcommand{\sob}{\sigma_\mathrm{b}}
\newcommand{\celsius}{^\circ\mathrm{C}}

\newtheorem{theorem}{Theorem}

\newtheorem{corollary}{Corollary}

\theoremstyle{definition}
\newtheorem{remark}{Remark}

\begin{document}
	\latintext
	\title{Splay ostracism on ferroelectric nematic shells}
	\author{Epifanio G. Virga}
	\email{eg.virga@unipv.it}
	\affiliation{Department of Mathematics, University of Pavia, Via Ferrata 5, 27100 Pavia, Italy}

	\date{\today}

	\begin{abstract}
\emph{Ostracism} was the practice in use in Athens and other ancient Greek cities to send into exile citizens whose influence was deemed to be dangerous. Quite similarly, \emph{splay} ostracism is here the approximation that bans splay distortion from thin films of ferroelectric nematic liquid crystals on a curved shell in space. This reduces to a mere geometric problem the quest for equilibria of director \emph{vortexes} and the \emph{walls} isolating them, as splay ostracism requires the director field lines to be everywhere orthogonal to systems of \emph{geodesics} on the shell. Consequences of this theorem are explored for shells with Gaussian curvature of either sign. The resulting quasi-static dynamics reveals both cases where walls escape from the shell and cases where they are trapped on it, with possibly different reflections on the vortex dynamics.
	\end{abstract}

	\maketitle

\section{Introduction}\label{sec:intro}
Spontaneous polarization is a property encountered in \emph{solid} ferroelectric materials, but there the polarization vector $\pol$ is constrained by the symmetry of the elementary unit cell in the crystal lattice. What makes the newly discovered ferroelectric nematic \emph{liquid crystals} so special is that they are capable of flowing, and so $\pol$ is not only unconstrained, but also easily acted upon by external electric fields.\footnote{The earlier discovery of a ferroelectric \emph{smectic} phase dates back to the 1970's \cite{meyer:ferroelectric}. The appeal of the new findings is that they involve the \emph{nematic} phase.}

A number of reviews are already available with focus on diverse aspects of these new phases. If \cite{mandle:supramolecular} attempts to classify the materials exhibiting the ferroelectric nematic phase, \cite{sebastian:ferroelectric}
focuses on their theoretical modeling, while \cite{cruickshank:emergence} is more concerned with synthesis and structural characterization of these materials and \cite{song:updated} with their applications to optoelectronics and display technologies. In a similar vein, a review more oriented to electroptic applications can be found in \cite{khan:reviewing}. 

Beyond the new technological applications that their distinctive features promise to achieve, these new phases also pose theoretical challenges, which are still partly unmet.

Phenomenological theories have been proposed to describe the transition from the ordinary nematic phase to the ferroelectric one \cite{zou:extended,vats:bulk}, which however leave out the effects of depolarization that have so striking experimental manifestations \cite{caimi:fluid}. These effects include domain formation and concentration of bound charges on interfaces. To account properly for them calls for a \emph{nonlocal} variational theory. In Sect.~\ref{sec:theory}, we recall the existing theories and phrase them in a unifying formalism. There, we also introduce the notion of \emph{splay ostracism} as a draconian approximation to treat depolarizing effects.

In Sect.~\ref{sec:theorem}, which the heart of the paper,\footnote{The reader acquainted with the way depolarization effects are treated in the literature can jump ahead there.} we show how this approximation applied to ferrorelectric nematic \emph{shells}, which are thin films covering curved surfaces, reduces the determination of \emph{domain walls} around defect \emph{vortexes} to a purely geometric problem. This problem is solved exactly in Sect.~\ref{sec:walls} for two exemplary surfaces, with Gaussian curvature positive in one case and negative in the other. Estimating the energy of pinned vortexes through the \emph{length} of the wall separating them, one witnesses two  complementary phenomena: walls can either be trapped on local minima on the shell or escape altogether from it, exposing concealed \emph{virtual} defects as they evaporate.

The conclusions of this study are summarized in Sect.~\ref{sec:conclusions} and a number of ancillary mathematical results are collected in two closing appendices.

\section{Variational Theory}\label{sec:theory}
The first theory for ferroelectric nematics war arguably proposed long ago by Born \cite{born:anisotrope}, but it clashed with the lack of experimental evidence supporting the emergence of a polar, macroscopically ordered phase. Most liquid crystal molecules are indeed polar, with a permanent dipole along their long axis, but when the ordered nematic phase condenses the molecular dipoles average out to naught, while the average orientation of molecular axes, incarnated by the nematic director $\n$, becomes a well defined order parameter.

The existence of a new nematic \emph{ferroelectric} phase has recently been established on firm grounds \cite{chen:first}, after a number of earlier suggestive anticipations \cite{mandle:nematic,mandle:rational,nishikawa:fluid,mertelj:splay} in what appears to have been a truly collective pursuit \cite{lavrentovich:ferroelectric} (see also the penetrating perspective offered in \cite{sebastian:ferroelectric}).

The microscopic origin of the ferroelectric phase in nematic liquid crystals is not yet fully understood. A model was proposed in \cite{madhusudana:simple},\footnote{Other attempts followed in \cite{etxebarria:generalization,emelyanenko:emergence,szydlowska:ferroelectric}.} which attributes the tendency of rod-like molecules to be aligned with their permanent dipoles parallel to one another to the presence of \emph{unevenly} distributed charges  on the lateral surface of (globally neutral) molecules. If the charge picks at both ends are depressed relative to the charge picks in the middle, the orientation of rods with \emph{parallel} dipoles can be energetically favourable when the inter-rod separation is sufficiently small. On the contrary, for evenly distributed charge picks, the \emph{antiparallel} orientation of dipoles is instead favoured when the rods are sufficiently close to one another in the side-by-side configuration, whereas the parallel configuration is favoured in the bottom-on-top configuration, as predicted by classical electrostatics in the absence of surface charge modulation.\footnote{This is why steric interactions were predicted to prompt a ferroelectric phase in \emph{disk-like} molecules (with permanent dipole orthogonal to the disk), as these would be close-packed more easily in the bottom-on-top configuration than in the side-by-side one \cite{paik:flexoelectricity,bisi:steric}. However, the newly discovered ferroelectric phases are \emph{not} stabilized by steric interactions.}  

In thermotropic liquid crystals, the ferroelectric nematic phase, if existing, is established upon cooling from the standard apolar nematic phase. Some materials also exhibit a distinct, intermediate phase that, which acts as a precursor of the polar  phase. Such a precursory phase is known under many names in the literature \cite{sebastian:ferroelectric-ferroelastic,mandle:twist,gleeson:freedericksz,emelyanenko:emergence,chen:smectic,rupnik:antiferroelectric}; it is called $\NX$ in \cite{fagg:elastic}, which, as most papers, also uses $\NF$ for ferroelectric nematics and $\mathsf{N}$ for standard ones. The $\NX$ phase is characterized by an antiferroelectric organization of the macroscopic polarization, which acquires a mostly splayed spatial arrangement \cite{rosseto:theory,mandle:twist,selinger:director}. A model is proposed in \cite{fagg:elastic} to explain why splay is favoured in the $\NX$ phase: the key factor to understand such an anomalous proliferation of splay is \emph{flexoelectricity} \cite{meyer:structural}.

Here, we shall not be concerned with how the $\NF$ phase is heralded and established from the standard apolar phase. We shall rather imagine the material placed in in the deep $\NF$ phase, and we shall study its elastic and electrostatic properties. We shall also assume that in the condensed polar phase the polarization vector $\pol$ is parallel to the nematic director $\n$,
\begin{equation}
\label{eq:P_n}
\pol=P_0\n,
\end{equation}
where $P_0>0$ is a (constant) polarization strength.\footnote{The possibility that $\pol$ be \emph{not} aligned to $\n$ is also contemplated in \cite{paik:flexoelectricity}.} For definiteness, we recall from \cite{chen:first} that for $\mathrm{RM}734$, a typical $\NF$ material, $P_0\approx10^{-1}\,\mathrm{C/m^2}$.\footnote{The phase sequence of $\mathrm{RM}734$ is $\mathsf{I}-(188\,\celsius)\rightarrow\-\mathsf{N}-(133\,\celsius)\rightarrow\NF-(84\,\celsius)\rightarrow\mathsf{Cr}$, where $\mathsf{I}$ stands for isotropic and $\mathsf{Cr}$ for crystal \cite{chen:first}.}

As shown experimentally in \cite{chen:first}, a standard nematic phase with director $\n$,  upon entering the $\NF$ phase, can develop a polarization $\pol$ in either directions along the average molecular orientation. Once the $\NF$ phase is established, according to \eqref{eq:P_n}, we shall identify the nematic director with the unit vector associated with $\pol$, which thus becomes the leading order parameter. In uniformly oriented samples, at the $\mathsf{N}\rightarrow\NF$ transition, the material typically develops \emph{domains} of uniform orientation separated by domain \emph{walls}, across which $\pol$ makes a $\pi$-turn. By applying an external electric field $\Ele$ along the common orientation of the $\pol$ domains, one can easily induce domains with $\pol$ parallel to $\Ele$ to expand and domains with $\pol$ opposite to $\Ele$ to shrink \cite{chen:first}. 

Domain walls are thin regions in three-dimensional space where $\pol$ varies hugely, giving rise to a large concentration of \emph{bound charges}.
As we shall see, such a space distribution of  charges  will also entail an extra electrostatic cost outside domain walls.

\subsection{Free energy}\label{sec:energy}
In their deep condensed phase, ferroelectric nematics bear essentially two types of energy, elastic and electrostatic.
The density (per unit volume) of the former is given at a point $\x$ in space by the classical Oseen-Frank formula \cite{oseen:theory,frank:theory},
\begin{equation}\label{eq:Oseen-Frank}
	W_\mathrm{OF}(\x):=\tfrac12K_{11}(\diver\n)^2+\tfrac12K_{22}(\n\cdot\curl\n)^2+ \tfrac12K_{33}|\n\times\curl\n|^2+K_{24}[\tr(\nablas\n)^2-(\diver\n)^2],
\end{equation}
where $K_{11}$, $K_{22}$, $K_{33}$, and $K_{24}$ are the \emph{splay}, \emph{twist}, \emph{bend}, and \emph{saddle-splay} elastic constants.\footnote{Here $\tr$ denotes the trace of a second-rank tensor: in Cartesian components, $\tr(\nablas\n)^2=n_{i,j}n_{j,i}$, with the usual convention of summing over repeated indices.} In \eqref{eq:Oseen-Frank}, the nematic director $\n$ is related to $\pol$ through \eqref{eq:P_n}.

We know from the classical theory of electrostatics (see, for example, Sect.\,4.3 of \cite{jackson:classical}) that the bound charges induced by a non-inform polarization $\pol$ has (volume) density
\begin{equation}
	\label{eq:bound_charges_volume}
	\rob=-\diver\pol. 
\end{equation}	
Thus, in the presence of screening ions, following \cite{kumari:ferroelectric,lavrentovich:twist,paik:flexoelectricity,paik:spontaneous}, we give the polarization self-interaction energy density due to the distribution of bound charges \eqref{eq:bound_charges_volume} the Coulomb-Debye formula,
\begin{equation}\label{eq:Coulomb_Debye}
	W_\mathrm{CD}(\x):=\frac{1}{4\pi\epsilon}\int_{\body}\frac{\rob(\x)\rob(\y)\ex^{-|\y-\x|/\Lambda}}{|\y-\x|}\dd V(\y),
\end{equation}
where $\epsilon$ is the dielectric constant,\footnote{The dielectric constant $\epsilon$ is often written as $\epsilon=\epsilon_0\epsilon_\mathrm{r}$, where $\epsilon_0$ is the dielectric constant in vacuum and $\epsilon_\mathrm{r}$ is the \emph{relative} dielectric permittivity. As recalled in \cite{lavrentovich:deformed}, for ferroelectric nematics, $\epsilon_\mathrm{r}\sim10$.} $\Lambda$ is the Debye screening length,\footnote{$\Lambda$ depends on both temperature and ion concentration, $\Lambda=\sqrt{\epsilon k_\mathrm{B}T/(Ze)^2n}$, where $k_\mathrm{B}$ is the Boltzmann constant, $T$ is the absolute temperature, $Z$ the ion valence, $e$ the electron charge, and $n$ the number concentration of free ions in the material.} $V$ denotes the volume measure, and $\body$ is the region in space occupied by the material. Here, for simplicity, we shall take $\epsilon$ as both constant and isotropic.

A number of remarks are in order about $W_\mathrm{CD}$.

First, unlike $W_\mathrm{OF}$, $W_\mathrm{CD}$ is \emph{nonlocal} in nature, as the electrostatic energy at point $\x$ depends on the bound charge density $\rob$ not only at points nearby, but at \emph{all} points where bound charges are present; clearly, these latter have an influence that fades away with distance, but it goes far beyond what local gradients could estimate. Local approximations of $W_\mathrm{CD}$ featuring gradients of $\rob$ can be obtained, but they are only meaningful in the limit as $\Lambda\to0$.\footnote{Which  we are not interested in exploring. On the other hand, the limit as $\Lambda\to\infty$ corresponds to a fully unscreened, long-range Coulomb interaction, which is more difficult to handle.} 

Second, this approach to the electrostatics of polarized media in entirely based on the Coulomb potential generated by bound charges; free charges are neglected, only their screening effect is accounted for.

Third, as already anticipated in \cite{khachaturyan:development}, for $\Lambda$ sufficiently large, a natural \emph{twist} can spontaneously develop in the ground state of a polar nematic confined to an indefinite cylinder of radius $R$ with director  orthogonal to the cylinder's axis; the electrostatic energy and its nonlocal nature are responsible for this instability. For
\begin{equation}
	\label{eq:Lambda_c}
	\Lambda>\Lambda_\mathrm{c}:=\sqrt[3]{\frac{4K_{22}\epsilon R}{P_0^2}},
\end{equation}
$\n$ (and thus $\pol$) acquires a helical arrangement precessing along the cylinder's axis much as in a chiral nematic liquid crystal \cite{paik:flexoelectricity,paik:spontaneous}.\footnote{Correspondingly, the helical pitch in equilibrium is $p=2\pi\Lambda\Lambda_\mathrm{c}/\sqrt{\Lambda^2-\Lambda_\mathrm{c}^2}$. As appropriately noted in \cite{paik:flexoelectricity}, $p$ depends on the radius $R$ of the cylinder through $\Lambda_\mathrm{c}$; this is the subtle difference between the twisted ground state of polar nematics and that of chiral nematics. The helical pitch of the latter is a genuine, size-independent material property.}

In the following, we shall assume that \eqref{eq:Lambda_c} applies with $R$ interpreted as a typical size of the region $\body$.

Equation \eqref{eq:bound_charges_volume} for the density of bound polarization charges results from integrating the dipolar term in the multipolar expansion of the electrostatic potential over the region $\body$ occupied by the medium. When the latter is bounded, the very same reasoning shows that $\pol$, even if uniform in space, induces a distribution of uncompensated bound charges on the bounding surface $\partial\body$ with (area) density
\begin{equation}\label{eq:bound_charge_surface}
	\sob=\pol\cdot\normalB,
\end{equation}
where $\normalB$ is the \emph{outer} unit normal to $\partial\body$.

More generally, if $\pol$ is discontinuous in $\body$ along a two-dimensional \emph{jump set} $\jumpset$, which we identify with a regular surface, bound charges also arise on $\jumpset$ with density
\begin{equation}
	\label{eq:bound_charges_jumpset}
	\sob=\jump{\pol}\cdot\normal_{12},
\end{equation}
where $\jump{\pol}:=\pol_1-\pol_2$ and $\normal_{12}$ is the unit normal to $\jumpset$ oriented from the side upon which $\pol=\pol_1$ to the side upon which $\pol=\pol_2$. Clearly, equation \eqref{eq:bound_charge_surface} agrees with \eqref{eq:bound_charges_jumpset}; the former follows from the latter by setting $\pol_2=\zero$ and identifying $\pol_1$ with $\pol$. Letting $\partial\body$ be formally included in $\jumpset$, we shall use \eqref{eq:bound_charges_jumpset} as the general expression for the uncompensated bound charge density.

A surface $\wall\subset\body$ where $\pol$ suffers a jump is the simplest realization of a domain wall. The elastic distortion involved in turning $\pol$ from one side value, $\pol_1$, to the other, $\pol_2$, is squeezed in $\wall$ and the easiest way to account for it is to assume that it is proportional to the area of $\wall$, $w_0A(\wall)$, where $w_0>0$ is a phenomenological parameter and $A$ denotes the area measure. Henceforth, this will be our way of representing domain walls in ferroelectric liquid crystals; accordingly, we shall set $\jumpset=\wall\cup\partial\body$. 

While the elastic energy is simply written as 
\begin{equation}
\label{eq:free_energy_elastic}
	\freela[\n]:=\int_{\body}W_\mathrm{OF}(\x)\dd V(\x),
\end{equation}
where $W_\mathrm{OF}$ is as in \eqref{eq:Oseen-Frank}, the electrostatic polarization energy has a more composite form. It contains the bulk energy, $\freebulk$, which results from the integration over $\body$ of $W_\mathrm{CD}$ in \eqref{eq:Coulomb_Debye},
\begin{subequations}\label{eq:free_energy_pol}
\begin{equation}
	\label{eq:free_energy_pol_bulk}
	\freebulk[\n]:=\frac12\frac{1}{4\pi\epsilon}\int_{\body}\int_{\body}\frac{\rob(\x)\rob(\y)\ex^{-|\y-\x|/\Lambda}}{|\y-\x|}\dd V(\x)\dd V(\y),
\end{equation}
where $\n$ is related to $\pol$ through \eqref{eq:P_n}, the domain wall energy, including the contribution from the boundary,
\begin{equation}
	\label{eq:free_energy_pol_wall}
	\freewall[\n,\wall]:=	\frac12\frac{1}{4\pi\epsilon}\int_{\jumpset}\int_{\jumpset}\frac{\sob(\x)\sob(\y)\ex^{-|\y-\x|/\Lambda}}{|\y-\x|}\dd A(\x)\dd A(\y),
\end{equation}
and an energy accounting for the interaction between bulk and surface bound charges,
\begin{equation}\label{eq:free_energy_pol_int}
	\freeint[\n,\wall]:=\frac12\frac{1}{4\pi\epsilon}\int_{\jumpset}\int_{\body}\frac{\sob(\x)\rob(\y)\ex^{-|\y-\x|/\Lambda}}{|\y-\x|}\dd A(\x)\dd V(\y).
\end{equation}
\end{subequations}
In equations \eqref{eq:free_energy_pol}, the correcting factor $\frac12$ avoids double counting.

The total electrostatic polarization free energy is then $\freepol[\n,\wall]:=\freebulk[\n]+\freewall[\n,\wall]+\freeint[\n,\wall]$ and the total free energy becomes $\free[\n,\wall]:=\freela[\n]+\freepol[\n,\wall]$. The explicit dependence on $\wall$ in $\free$ makes it a \emph{free discontinuity} functional, as neither the surface $\wall$ is known a priori nor are the values of $\pol$ across it; they must both be determined by minimizing $\free$.

\subsection{Splay ostracism}\label{sec:ostracism}
Even leaving aside external actions, such as those prompted by applied electric fields, finding the minimizers of $\free$ is a formidable task, mainly because of the nonlocal nature of the functional. We call \emph{spay ostracism} a drastic approximation that will tame the nonlocality of $\free$, while leaving untouched its free discontinuity nature.

For $\Lambda>\Lambda_\mathrm{c}$, since $\freepol$ is expected to prevail over $\freela$, we may conjecture that $\free$ is minimized whenever $\freepol$ is. Leaving boundary conditions aside, this is realized by setting
\begin{equation}
	\label{splay_ostracism_charges}
	\rob\equiv0\quad \text{in}\quad\body\setminus\wall\qquad\text{and}\qquad\sob\equiv0\quad\text{on}\quad\wall.
\end{equation} 
By \eqref{eq:P_n}, \eqref{eq:bound_charges_volume}, and \eqref{eq:bound_charges_jumpset}, \eqref{splay_ostracism_charges} entails that minimizers $(\n,\wall)$ of $\free$ are minimizers of $\freela[\n]+w_0A(\wall)$, subject to
\begin{equation}
	\label{eq:splay_ostracism_director}
	\diver\n\equiv0\quad\text{in}\quad\body\setminus\wall\qquad\text{and}\qquad\jump{\n}\cdot\normalW\equiv0\quad\text{on}\quad\wall,
\end{equation}
where $\normalW$ is any orientation of the unit normal to $\wall$. The domains with uniform twist of the director, separated by thin walls, observed experimentally in \cite{lavrentovich:twist} seem to support the validity of both constraints in \eqref{eq:splay_ostracism_director}.

Within this approximation, in the following section, we shall conduct the search for energy minimizers in the special case of ferroelectric nematic shells, which are thin films of ferroelectric nematics covering a solid surface.




\section{Splay ostracism theorem}\label{sec:theorem}
A convincing heuristic argument presented in \cite{hedlund:freely} suggests that when a ferroelectric nematic is confined within a \emph{thin} layer (typically with thickness $h\approx4\text{-}10\,\mu\mathrm{m}$) placed on an \emph{isotropic} fluid (typically glycerin), a strong effective \emph{planar degenerate} anchoring acts on $\pol$ (and so also on $\n$) that prevents $\pol$ from having a component along the layer's normal. This anchoring results from neutralizing the depolarizing electric field that would be generated across the layer by the bound charges distributed on the layer's bounding surfaces, were $\pol$ allowed to have a component along the layer's normal.

In our mathematical model, we shall treat a ferroelectric nematic \emph{shell} as a smooth material surface $\surface$ (at least of class $C^2$) oriented by its unit normal field $\normal$. The nematic director $\n$, oriented like $\pol$, is constrained to be everywhere tangent to $\surface$,
\begin{equation}
	\label{eq:n_orthogonal_nu}
	\n\cdot\normal\equiv0,
\end{equation}
and free otherwise.\footnote{To be coherent with the notation employed in Sec.~\ref{sec:theory}, we should have denoted $\nu_{\surface}$ the unit normal to $\surface$. Dropping the subscript in $\nu_{\surface}$ avoids typographical clutter  without renouncing clarity.}

\begin{figure}
	\centering
	\includegraphics[width=.33\linewidth]{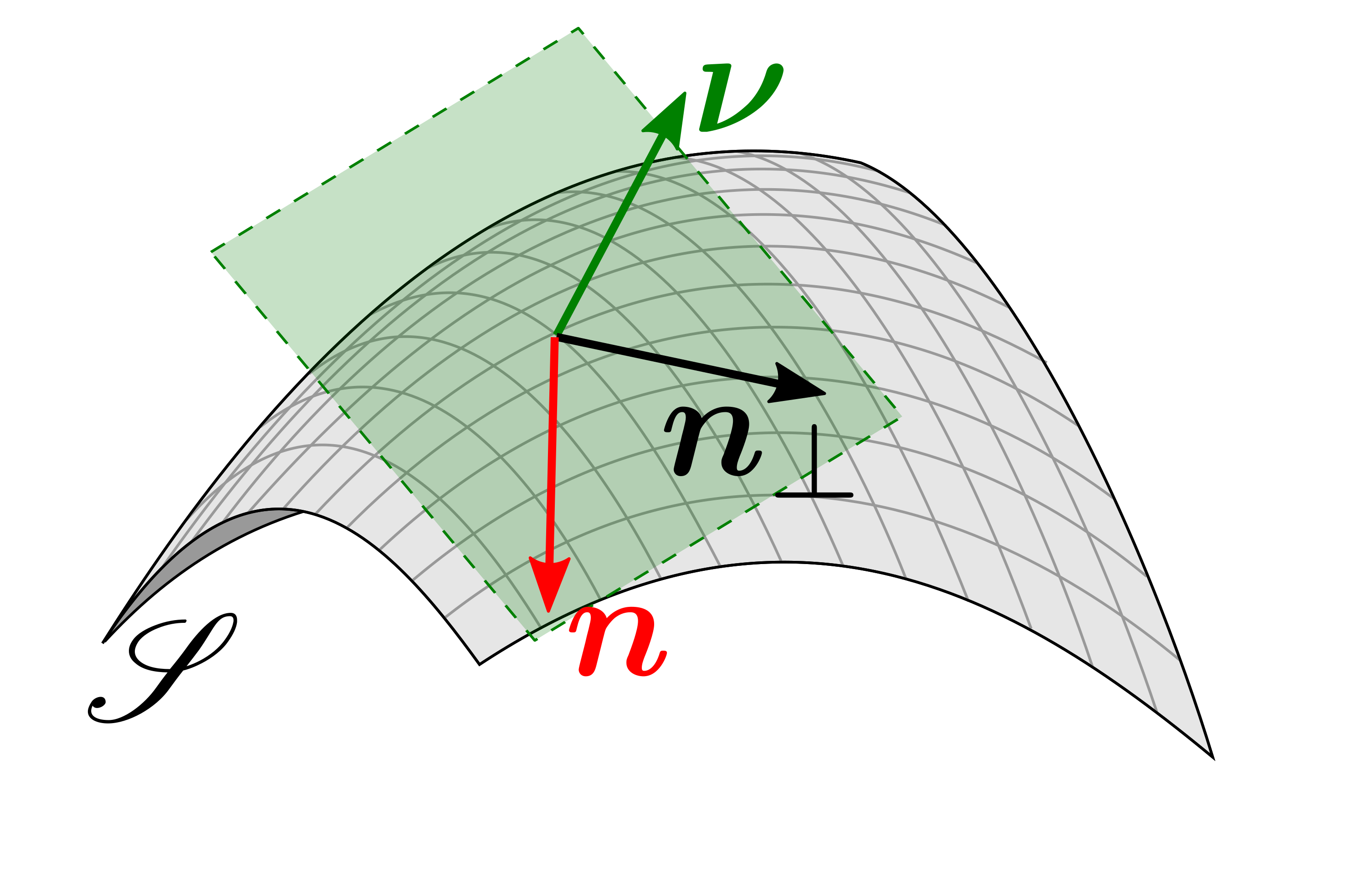}
	\caption{The moving frame $\framen$ gliding on $\surface$ according to the laws \eqref{eq:glidind_laws}.}
	\label{fig:moving_frame}
\end{figure}

For any smooth, tangent unit vector field $\n$ (at least of class $C^2$) prescribed on $\surface$, letting $\nperp:=\normal\times\n$, we introduce the \emph{moving frame} $\framen$ that glides on $\surface$ according to the (structural) equations
\begin{subequations}\label{eq:glidind_laws}
\begin{align}[left={\empheqlbrace}]
		&\nablas\n=\nperp\otimes\cv+\normal\otimes\dv_1,\label{eq:glidind_laws_1}\\
		&\nablas\nperp=-\n\otimes\cv+\normal\otimes\dv_2,\label{eq:glidind_laws_2}\\
		&\nablas\normal=-\n\otimes\dv_1-\nperp\otimes\dv_2,\label{eq:glidind_laws_3}
\end{align}
\end{subequations}
see also Fig.~\ref{fig:moving_frame}.
Here $\nablas$ denotes the \emph{surface} gradient, $\cv$ is the \emph{spin} connector, and $\dv_1$, $\dv_2$ are the \emph{curvature} connectors, which obey the identity
\begin{equation}
	\label{eq:d_identity}
	\dv_1\cdot\nperp=\dv_2\cdot\n
\end{equation}
to guarantee the symmetry of the curvature tensor $\nablas\normal$ (see, for example, \cite{pedrini:surface}).

The \emph{intrinsic}\footnote{In differential geometry, are called intrinsic the properties of a surface that are independent of its embedding in three-dimensional
space, and extrinsic those properties that depend on that \cite[p.\,11]{needham:visual}.} 
scalar measures of surface distortion for $\n$ are extracted from the \emph{covariant} gradient
\begin{equation}
	\label{eq:covariant_gradient}
	\nablac\n:=\proj\nablas\n=\nperp\otimes\cv,
\end{equation} 
where $\proj$ is the projection onto the local tangent plane $\tplane$, and use has been made of \eqref{eq:glidind_laws_1}.\footnote{The covariant gradient reduces $\nablas\n$ to an action onto $\tplane$ , and so it can be considered as being totally inherent to the surface (or intrinsic), insensitive to the surrounding space \cite[p.\,240]{needham:visual}.}
These measures are
\begin{equation}
\label{eq:distorion_measures}
\begin{cases}
	S:=\divc\n=\cv\cdot\nperp,\\
	B:=\n\times\curlc\n\cdot\nperp=-\curlc\n\cdot\normal=-\cv\cdot\n,\\
	T:=\n\cdot\curlc\n\equiv0,
\end{cases}
\end{equation}
where $\curlc$ denotes the covariant curl; these are the measures of splay, bend, and twist, respectively. Equations \eqref{eq:distorion_measures} show that a tangential nematic field bears \emph{no} twist,\footnote{Since $\tr(\nablac\n)^2-(\divc\n)^2=0$, as seen immediately from \eqref{eq:covariant_gradient}, no intrinsic saddle-splay can either be associated with a tangential nematic field $\n$.} and its splay and bend are the components in the moving frame $\framen$ of the spin connector.\footnote{In particular, as shown in \cite{pedrini:surface}, this justify calling \emph{uniform} a nematic field tangential to $\surface$ whose spin connector $\cv$ has constant components in the moving frame $\framen$.}

For a ferroelectric nematic shell, a domain wall $\wall$ becomes a line $\Line$, which we shall still call a wall, and by precisely the same reasoning that in Sec.~\ref{sec:theory} led us to justify the splay ostracism approximation, we require that
\begin{equation}
	\label{eq:splay_ostracism_shell}
	S\equiv0\quad\text{on}\quad\surface\setminus\Line\quad\text{and}\quad\jump{\n}\cdot\normalL\equiv0\quad\text{on}\quad\Line,
\end{equation}
where $\normalL$ is a conormal of $\Line$.\footnote{That is, a unit normal to $\Line$ on the local tangent plane $\tplane$ to $\surface$.}
By using the method presented in \cite{pedrini:surface}, we can prove the following theorem.
\begin{theorem}[Splay ostracism]\label{thm:splay}
A smooth field $\n:\surface\to\sphere$ for which $\n\cdot\normal\equiv0$ is such that $S\equiv0$, if and only if 
\begin{equation}
	\label{eq:theorem_statement}
	\n\cdot\tangent\equiv0,
\end{equation}
where $\tangent$ is a unit tangent field to a \emph{system} of geodesics of $\surface$.\footnote{These are geodesic lines that together with their orthogonal lines constitute a local system of coordinates.}
\end{theorem}
\begin{proof}
	The proof is deferred to Appendix~\ref{sec:proof}.
\end{proof}
\begin{corollary}[Bisection rule]\label{cor:bisection}
Two fields $\n_1$ and $\n_2$ as in Theorem~\ref{thm:splay} are such that $\jump{\n}\cdot\normalL=0$ along a smooth line $\Line$ on $\surface$, if and only if the direction of the unit tangent $\bm{\tau}$ to $\Line$ \emph{bisects} the directions $\tangent_1$ and $\tangent_2$ of the geodesics of $\surface$ that convey $\n_1$ and $\n_2$ according to \eqref{eq:theorem_statement}. 
\end{corollary}
\begin{proof}
	Since $\n_1$ and $\n_2$ are orthogonal to $\tangent_1$ and $\tangent_2$, respectively, the direction of $\normalL$ bisects the directions of the former pair, if and only if the direction of $\bm{\tau}$ bisects the directions of the latter.
\end{proof}
A number of remarks are suggested by these results.
\begin{remark}
	\label{rmk:vortex}
	Since for polar nematics $\n$ is \emph{not} equivalent to $-\n$, \emph{point} defects of $\n$ may only  have an \emph{integer} topological charge, $m\in\mathbb{Z}$ \cite{lavrentovich:deformed}. We call \emph{vortex} a point defect of topological charge $m=+1$. Vortexes may differ by the sense of circulation  (clockwise or anticlockwise) around the normal $\normal$, while retaining the same topological charge. Splay ostracism on $\surface$ requires the field lines of a point defect  to be orthogonal to geodesics. Thus, point defects compatible with splay ostracism can only be vortexes. Splay ostracism  acts as a selection rule for admissible point defects, thus revealing the energetic nature hidden in  its inception. 
\end{remark}
\begin{remark}
	\label{rmk:circles}
	In particular, if $\surface$ is part of a plane, splay ostracism requires the field lines of a vortex to be concentric \emph{circles}, as all geodesics emanating from a point are \emph{straight} lines. No  point defect other than a vortex could be compatible with splay ostracism in the plane. For completeness, in Appendix~\ref{sec:caustic}, we also show how systems of geodesics in the plane can be used to construct \emph{caustics}, where defects are densely distributed in a region of plane with positive area.
\end{remark}
\begin{remark}
	\label{rmk:conical_sections}
	A pencil of \emph{parallel} straight lines in the plane is a defect free system of geodesics; correspondingly, we call \emph{uniform} a director with parallel field lines. Thus, leaving caustics aside, splay ostracism in the plane  dictates that a wall $\Line$ can separate either two vortexes or a vortex and a uniform director field.\footnote{In principle, two different uniform director fields could  also be separated by $\Line$; this case will not be treated explicitly, as it can easily be retrieved as a limiting case of the two contemplated here.} In either case, the bisection rule prescribes $\Line$ to be a \emph{conical section}, a hyperbola or a parabola, respectively, as predicted and beautifully confirmed experimentally in \cite{lavrentovich:twist,hedlund:freely}. 
\end{remark}
\begin{remark}
	\label{rmk:bend_ostracism}
	By \eqref{eq:glidind_laws_2}, the unit vector field $\nperp$, which is tangential to $\surface$ precisely as $\n$, has covariant gradient
	\begin{equation}
		\label{eq:covariant_gradient_perp}
		\nablac\nperp=-\n\otimes\cv,
	\end{equation}
	which plays for $\nperp$ the same role played by \eqref{eq:covariant_gradient} for $\n$. Thus, precisely as in \eqref{eq:distorion_measures}, the intrinsic measures of distortion for $\nperp$ read as
	\begin{equation}
		\label{eq:distorion_measures_perp}
		\begin{cases}
			S_\perp:=\divc\nperp=-\cv\cdot\n=-B,\\
			B_\perp:=-\nperp\times\curlc\nperp\cdot\n=-\cv\cdot\nperp=-S,\\
			T_\perp:=\nperp\cdot\curlc\nperp\equiv0.
		\end{cases}
	\end{equation}
These equations show that a field $\n$ obeys $S\equiv0$, whenever $\nperp$ obeys $B_\perp\equiv0$. Thus, if $\n$ is a splay ostracism field then $\nperp$ is a \emph{bend ostracism} field. By Theorem~\ref{thm:splay}, a bend ostracism field is characterized by being \emph{tangent} to a system of geodesics of $\surface$.
\end{remark}
In the following section, we shall construct explicitly systems of geodesics on curved surfaces with constant Gaussian curvature that convey field lines of vortexes and we shall investigate different configurations for the  walls allowed to separate them.

\section{Wall escaping and trapping}\label{sec:walls}
In this section, we flesh out the geometric construction of divergence free director fields on surfaces presented in Sec.~\ref{sec:theorem}. We shall concentrate on the length $L(\Line)$ of walls $\Line$ on a shell $\surface$, as we reckon that 
\begin{equation}
	\label{eq:free_energy_dominant}
	\free_\mathrm{dom}:=w_0hL(\Line)
\end{equation}
is the \emph{dominant} free energy whenever vortexes of $\n$ are \emph{pinned} on $\surface$. In \eqref{eq:free_energy_dominant}, $w_0>0$ is the strength of the wall energy introduced in Sec.~\ref{sec:energy} and $h$ measures the thickness of the material film comprising the shell.

We shall be interested in equilibrium configurations of $\Line$, for fixed vortexes. We shall show settings where $\Line$  either escapes from $\surface$, entraining a change in the defect landscape, or it remains trapped on it.  Wall escaping and trapping is seen to occur on surfaces $\surface$ with positive and negative Gaussian curvature, respectively: we shall work out the details for both $\surface$ a sphere and a pseudosphere.

\subsection{Spherical shell}\label{sec:sphere}
For definiteness, we shall take for $\surface$ the unit sphere $\sphere$ in three-dimensional space. For polar nematics, a theorem of Poincar\'e \cite{poincare:courbes} prescribed the total topological charge $M$ of defects of a director field $\n$ tangent to a surface of genus $g=0$, such as $\sphere$, to be $M=+2$ (see also \cite[p.\,206]{needham:visual}). Two vortexes of charge $m=+1$  thus conjure the least distorted configuration compatible with this topological restriction. Splay ostracism, in addition, requires $\n$ to be orthogonal to systems of geodesics on $\sphere$; these are pencils of \emph{great circles} meeting at \emph{antipodal} points.

\begin{figure}
		\centering
		\begin{subfigure}[c]{0.45\linewidth}
			\centering
			\includegraphics[width=.8\linewidth]{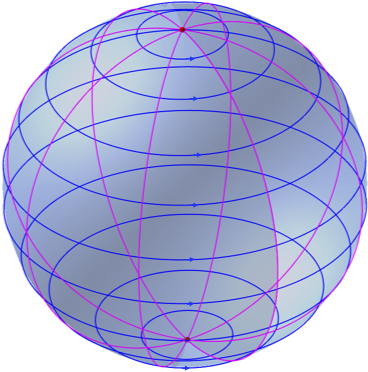}
			\caption{Vortexes at the poles.}
			\label{fig:antipodal_pole}
		\end{subfigure}
		\qquad
		\begin{subfigure}[c]{0.45\linewidth}
			\centering
	\includegraphics[width=.8\linewidth]{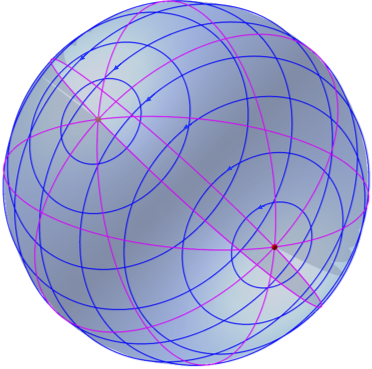}
	\caption{Vortexes on the equator.}
	\label{fig:antipodal_equator}
	\end{subfigure}
	\caption{Vortexes on $\sphere$: geodesics are magenta, while the field lines of $\n$ are blue; arrows indicate the orientation of $\n$. Relative to the outer normal $\normal$, $\n$ swirls clockwise in one vortex and anticlockwise in the other.}
	\label{fig:antipodal}
\end{figure}

Figure~\ref{fig:antipodal} shows two such arrangements of geodesics with vortexes at the poles and on the equator, respectively.
In each arrangement, the director field $\n$ is oriented clockwise relative to the outer unit vector $\normal$ to $\sphere$ at one vortex, and anticlokwise at the other. In accordance with the jump condition \eqref{eq:splay_ostracism_shell},
either of the vortexes in Fig.~\ref{fig:antipodal_pole} can coexist with an appropriate  vortex in Fig.~\ref{fig:antipodal_equator}.
We now construct all walls $\Line$ allowed in both cases of possible vortex coexistence: the director field will bear total topological charge $M=+2$, with one vortex at a pole and the other on the equator. We shall assume throughout that the vortexes are pinned on the surface, while the wall separating them can move to adjust its length.

To afford a detailed description of the admissible walls, we map $\sphere$ on the equatorial plane by the \emph{stereographic} projection, defined as
\begin{equation}\label{eq:stereographic_projection}
X=\frac{2x}{x^2+y^2+1},\quad Y=\frac{2y}{x^2+y^2+1},\quad Z=\frac{x^2+y^2-1}{x^2+y^2+1},
\end{equation}
where $(X,Y,Z)$ are the (Cartesian) coordinates of a point on $\sphere$ and $(x,y)$ the coordinates of its image (see also \cite[p.\,158]{needham:visual_complex}). This projection is effected by rays emanating from the north pole $(0,0,1)$ and maps the equator onto itself; it is \emph{conformal}, as it preserves angles cut by curves on $\sphere$.

\begin{figure}
	\centering
	\begin{subfigure}[c]{0.45\linewidth}
		\centering
		\includegraphics[width=\linewidth]{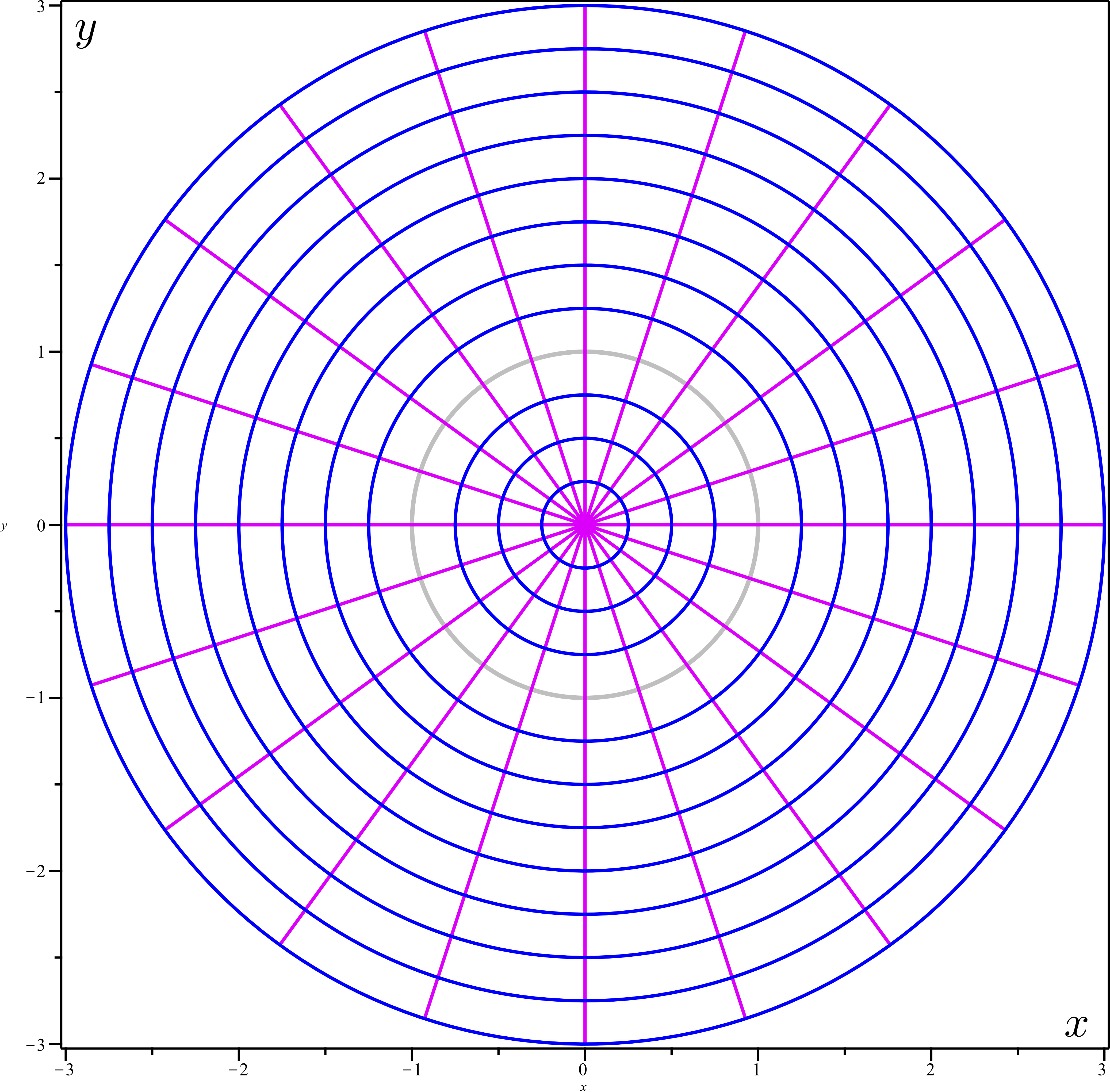}
		\caption{Straight (magenta) lines emanating from the image of the south pole $(0,0,-1)$ are images of geodesics of $\sphere$; the corresponding projection of field lines of $\n$ are orthogonal (blue) circles.}
		\label{fig:antipodal_pole_lines}
	\end{subfigure}
	\qquad
	\begin{subfigure}[c]{0.44\linewidth}
		\centering
		\includegraphics[width=\linewidth]{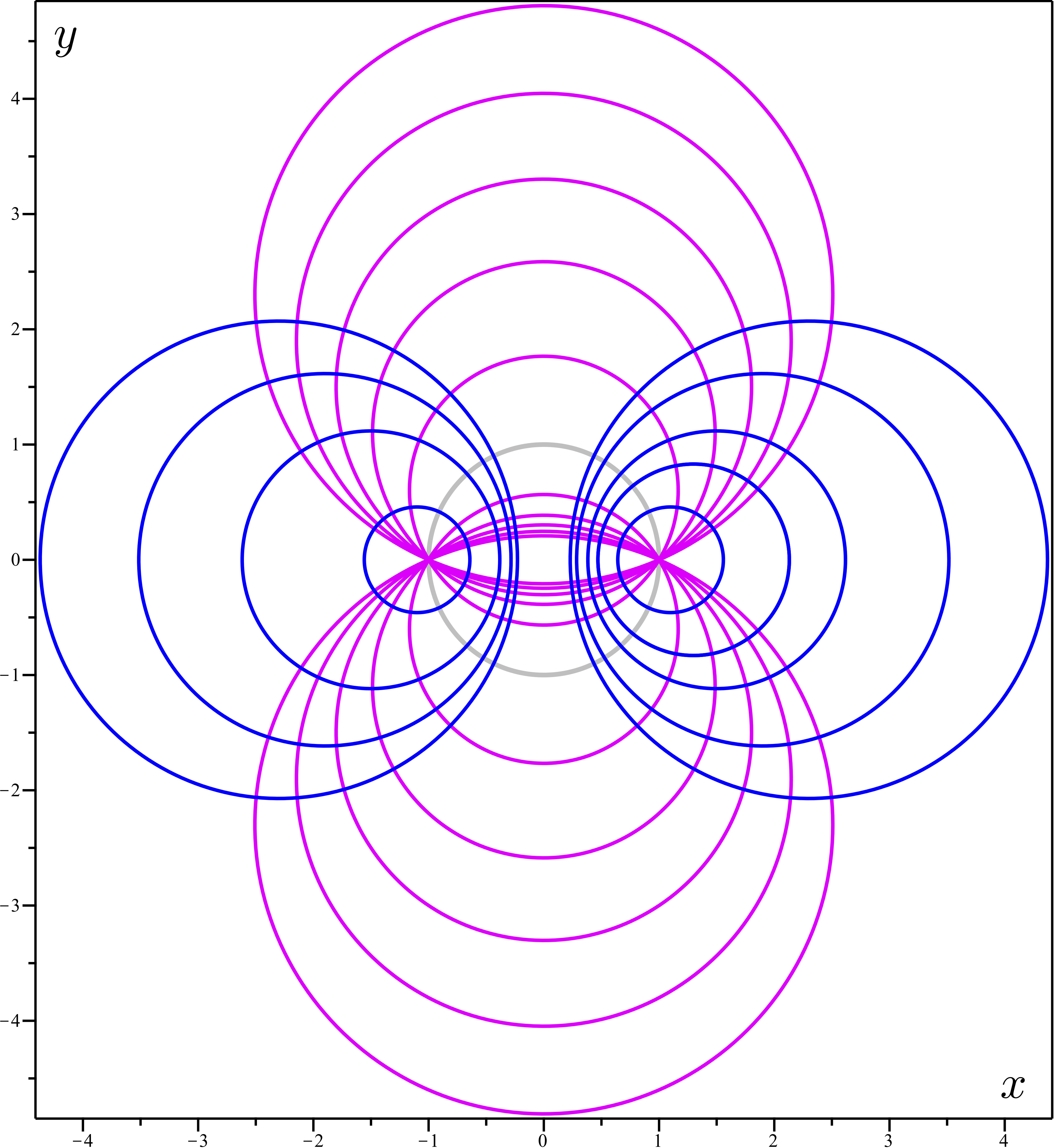}
		\caption{Pencils of orthogonal Apollonian circles; they are geodesics (magenta circle) and director field lines (blue circles).}
		\label{fig:antipodal_equator_lines}
	\end{subfigure}
	\caption{Stereographic projections according to \eqref{eq:stereographic_projection} of both geodesics and field lines depicted in Fig.~\ref{fig:antipodal}. In both panels the gray circle is the unit sphere's equator, which is mapped into itself by \eqref{eq:stereographic_projection}.}
	\label{fig:antipodal_lines}
\end{figure}

Figure~\ref{fig:antipodal_lines} shows the stereographic projections of both geodesics (magenta) and field lines (blue) depicted in Fig.~\ref{fig:antipodal}.
Geodesics and field lines in Fig.~\ref{fig:antipodal_equator_lines} are Apollonian circles; the former form an \emph{elliptic} pencil, while the latter form a \emph{hyperbolic} pencil  with \emph{limiting} points on the equator of $\sphere$.\footnote{In the plane, there are three types of pencils of coaxial circles: \emph{elliptic} (intersecting), \emph{hyperbolic} (non-intersecting), and \emph{parabolic} (tangent), see \cite[$\S$\,6.5]{coxeter:geometry}. Apollonius of Perga (ca. 260--190 BC) discovered that the locus of points whose distances from two fixed points, $A$ and $A'$, are in constant ratio $\mu>0$ is a \emph{circle}. When $\mu$ varies while $A$ and $A'$ remain fixed, the circles of Apollonius form a hyperbolic pencil with limiting points $A$ and $A'$. In this case,  Apollonian limiting points correspond to defects on the equator of $\sphere$.}

\begin{figure}
		\centering
		\includegraphics[width=.5\linewidth]{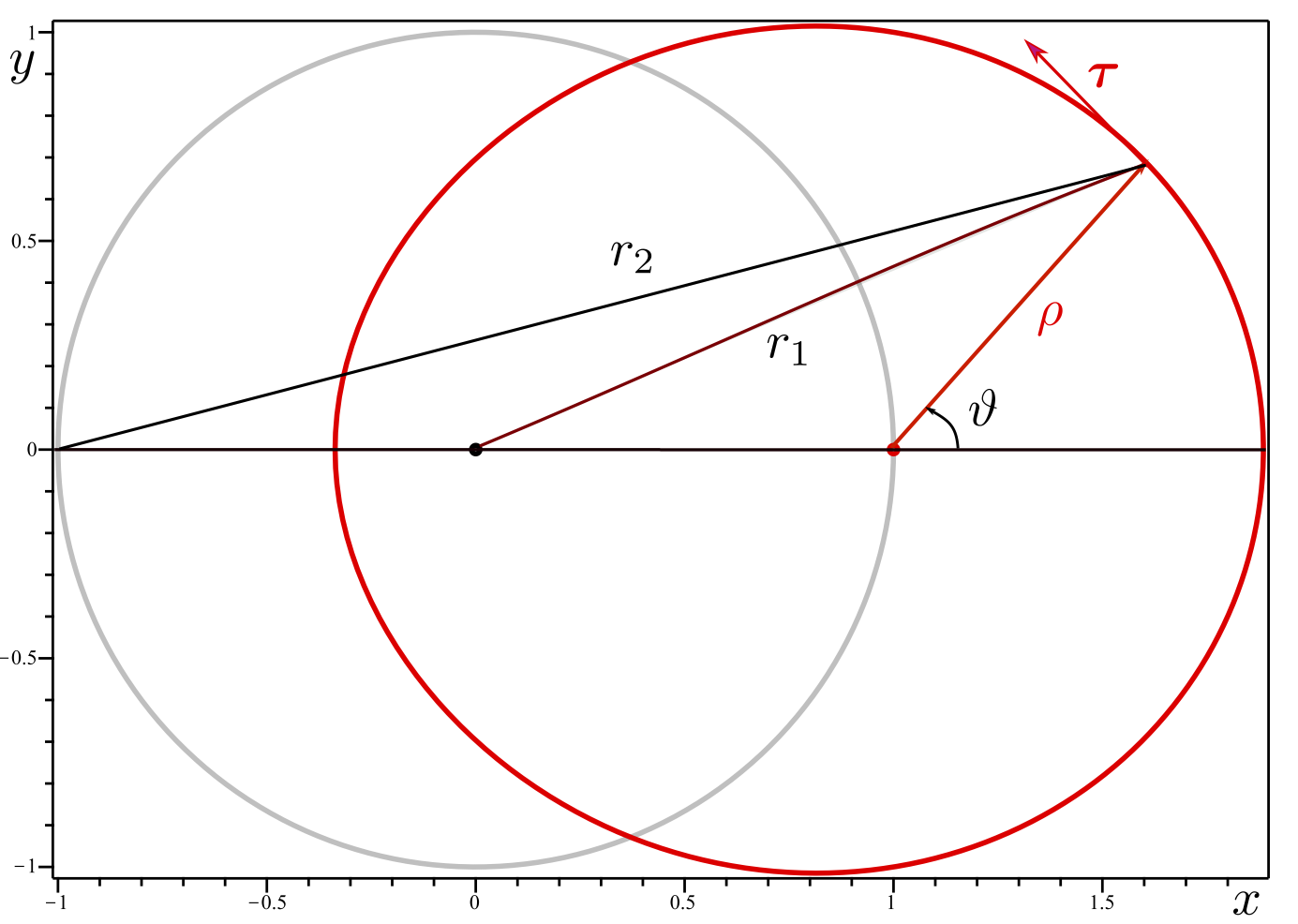}
		\caption{Stereographic projection of a wall (red curve) represented in polar coordinates $(\rho,\vt)$ by a smooth function $\rho(\vt)$. The unit tangent vector $\bm{\tau}$ is given by \eqref{eq:wall_tangent} in the polar frame $(\e_\rho,\e_\vt)$, while $r_1$ and $r_2$, expressed by \eqref{eq:wall_r_1_2}, are the distances of the point $(\rho(\vt),\vt)$ from the center of the equatorial circle and from the twin limiting point, respectively.}
		\label{fig:wall}
	\end{figure}

We now build an admissible wall around one limiting point; it will be described as a closed curve in the $(x,y)$ plane, the image through \eqref{eq:stereographic_projection} of a closed curve on $\sphere$. With a (slight) abuse of language, we shall denote both curves by $\Line$, as they can be identified with one another. In polar coordinates $(\rho,\vt)$ in the $(x,y)$ plane with pole at $x=1$, $y=0$, $\Line$ is represented by a smooth function $\rho=\rho(\vt)$, see Fig.~\ref{fig:wall}.
In the polar frame $(\e_\rho,\e_\vt)$, the unit tangent $\bm{\tau}$ to $\Line$ is given by
\begin{equation}\label{eq:wall_tangent}
	\bm{\tau}=\frac{1}{\sqrt{\rho^2+\rho'^2}}(\rho'\e_\rho+\rho\e_\vartheta),
\end{equation}	 
where a prime $'$ denotes differentiation with respect to $\vt$. Moreover, the distance $r_1$ between the current point $(\rho(\vt),\vt)$ of $\Line$ and the center of the equatorial circle and the distance $r_2$ between $(\rho(\vt),\vt)$ and the twin limiting point at $x=-1$, $y=0$ are given by
\begin{equation}
	\label{eq:wall_r_1_2}
	r_1=\sqrt{1+2\rho\cos\vt+\rho^2},\qquad
	r_2=\sqrt{4+4\rho\cos\vt+\rho^2}.
\end{equation}

We denote by $\nS$ and $\nA$ the director fields in the $(x,y)$ plane whose field lines are depicted in Fig.~\ref{fig:antipodal_pole_lines} and Fig.~\ref{fig:antipodal_equator_lines}, respectively; they are oriented coherently with the vortexes depicted in Fig.~\ref{fig:antipodal} around the poles and the antipodal pair on the equator corresponding to the Apollonian limiting points in Fig.~\ref{fig:antipodal_equator_lines}. In the $(x,y)$ coordinates, $\nS$ and $\nA$ are explicitly expressed as
\begin{equation}
	\label{eq:n_S_n_A_cartesian}
	\nS=\frac{-y\e_x+x\e_y}{\sqrt{x^2+y^2}},\qquad\nA=\frac{2xy\e_x+(y^2-x^2-1)\e_y}{\sqrt{4x^2y^2+(y^2-x^2-1)^2}},
\end{equation}
where $(\e_x,\e_y)$ denotes the Cartesian frame in the $(x,y)$ plane. Along the curve $\Line$ on the plane,
\begin{equation}
	\label{eq:wall_coordinates}
	\begin{cases}
		x=1+\rho(\vt)\cos\vt,\\
		y=\rho(\vt)\sin\vt,
	\end{cases}
\end{equation}	
and the fields $\nS$ and $\nA$ acquire the form
\begin{equation}
	\label{eq:n_S_n_A_polar}
\nS=\frac{1}{r_1}[\sin\vt\e_r+(\rho(\vt)+\cos\vt)\e_\vt],\qquad
		\nA=\frac{1}{r_2}[\rho(\vt)\sin\vt-(2+\rho(\vt)\cos\vt)\e_\vt].
\end{equation}

If $\nS$ lies one one side of $\Line$ while $\nA$ lies on the other, the jump condition in \eqref{eq:splay_ostracism_shell} reads equivalently as
\begin{equation}
	\label{eq:wall_jump_condition}
	\bm{\tau}\cdot(\nS+\nA)=0,
\end{equation}
which by \eqref{eq:n_S_n_A_polar} and \eqref{eq:wall_tangent} \label{sec:pseudo} becomes the following differential equation for $\rho(\vt)$,
\begin{equation}
	\label{eq:wall_equation}
	\frac{\dd\rho}{\dd\vt}\sin\vt(\rho r_1+r_2)+\rho[(\rho+\cos\vt)r_2-(2+\rho\cos\vt)r_1]=0,
\end{equation}
later also referred to as the \emph{wall equation}.
\begin{remark}
	\label{rmk:conformality}
	Equation \eqref{eq:wall_jump_condition}, which  involves planar vector fields, is equivalent to  \eqref{eq:wall_tangent} because the stereographic projection is conformal: the two equations thus embody the same bisection rule. 
\end{remark}	
\begin{remark}
	\label{rmk:symmetry}
	Clearly, \eqref{eq:wall_jump_condition} is symmetric under the exchange of $\nS$ and $\nA$. This implies that the same curve $\Line$ could enclose within either $\nA$ or $\nS$, leaving $\nS$ or $\nA$ without. For definiteness, we shall refer to the former as the direct case and to the latter as the inverse one.
\end{remark}
\begin{remark}
	\label{rmk:symmetry_shape}
	The wall equation \eqref{eq:wall_equation} is symmetric under the transformation $\vt\mapsto-\vt$, $\rho(\vt)\mapsto\rho(-\vt)$, and so all admissible walls are symmetric relative to the $x$ axis.
\end{remark}
\begin{remark}
	\label{rmk:circle}
	It is easy to check by direct inspection that there is a single constant solution of \eqref{eq:wall_equation}, that is, $\rho\equiv\sqrt{2}$. It represents a circle in the $(x,y)$ plane that is mapped by \eqref{eq:stereographic_projection} onto a great circle of $\sphere$ at $45^\circ$ with the equatorial plane, cutting symmetrically through the pairs of antipoldal defects in Fig.~\ref{fig:antipodal}.
\end{remark}
Since the line element $\dd\ell$ of $\Line$ on $\sphere$ is related to the line element $\dd s$ of $\Line$ on the equatorial plane by (see, for example, \cite[p.\,47]{needham:visual})
\begin{equation}\label{eq:line_elements}
	\dd\ell=\frac{2}{1+r_1^2}\dd s,
\end{equation}
the length $L$ of $\Line$ on $\sphere$ is given by
\begin{equation}
	\label{eq:sphere_length}
	L(\Line)=\int_0^\pi\frac{4\sqrt{\rho^2+\rho'^2}}{2+2\rho\cos\vt+\rho^2}\dd\vt,
\end{equation}
where use has been made of \eqref{eq:wall_r_1_2}, $\rho$ is a solution of \eqref{eq:wall_equation}, and a prime denotes differentiation with respect to $\vt$.

\begin{figure}
	\centering
	\begin{subfigure}[c]{0.33\linewidth}
		\centering
		\includegraphics[width=\linewidth]{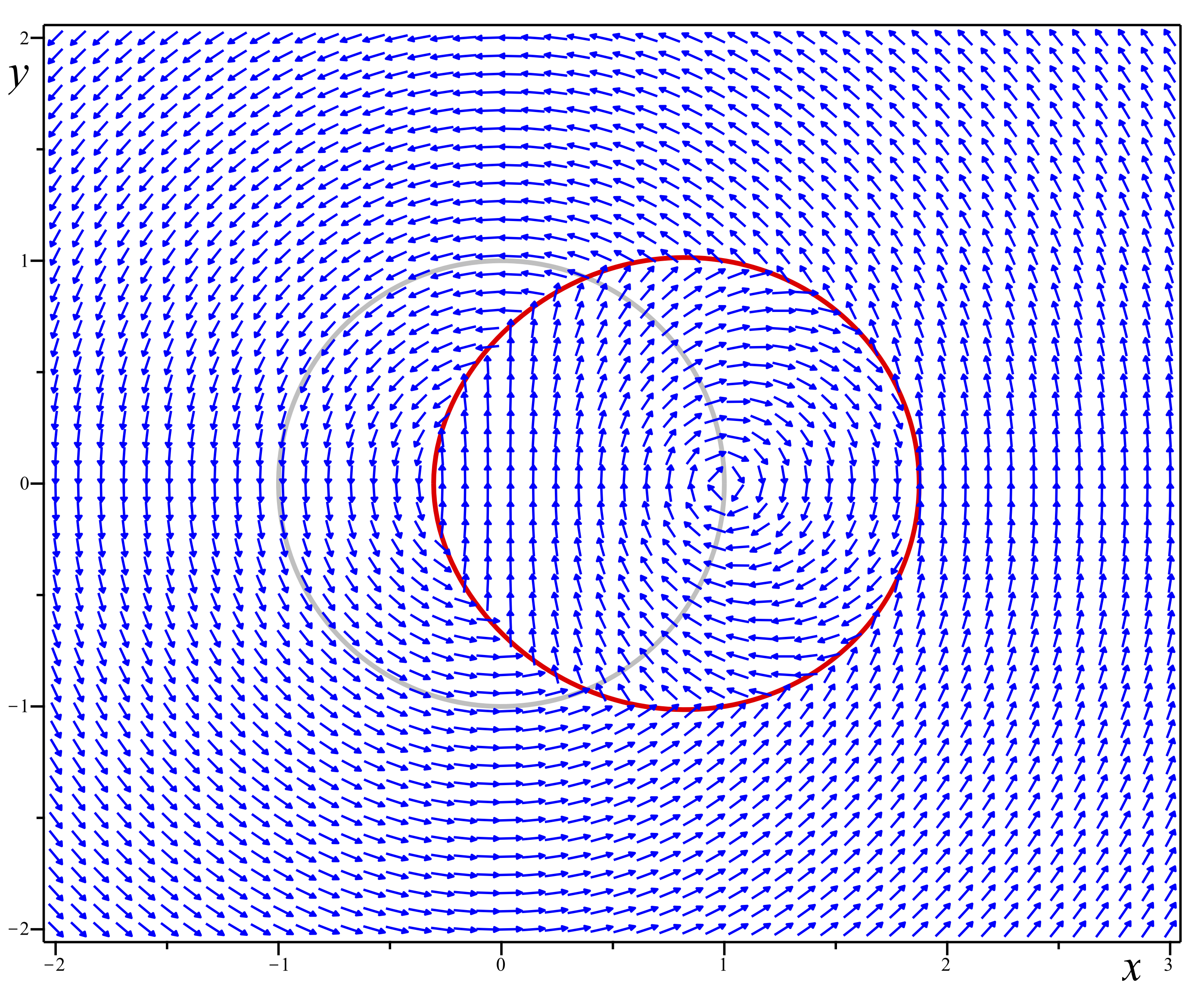}
		\caption{direct, $\rho_0=1$}
		\label{fig:sphere_director_rho0_1_direct}
	\end{subfigure}
	\qquad
	\begin{subfigure}[c]{0.33\linewidth}
	\centering
	\includegraphics[width=\linewidth]{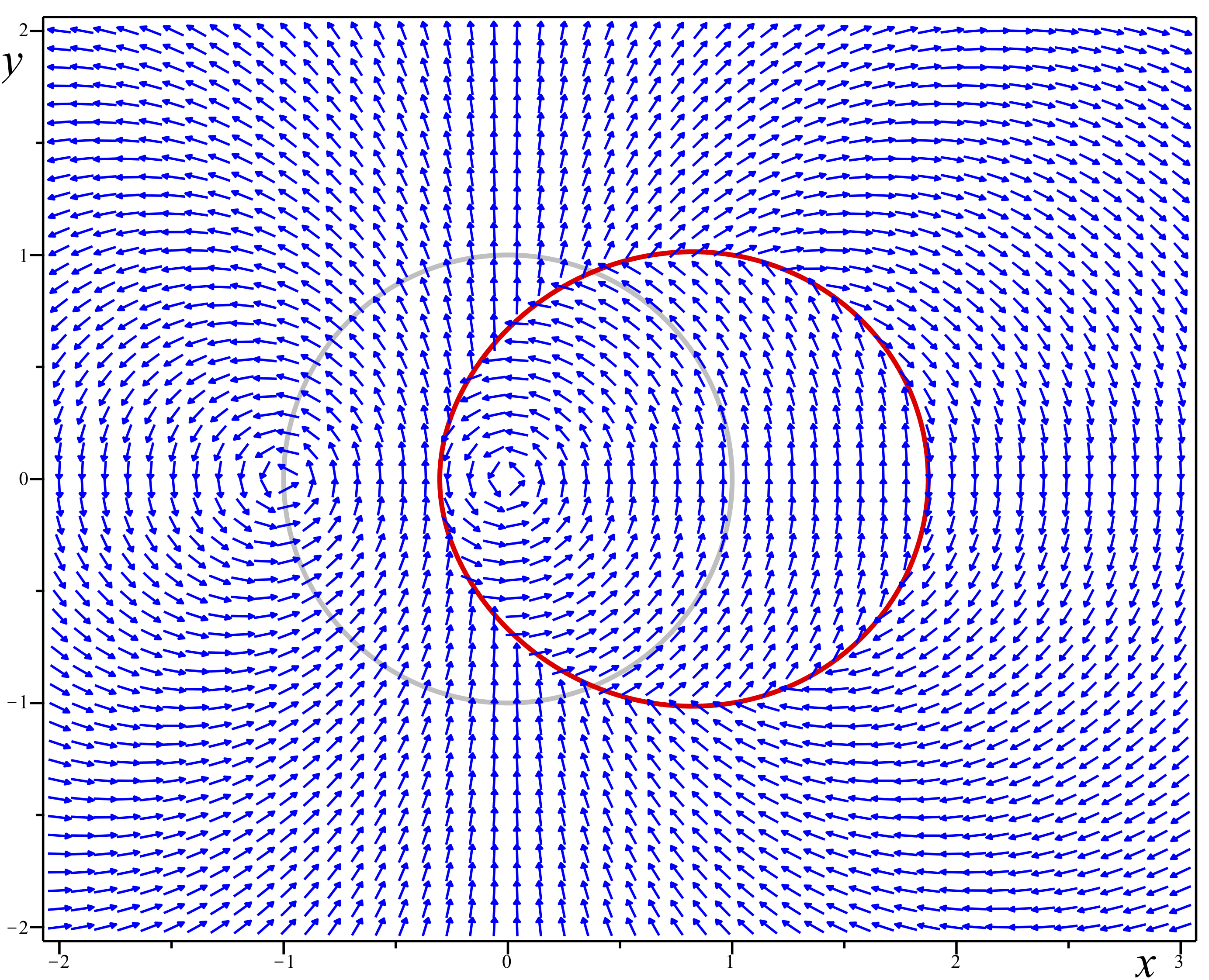}
	\caption{inverse, $\rho_0=1$}
	\label{fig:sphere_director_rho0_1_inverse}
\end{subfigure}
\begin{subfigure}[c]{0.33\linewidth}
	\centering
	\includegraphics[width=\linewidth]{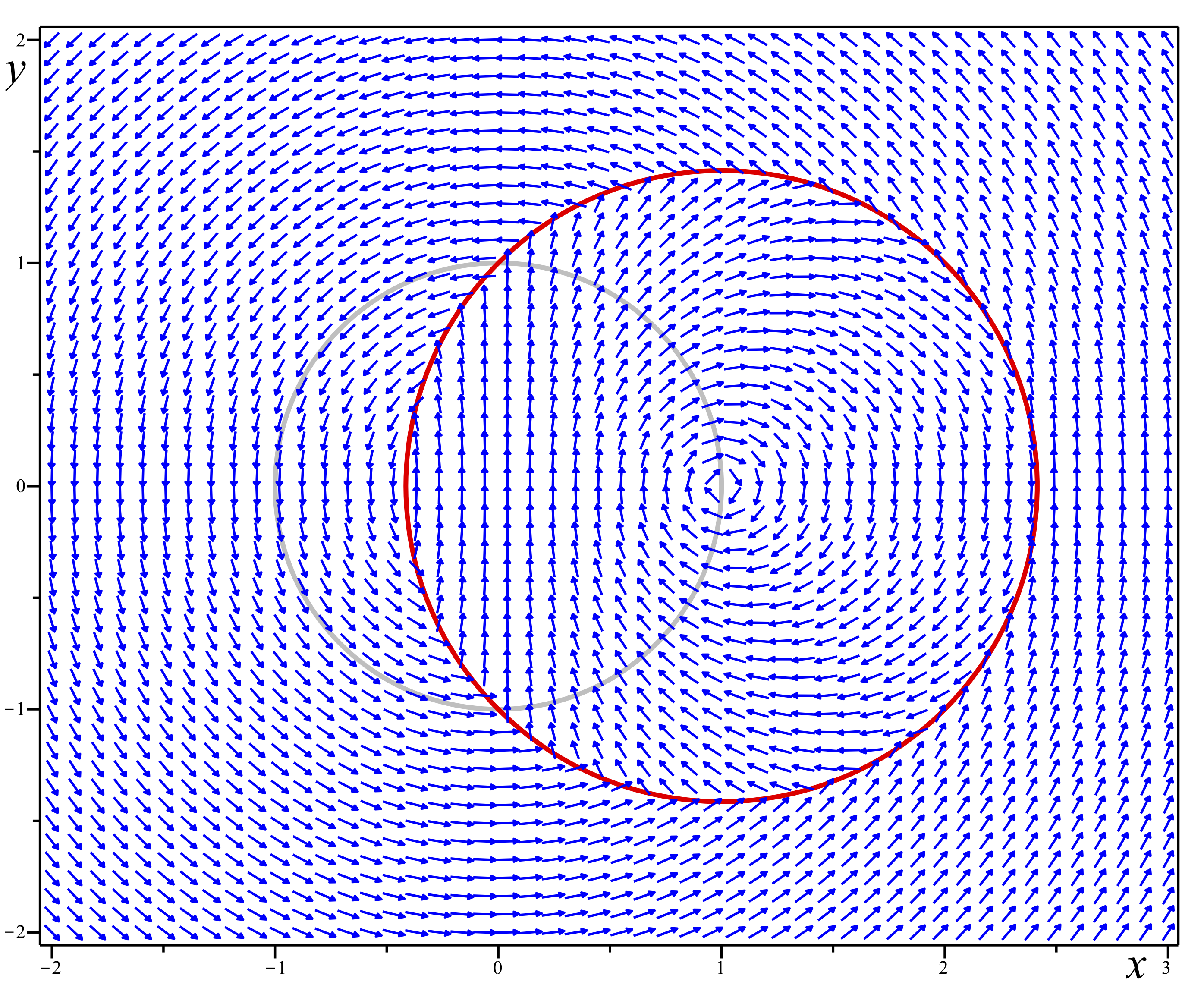}
	\caption{direct, $\rho_0=\sqrt{2}$}
	\label{fig:sphere_director_rho0_sqrt(2)_direct}
\end{subfigure}
\qquad
\begin{subfigure}[c]{0.33\linewidth}
	\centering
	\includegraphics[width=\linewidth]{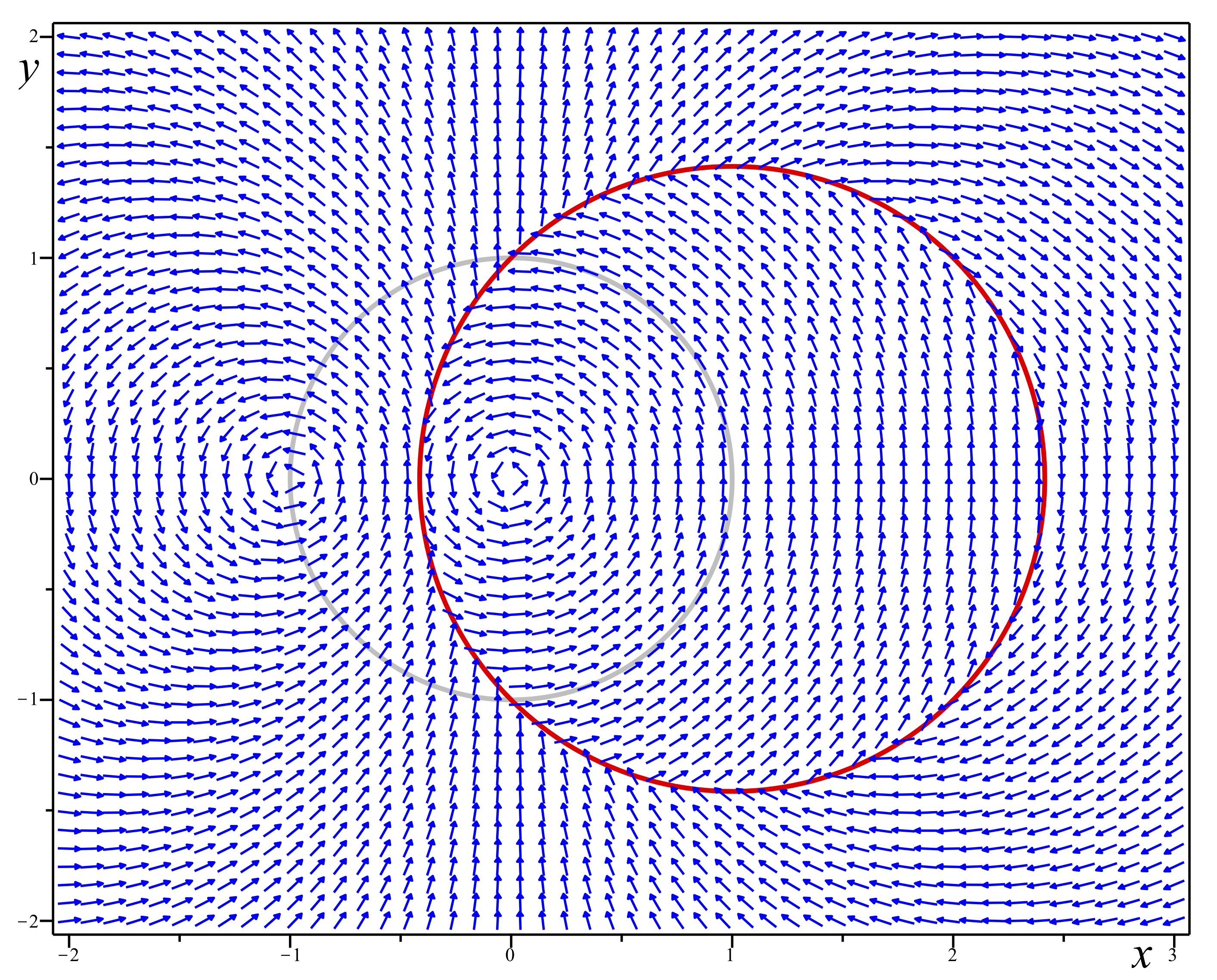}
	\caption{inverse, $\rho_0=\sqrt{2}$}
	\label{fig:sphere_director_rho0_sqrt(2)_inverse}
\end{subfigure}
\begin{subfigure}[c]{0.33\linewidth}
	\centering
	\includegraphics[width=\linewidth]{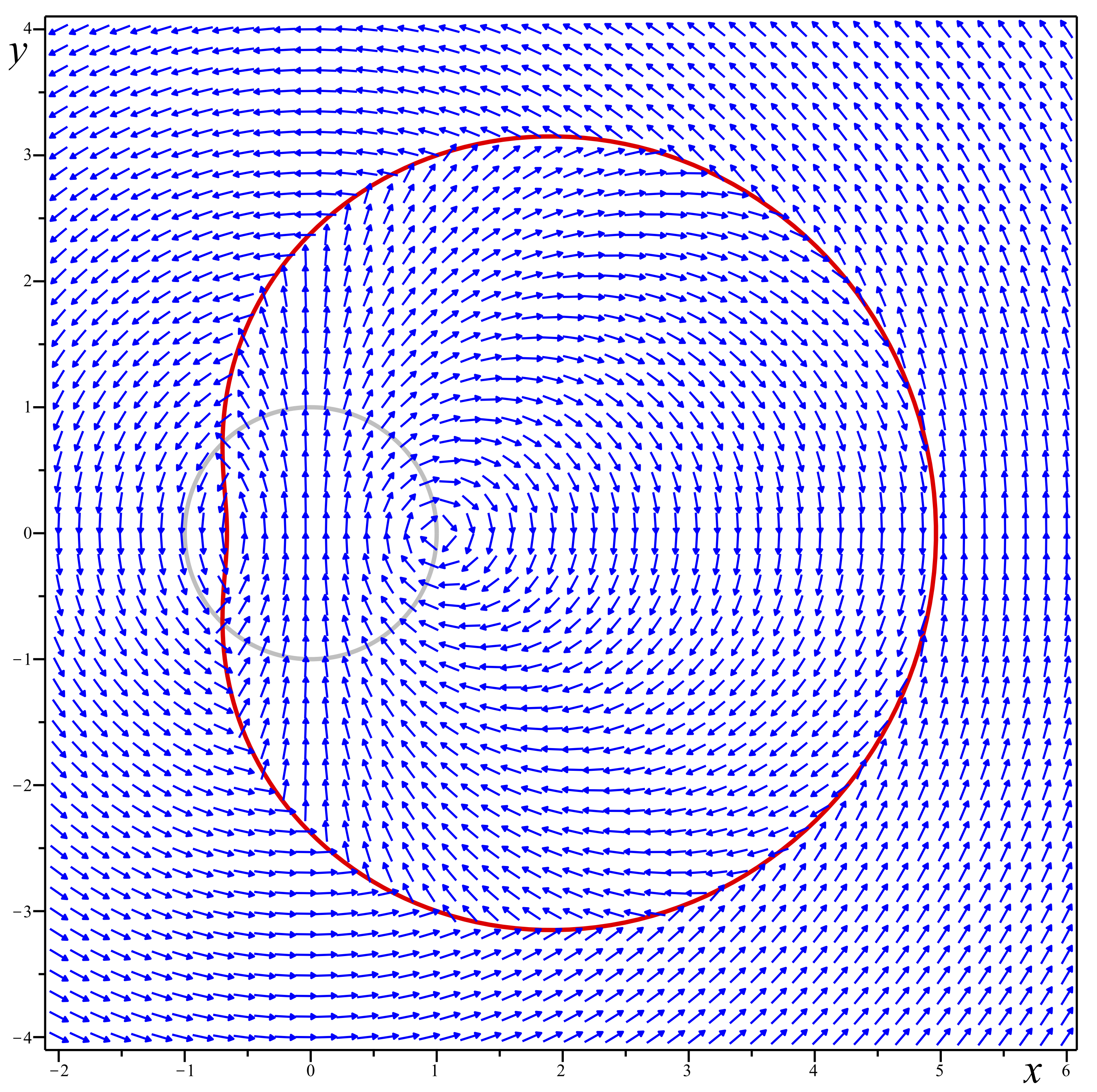}
	\caption{direct, $\rho_0=3$}
	\label{fig:sphere_director_rho0_3_direct}
\end{subfigure}
\qquad
\begin{subfigure}[c]{0.33\linewidth}
	\centering
	\includegraphics[width=\linewidth]{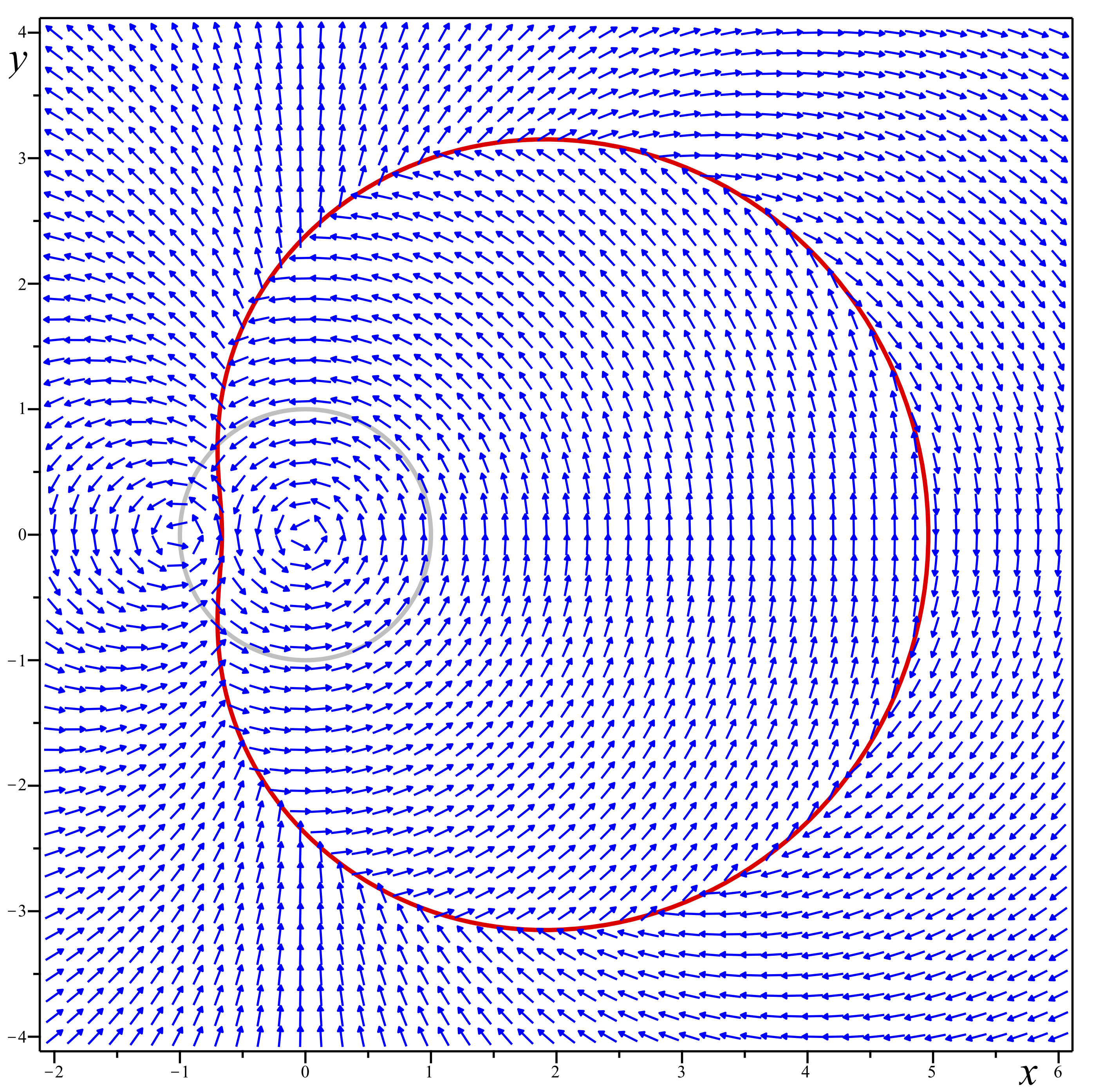}
	\caption{inverse, $\rho_0=3$}
	\label{fig:sphere_director_rho0_3_inverse}
\end{subfigure}
\begin{subfigure}[c]{0.33\linewidth}
	\centering
	\includegraphics[width=\linewidth]{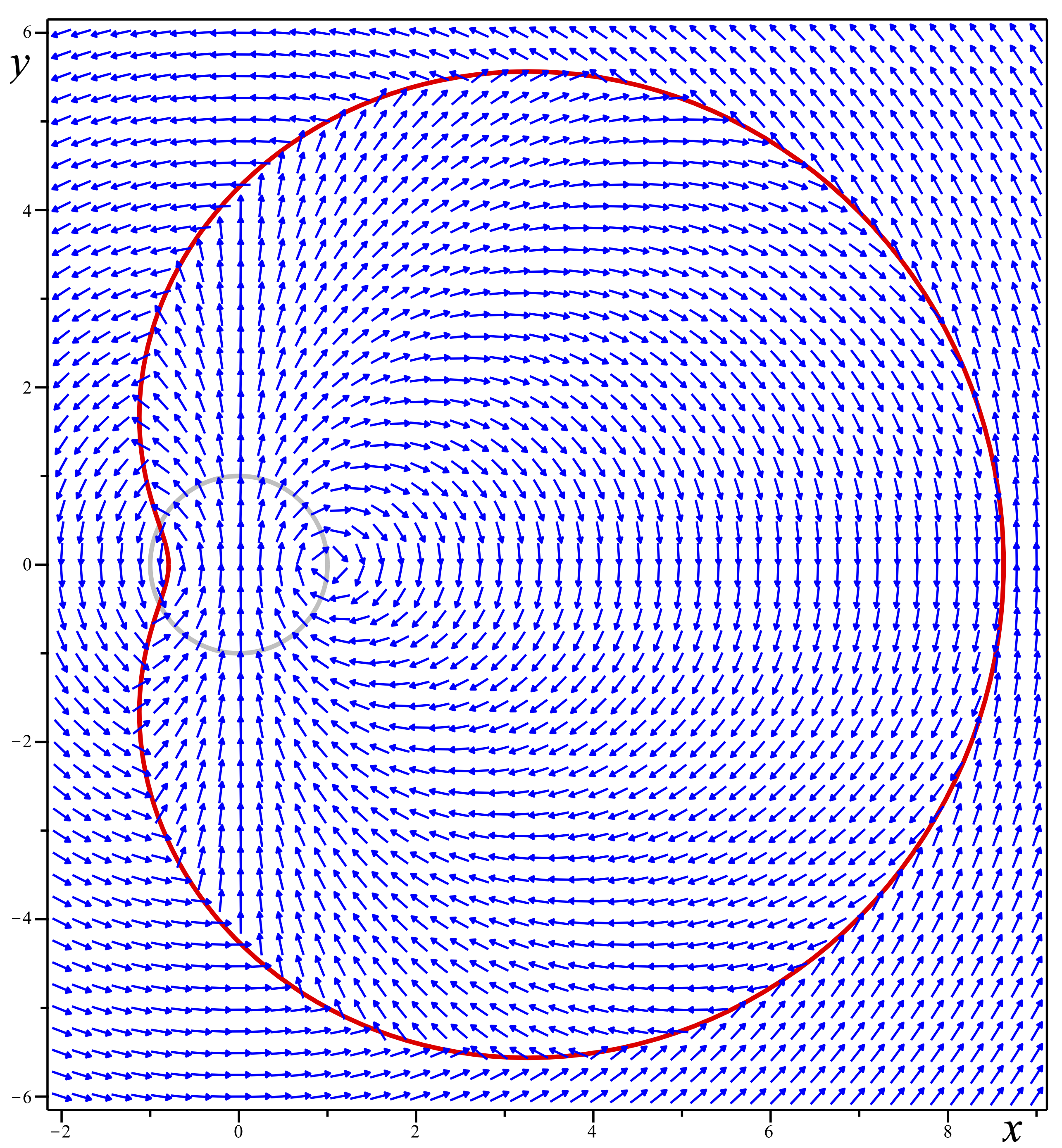}
	\caption{direct, $\rho_0=5$}
	\label{fig:sphere_director_rho0_5_direct}
\end{subfigure}
\qquad
\begin{subfigure}[c]{0.33\linewidth}
	\centering
	\includegraphics[width=\linewidth]{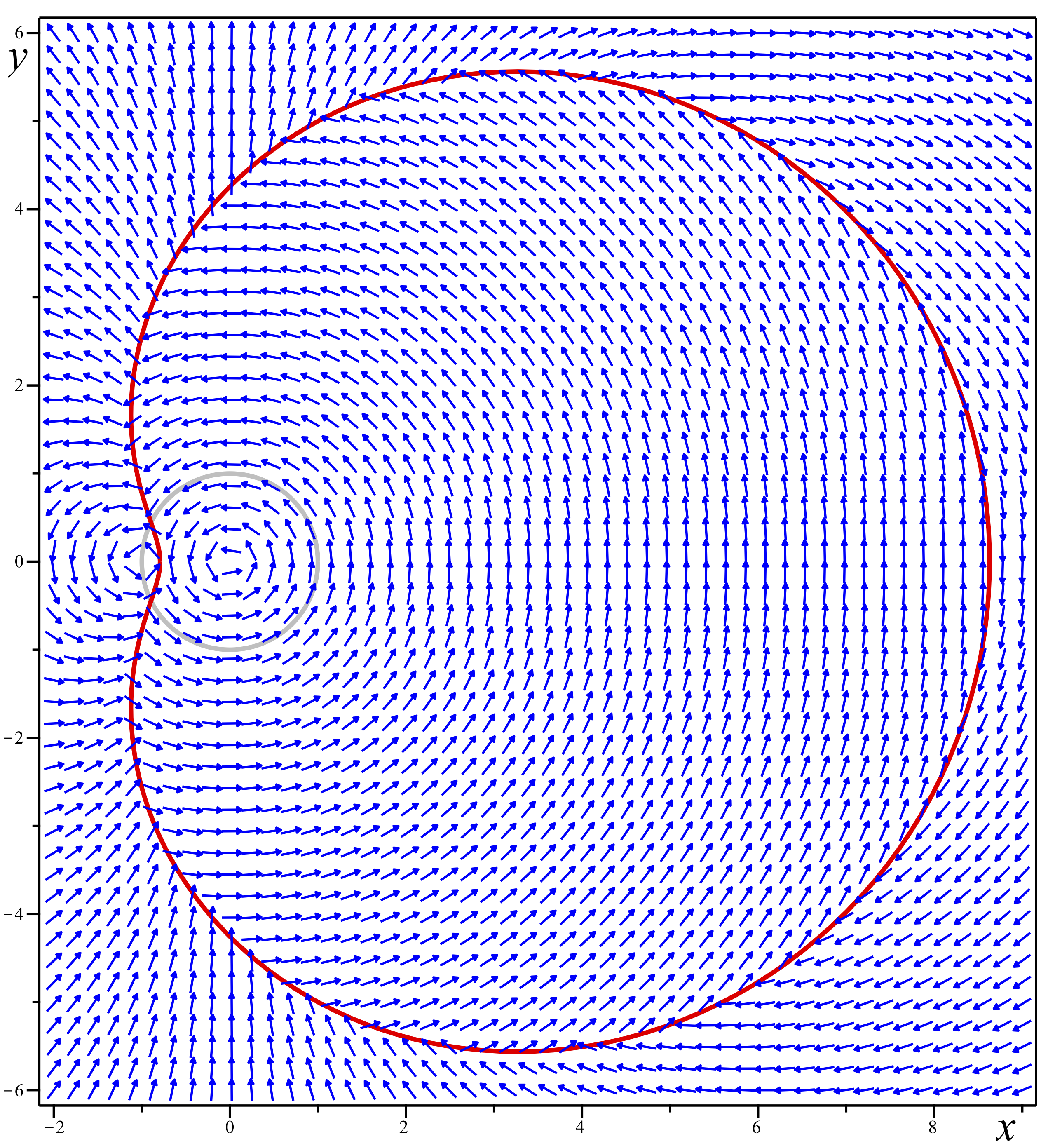}
	\caption{inverse, $\rho_0=5$}
	\label{fig:sphere_director_rho0_5_inverse}
\end{subfigure}
\caption{Admissible wall $\Line$ in the $(x,y)$ plane corresponding to different values of the parameter $\rho_0$ in \eqref{eq:wall_intial_condition}. Both direct and inverse configurations are shown. As pointed out in Remark~\ref{rmk:circle}, for $\rho_0=\sqrt{2}$, $\Line$ is a circle. The gray circle is the equator of $\sphere$ in all panels.}
\label{fig:sphere_director_fields}
\end{figure}

We now illustrate a number of solutions of the wall equation obtained numerically with routines built in Maple 2021. They are indexed in the parameter $\rho_0>0$ defined by the following initial condition for \eqref{eq:wall_equation},
\begin{equation}
	\label{eq:wall_intial_condition}
	\rho(\tfrac\pi2)=\rho_0.
\end{equation}
Figure~\ref{fig:sphere_director_fields} shows director fields in the $(x,y)$ plane in both direct and inverse configurations for several values of $\rho_0$.
In the direct case, the director field on $\sphere$ has a vortex at the north pole and another on the equator, whereas in the inverse case vortexes are swapped with their antipodal twins. 

\begin{figure}
	\centering
	\includegraphics[width=.33\linewidth]{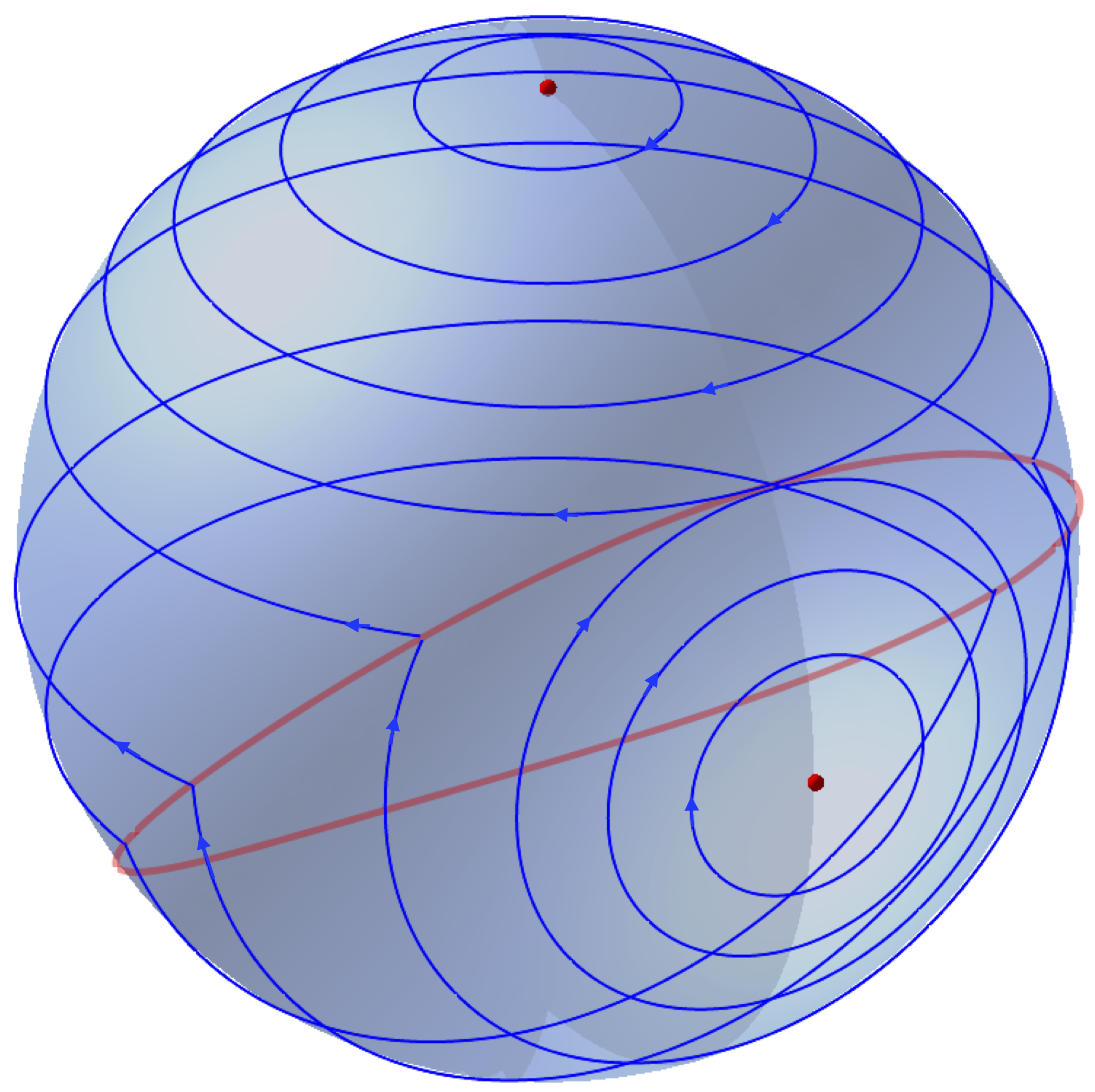}
	\caption{Director field lines on $\sphere$ are depicted as oriented blue circles bisected by the admissible wall (red curve) for $\rho_0=1$. The vortexes of the field are marked as red dots.}
	\label{fig:sphere_director_lines}
	\end{figure}

\begin{figure}
	\centering
	\begin{subfigure}[c]{0.25\linewidth}
		\centering
		\includegraphics[width=\linewidth]{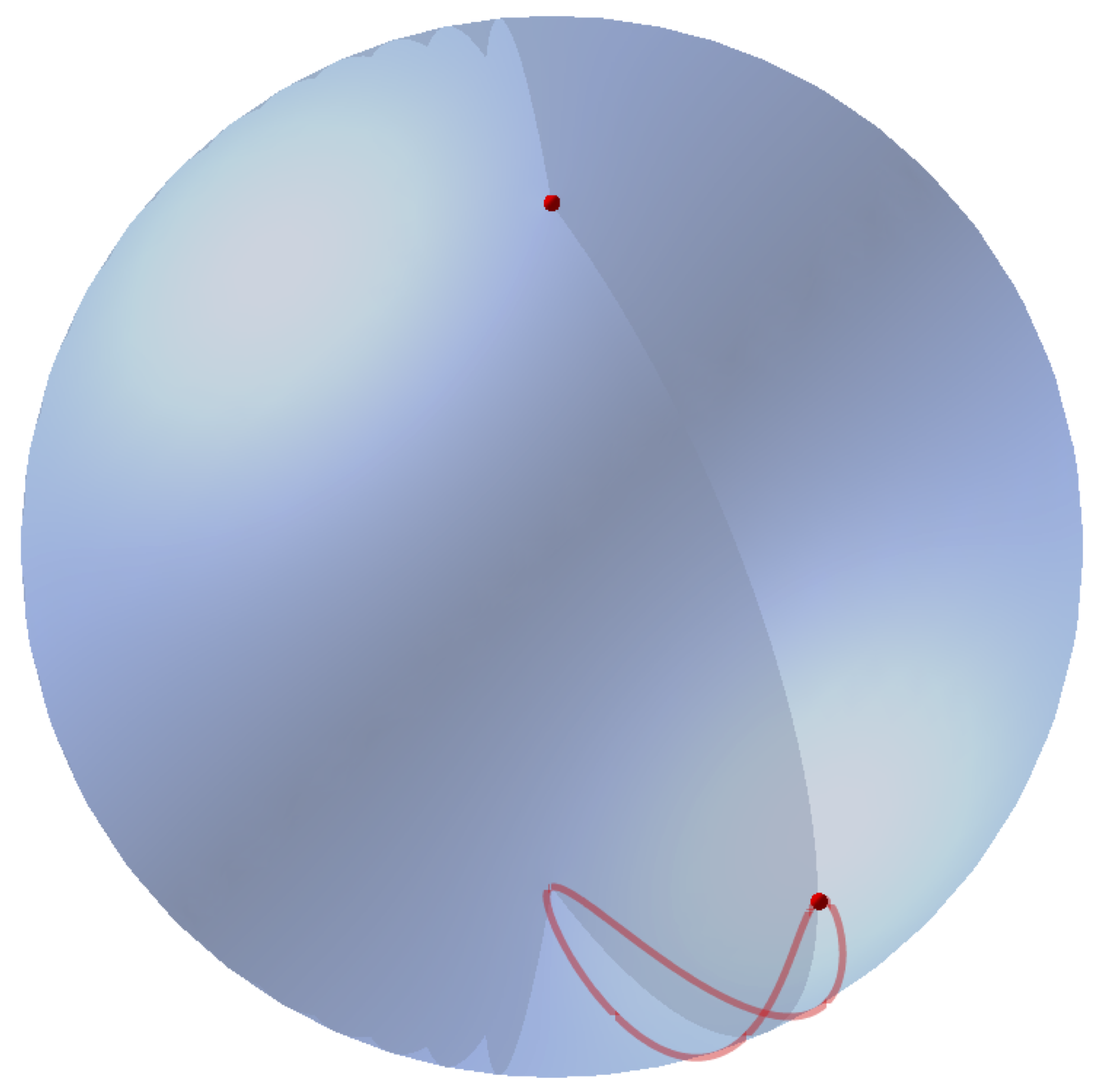}
		\caption{$\rho_0=1/100$}
		\label{fig:sphere_wall_rho0_1o100}
	\end{subfigure}
	\qquad
		\begin{subfigure}[c]{0.25\linewidth}
		\centering
		\includegraphics[width=\linewidth]{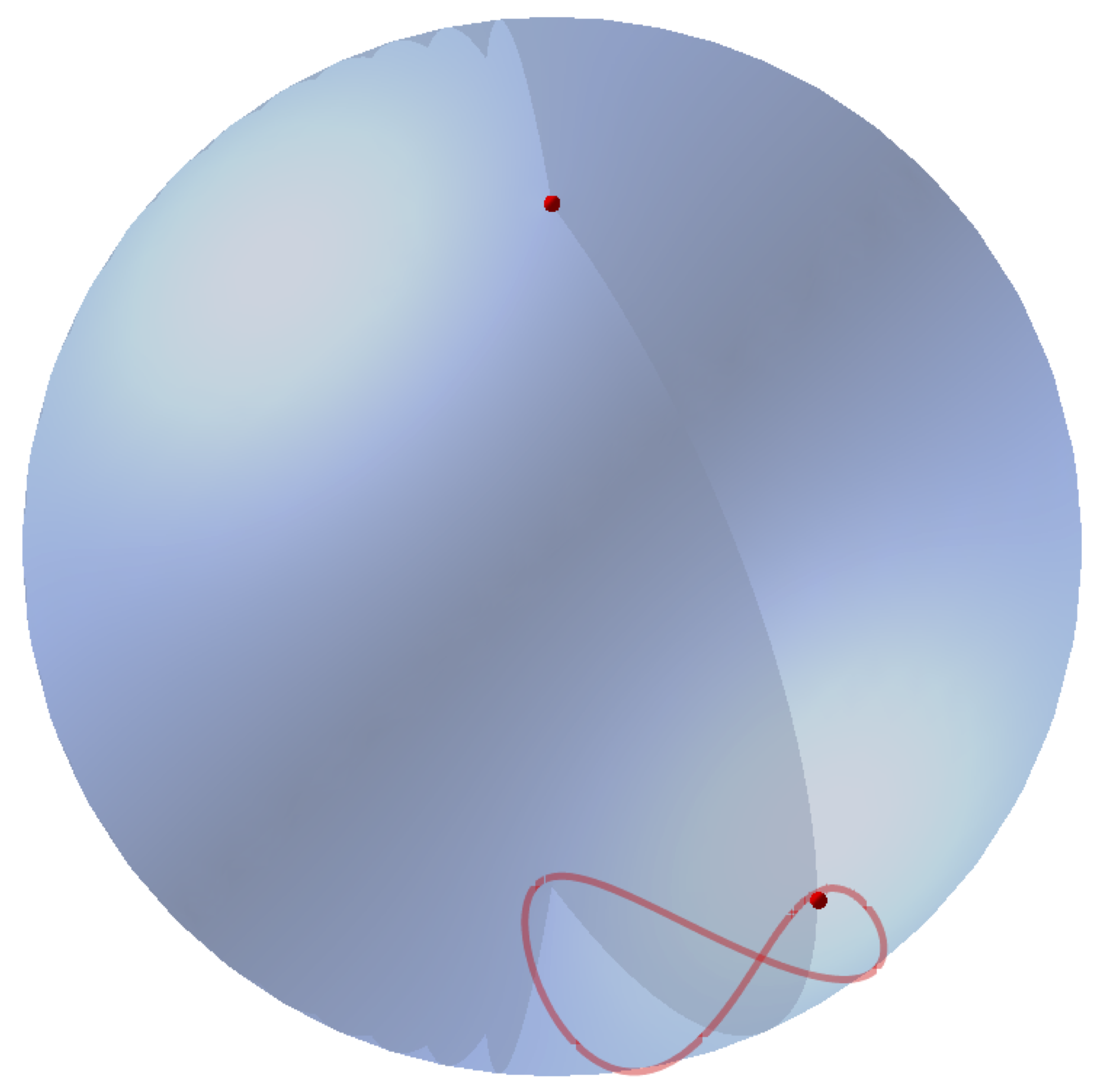}
		\caption{$\rho_0=1/20$}
		\label{fig:sphere_wall_rho0_1o20}
	\end{subfigure}
	\\
		\begin{subfigure}[c]{0.25\linewidth}
		\centering
		\includegraphics[width=\linewidth]{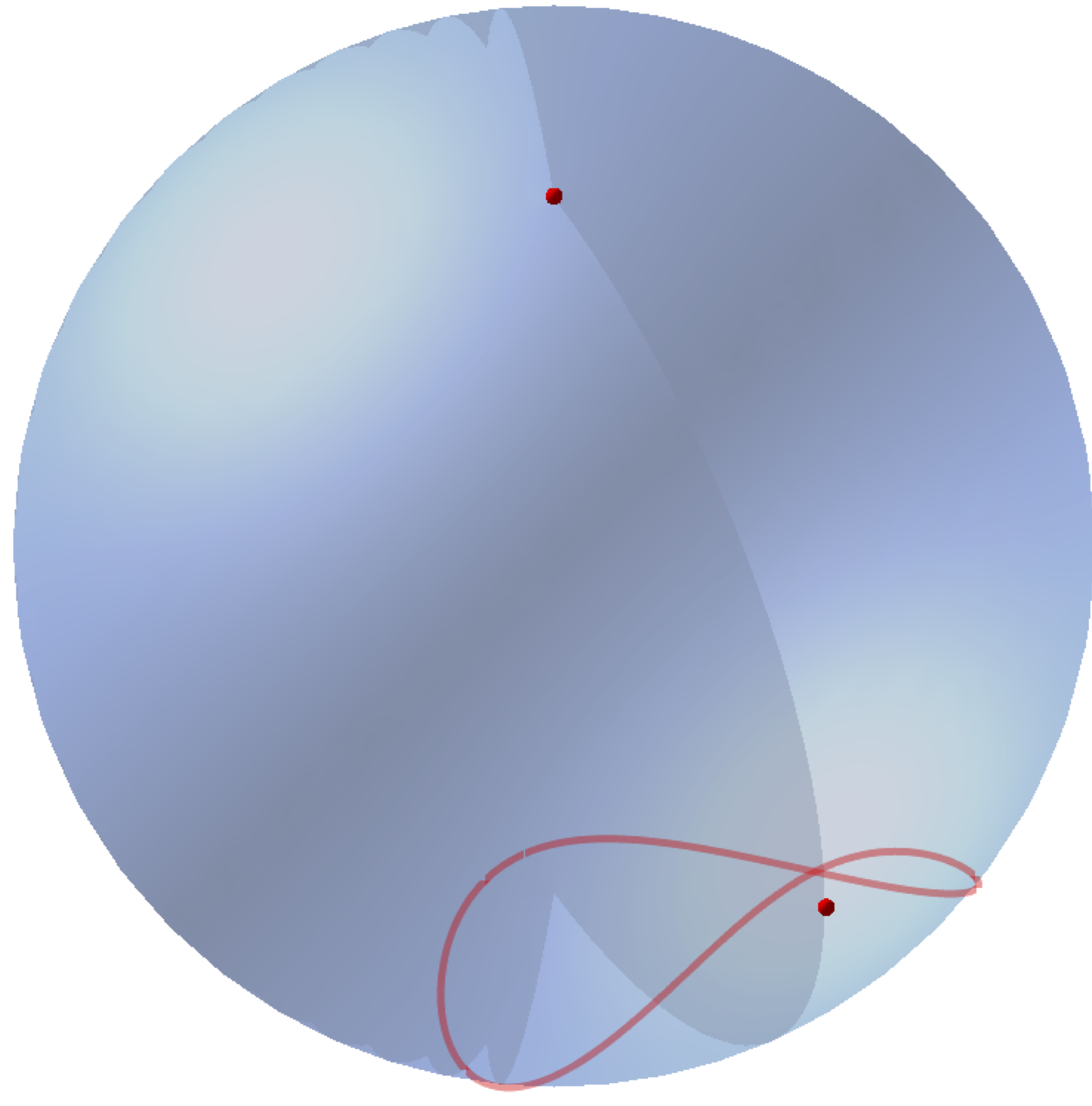}
		\caption{$\rho_0=1/5$}
		\label{fig:sphere_wall_rho0_1o5}
	\end{subfigure}
	\qquad
	\begin{subfigure}[c]{0.25\linewidth}
		\centering
		\includegraphics[width=\linewidth]{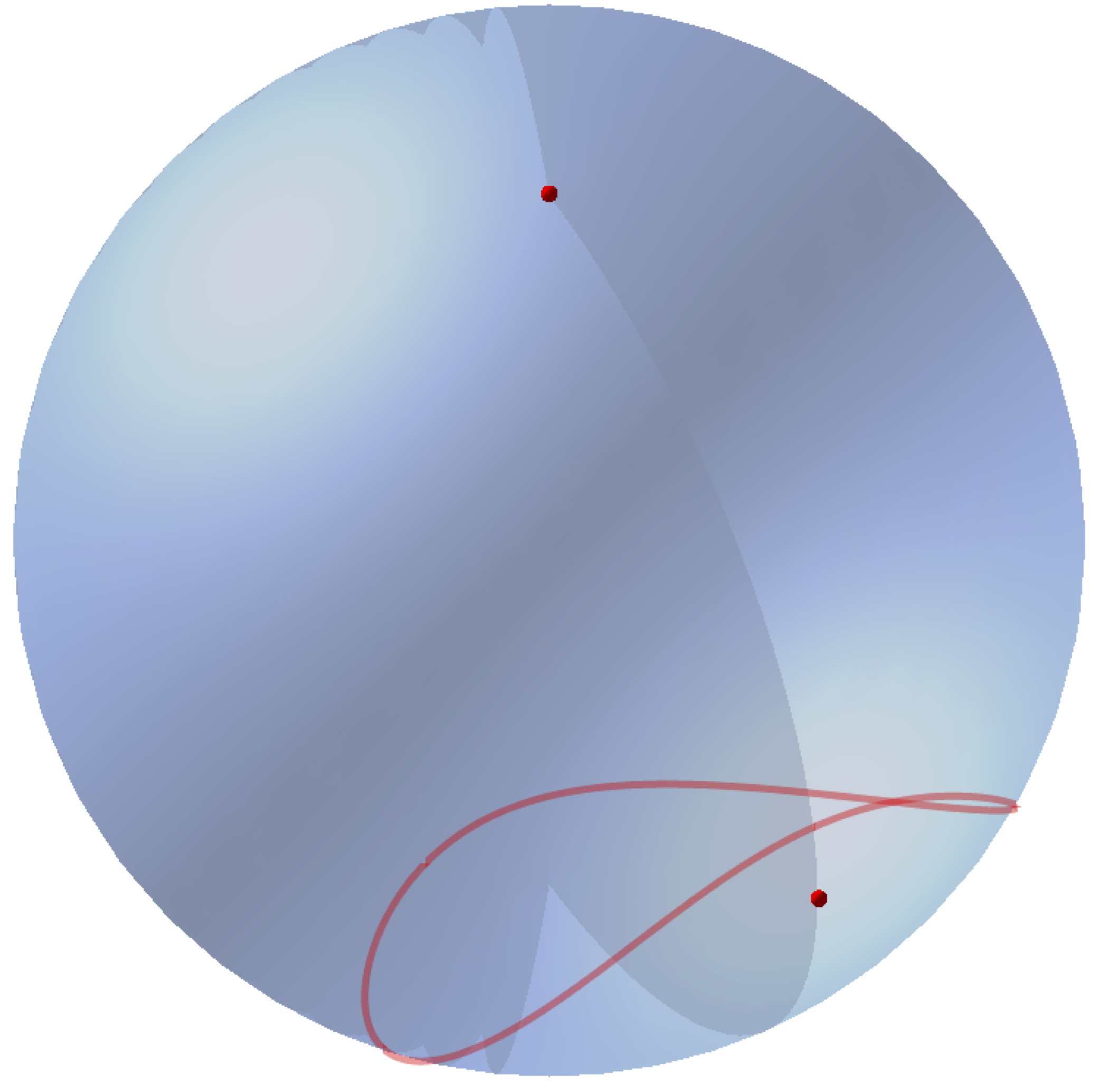}
		\caption{$\rho_0=1/3$}
		\label{fig:sphere_wall_rho0_1o3}
	\end{subfigure}
	\\
		\begin{subfigure}[c]{0.25\linewidth}
		\centering
		\includegraphics[width=\linewidth]{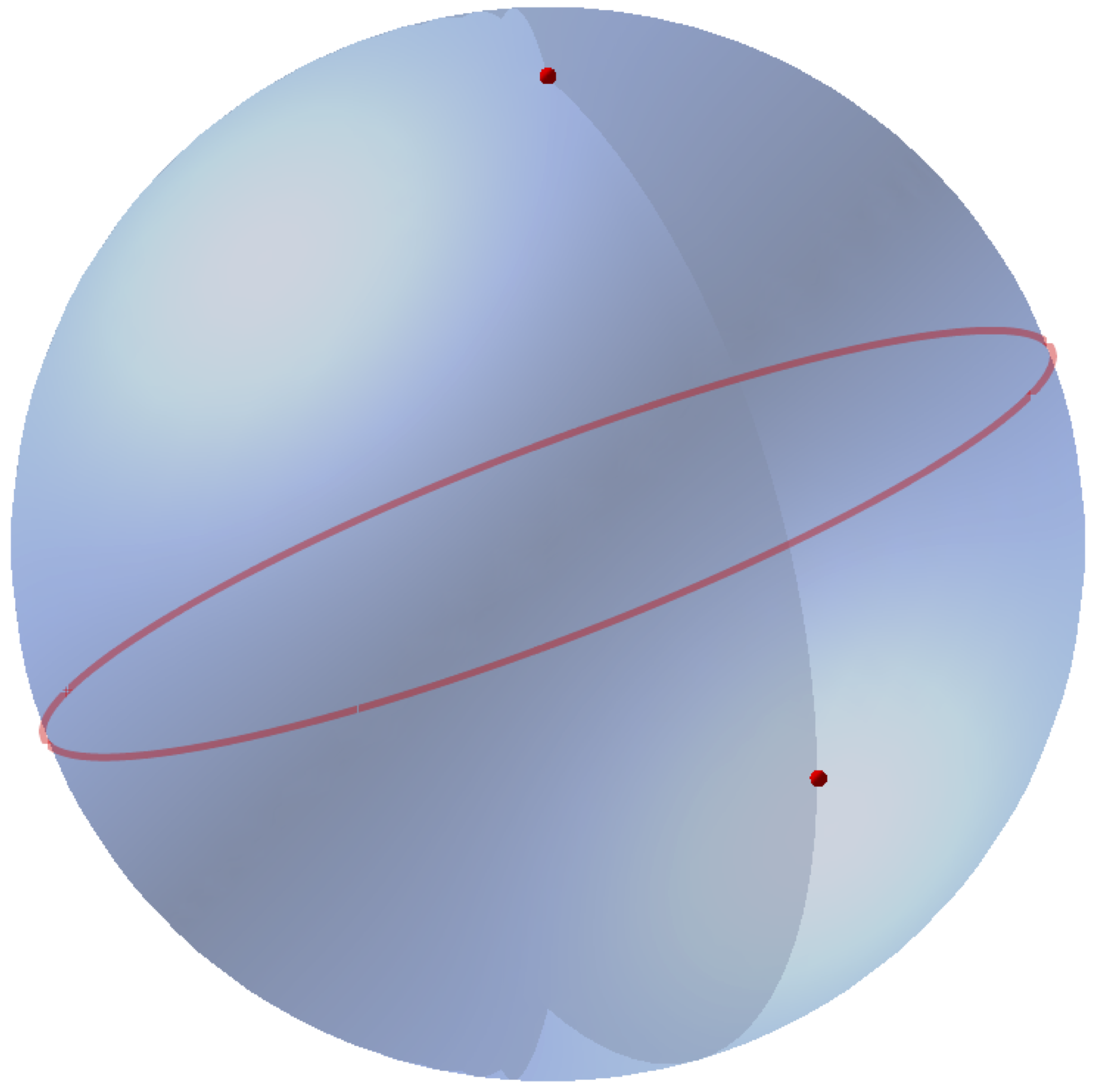}
		\caption{$\rho_0=\sqrt{2}$}
		\label{fig:sphere_wall_rho0_sqrt(2)}
	\end{subfigure}
	\\
		\begin{subfigure}[c]{0.25\linewidth}
		\centering
		\includegraphics[width=\linewidth]{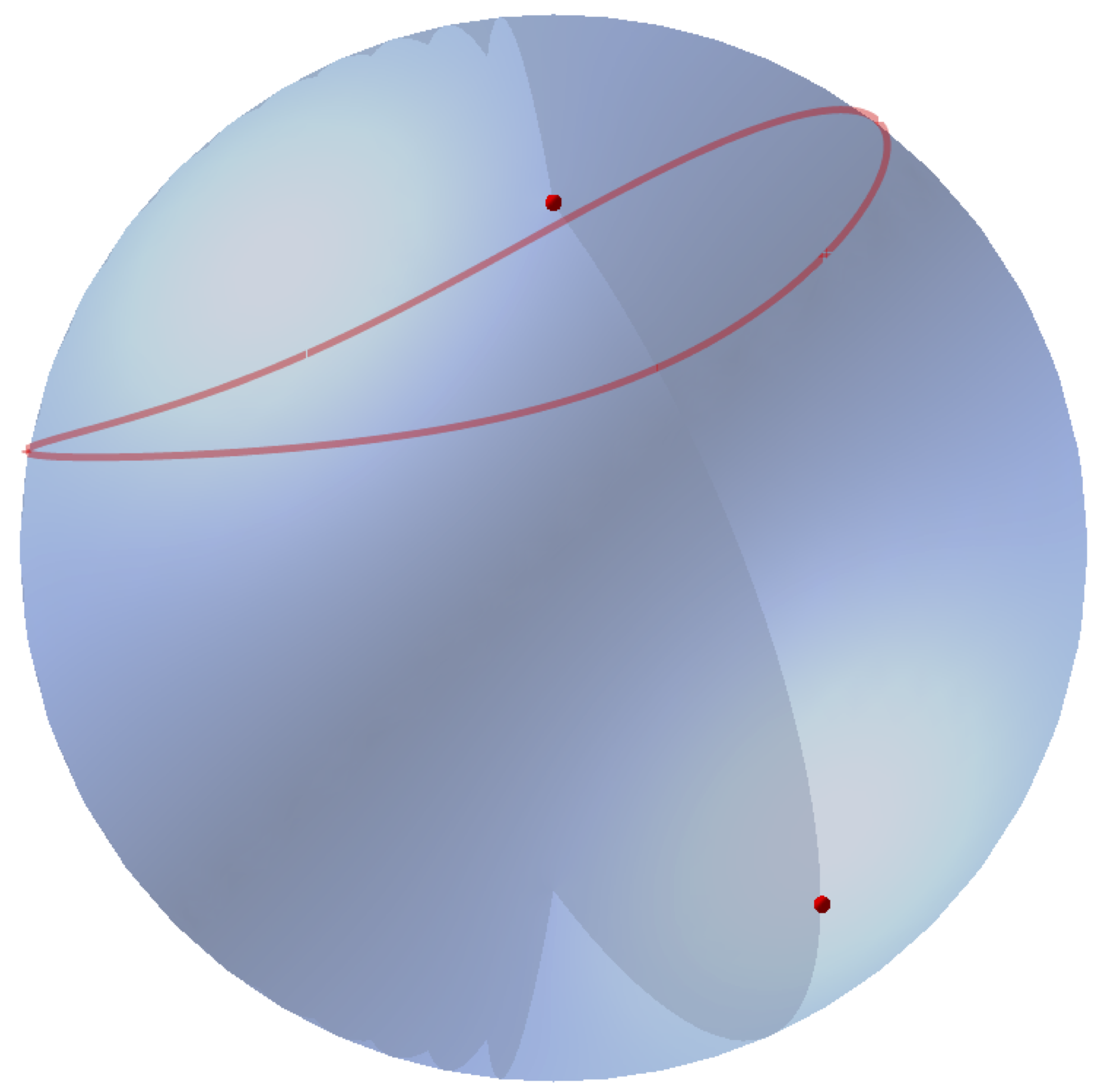}
		\caption{$\rho_0=3$}
		\label{fig:sphere_wall_rho0_3}
	\end{subfigure}
	\qquad
	\begin{subfigure}[c]{0.25\linewidth}
		\centering
		\includegraphics[width=\linewidth]{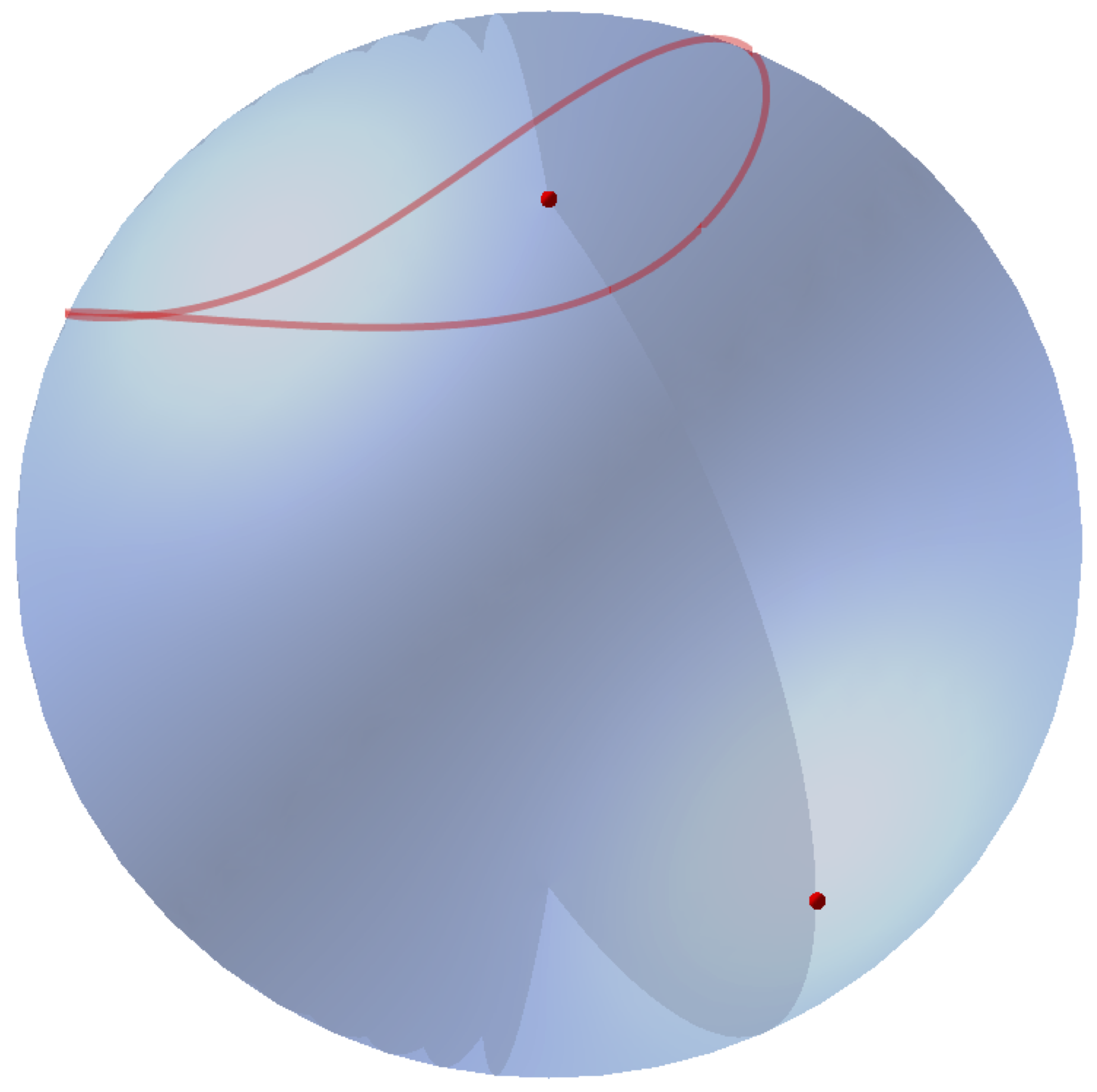}
		\caption{$\rho_0=5$}
		\label{fig:sphere_wall_rho0_5}
	\end{subfigure}
	\\
		\begin{subfigure}[c]{0.25\linewidth}
		\centering
		\includegraphics[width=\linewidth]{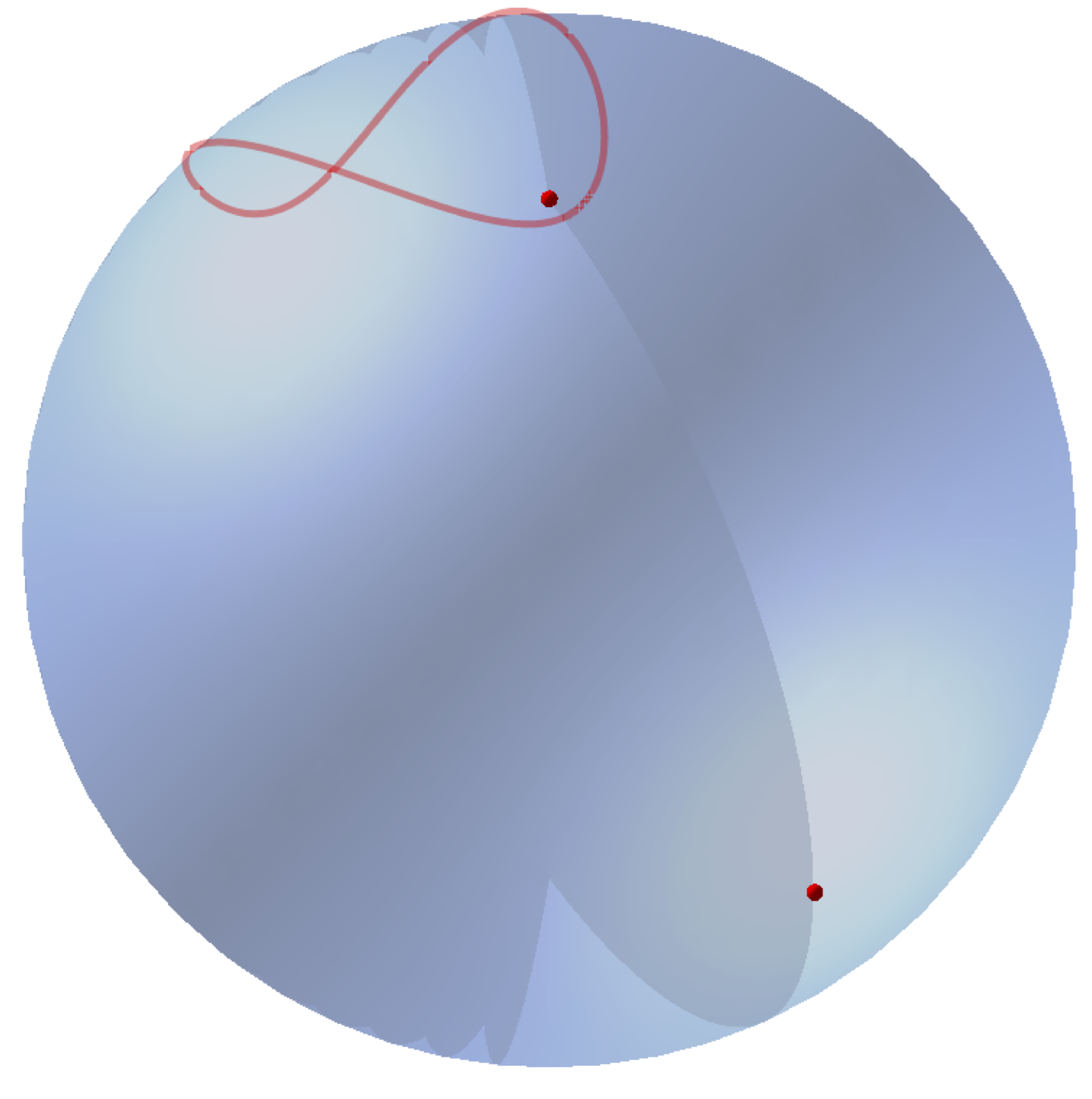}
		\caption{$\rho_0=20$}
		\label{fig:sphere_wall_rho0_20}
	\end{subfigure}
	\qquad
	\begin{subfigure}[c]{0.25\linewidth}
		\centering
		\includegraphics[width=\linewidth]{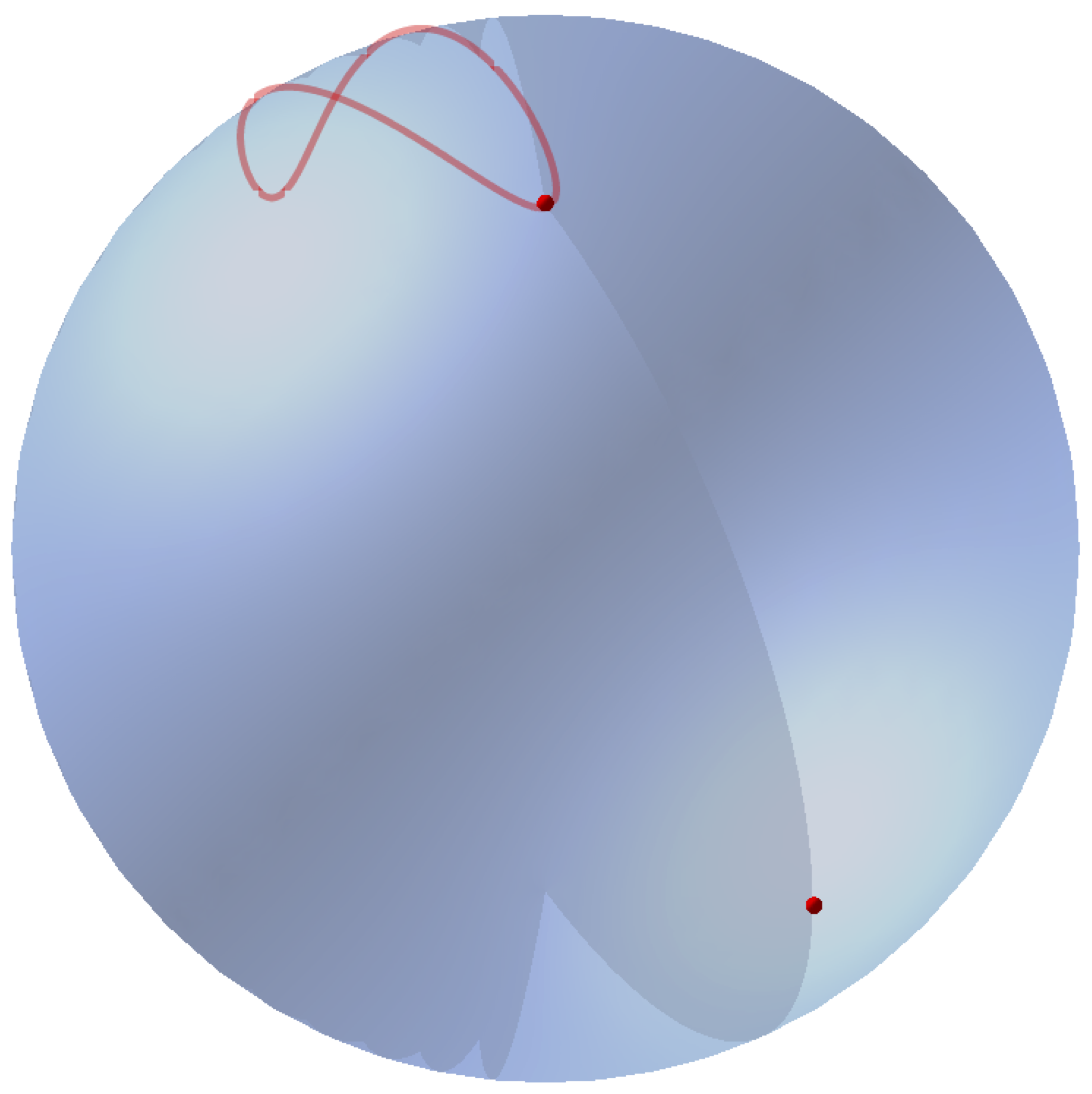}
		\caption{$\rho_0=100$}
		\label{fig:sphere_wall_rho0_100}
	\end{subfigure}
	\caption{Admissible walls on $\sphere$ for fixed vortexes and several values of the parameter $\rho_0$. As $\rho_0$ increases, the wall shrinks around the north pole and the point antipolad to the vortex on the equator. As $\rho_0$ decreases, the wall shrinks around the vortex on the equator and the south pole. The wall corresponding to $\rho=\sqrt{2}$ is a great circle in the plane of mirror symmetry for the vortexes.}
	\label{fig:sphere_walls}
\end{figure}

The director field lines on $\sphere$ are shown in Fig.~\ref{fig:sphere_director_lines} for a direct case, while only the walls (the same for both cases) are shown in Fig.~\ref{fig:sphere_walls}.

\begin{figure}
	\centering
	\includegraphics[width=.5\linewidth]{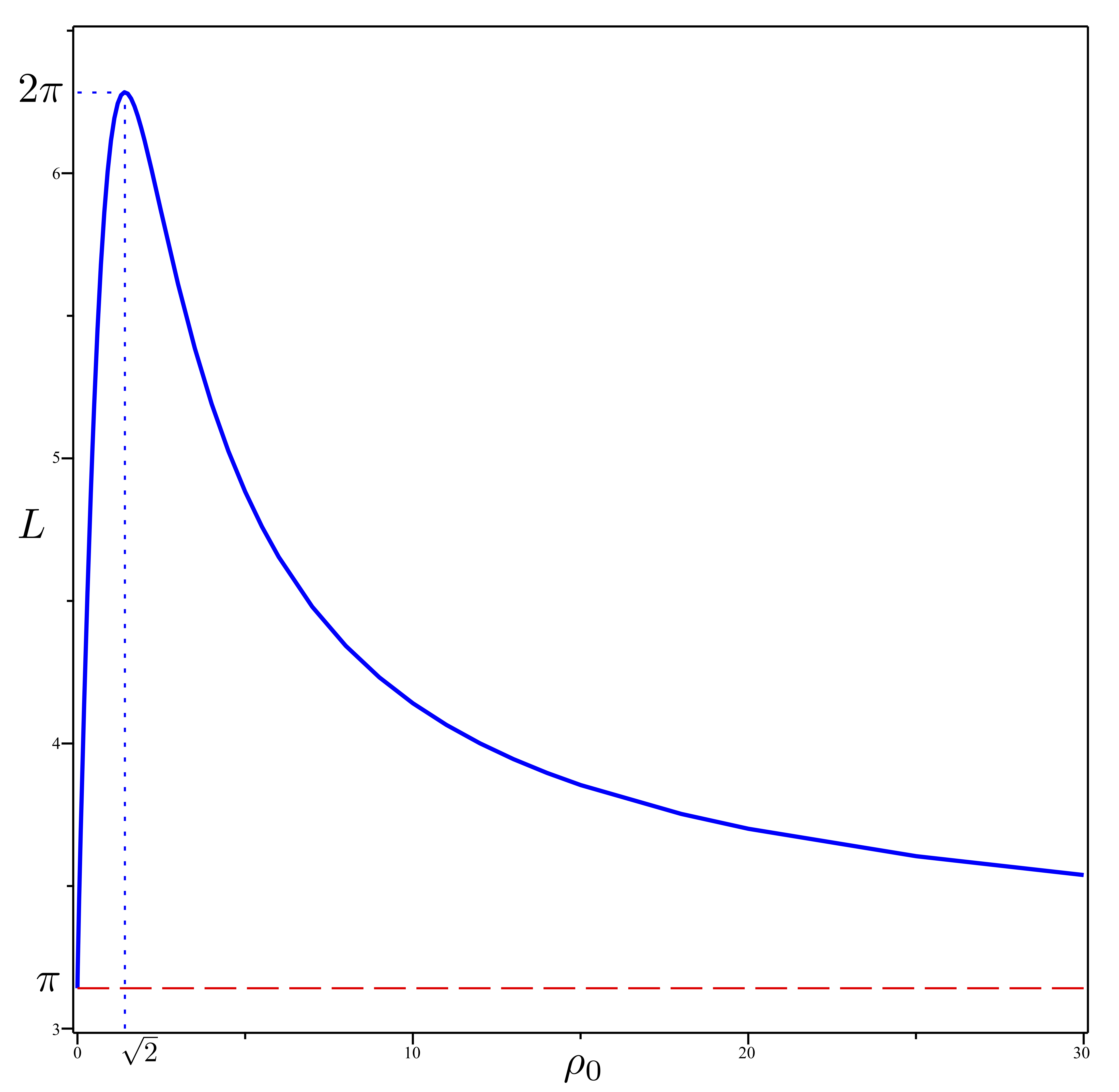}
	\caption{The length $L$ in \eqref{eq:sphere_length} of walls $\Line$ corresponding to solutions of \eqref{eq:wall_equation} plotted as a function of $\rho_0$ in \eqref{eq:wall_intial_condition}.}
	\label{fig:sphere_length}
\end{figure}

The length $L(\Line)$ given by \eqref{eq:sphere_length} is plotted in Fig.~\ref{fig:sphere_length} as a function of $\rho_0$.
It exhibits the same limit $L_\mathrm{min}=\pi$ for both $\rho_0\to0$ and $\rho_0\to\infty$. Correspondingly, $\Line$ shrinks around the vortex on the equator and the south pole or around the point antipodal to the vortex on the equator and the north pole: in both limits, the wall \emph{escapes} from $\sphere$, leaving behind only vortexes at the poles or vortexes on the equator, respectively. The maximum $L_\mathrm{max}=2\pi$ is attained at $\rho_0=\sqrt{2}$; the corresponding wall is a great circle of $\sphere$.

Thus, in a quasi-static evolution governed by the dominant free energy $\free_\mathrm{dom}$ in \eqref{eq:free_energy_dominant}, one of the defect configurations shown in Fig.~\ref{fig:antipodal} is reached in finite time, with either antipodal vortexes at the poles or on the equator, depending on whether the wall starts below or above the great circle of defect symmetry, respectively. This circle maximizes the wall's length, and so it is \emph{unstable}. In both cases, no wall survives in the end and a \emph{virtual} defect, which had been \emph{covered}, is exposed as the wall evaporates.

In the following section, we shall see that a local minimum of $\free_\mathrm{dom}$ can be attained on a surface with (constant) negative Gaussian curvature, where a wall can thus be trapped and survive.

\subsection{Pseudospherical shell}
\begin{figure}
	\centering
	\includegraphics[width=.33\linewidth]{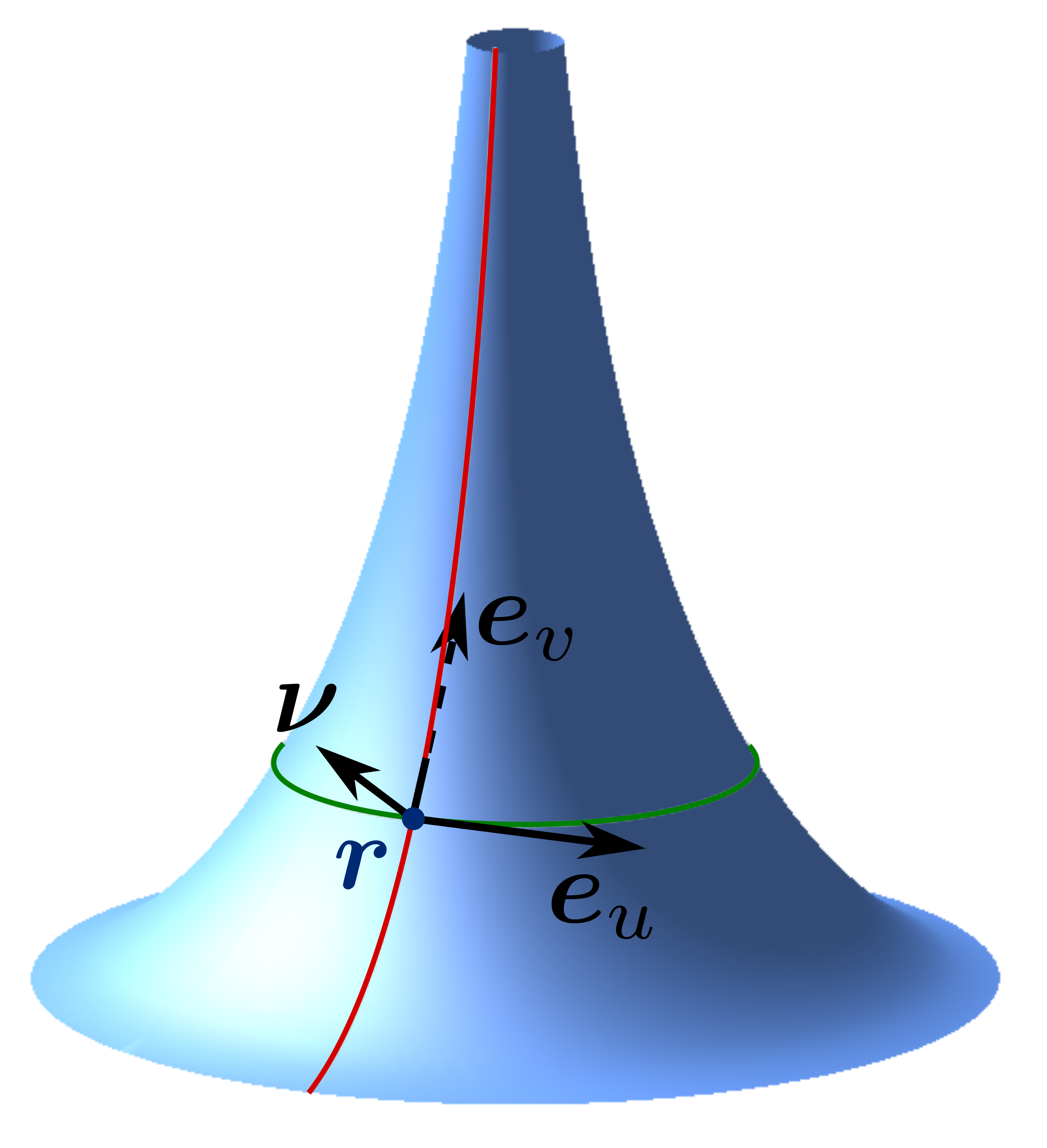}
	\caption{Beltrami's pseudosphere. Meridians (like the red line) are geodesics and principal lines of curvature, whereas the orthogonal circles (like the green one) are principal lines of curvature, but not geodesics.}
	\label{fig:Beltrami}
\end{figure}

Beltrami's pseudosphere of radius $R=1$ is represented in a Cartesian frame $\frameC$ as $\rv(u,v)=X(u,v)\e_x+Y(u,v)\e_y+Z(u,v)\e_z$, where 
\begin{equation}
	\label{eq:pseudosphere_representation}
	X(u,v)=\frac{1}{v}\cos u,\quad
Y(u,v)=\frac{1}{v}\sin u,\quad
Z(u,v)=\ln(v+\sqrt{v^2-1})-\frac{\sqrt{v^2-1}}{v},
\end{equation}
whose parameters $(u,v)$ range in the strip $[0,2\pi]\times[1,\infty)$. As shown, for example in \cite[pp.\,54-55]{needham:visual}, the mapping \eqref{eq:pseudosphere_representation} is conformal and the line element $\dd\ell$ on the pseudosphere is given by
\begin{equation}
	\label{eq:pseudosphere_line_element}
	\dd\ell=\frac{1}{v}\sqrt{\dd u^2+\dd v^2}.
\end{equation}
Figure~\ref{fig:Beltrami} represents this surface alongside the principal frame $(\e_u,\e_v,\normal)$. 
As for all surfaces of revolution, meridians are geodesics. All other systems of geodesics are generated by pencils of circles in the $(u,v)$ plane with centers on the $u$ axis (see, for example, \cite[p.\,230]{pressley:elementary}).

\begin{figure}
	\centering
	\includegraphics[width=.5\linewidth]{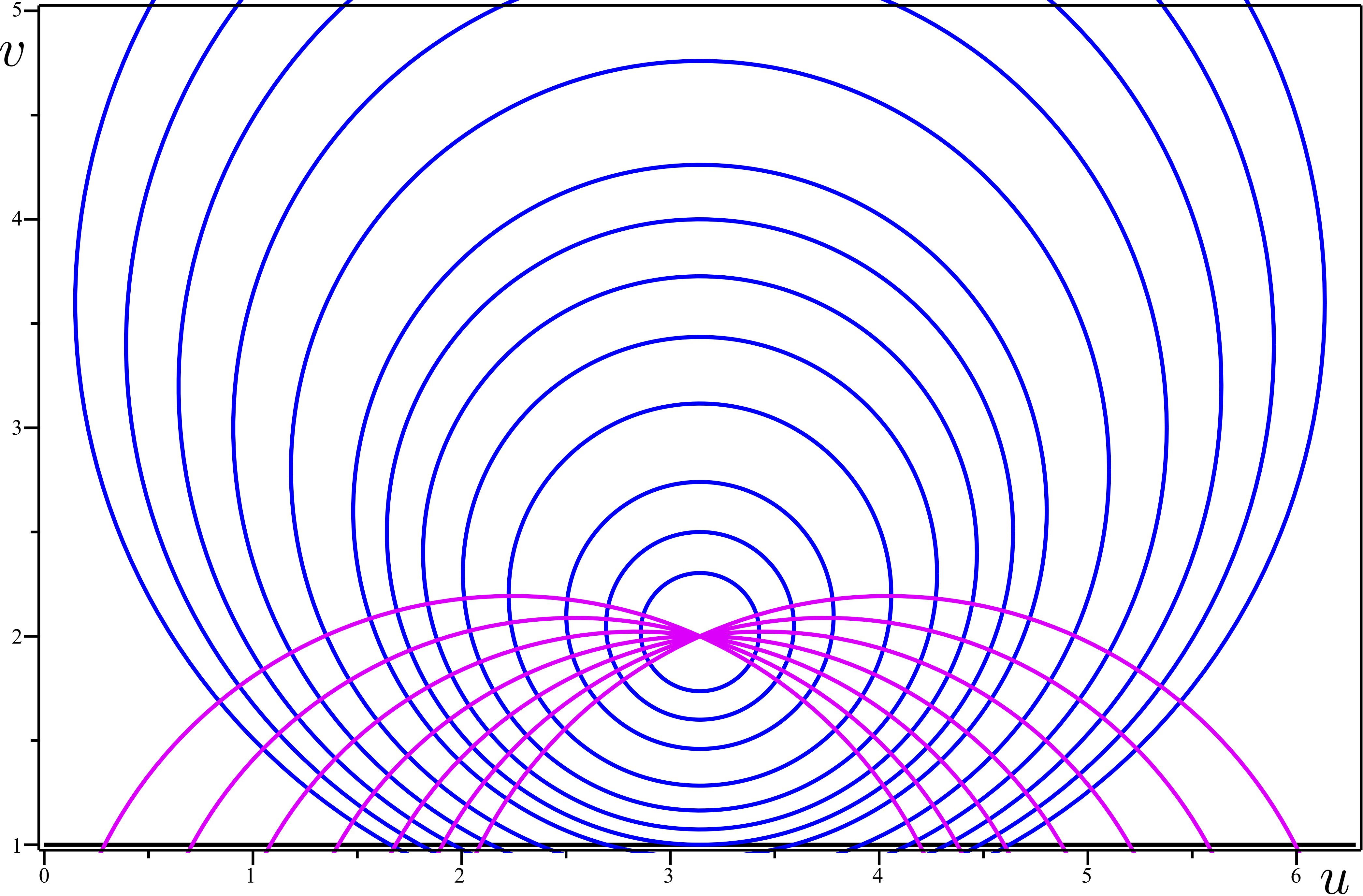}
	\caption{Geodesic lines (magenta) and corresponding director field lines (blue) for a vortex pinned at $(u_0,v_0)$ in the $(u,v)$ plane. Here $u_0=\pi$ and $v_0=2$.}
	\label{fig:Apollonian}
\end{figure}

Thus, for a vortex sitting at $(u_0,v_0)$, the splay ostracism theorem requires the director field lines (and  associated orthogonal geodesics) to be Apollonian circles in the $(u,v)$ plane, as those shown in Fig.~\ref{fig:Apollonian} (adapted from Fig.~\ref{fig:antipodal_equator_lines}).
In the $(u,v)$ plane, the corresponding director field $\nA$ is represented as
\begin{equation}
\label{eq:pseudosphere_n_A}
\nA=\frac{(u-c)\e_u+v\e_v}{\sqrt{(u-c)^2+v^2}},
\end{equation}	
where $c$, which is the $u$ coordinate on the line $v=0$ of the center of a geodesic, is expressed by
\begin{equation}
	\label{eq:pseudosphere_c}
	c=\frac{u_0^2+v_0^2-u^2-v^2}{2(u_0-u)}.
\end{equation}

On the pseudosphere, a vortex may coexist with the uniform director field $\nU\equiv\e_u$, everywhere orthogonal to meridians. Letting the separating wall $\Line$ be described in polar coordinates $(\rho,\vt)$ on the $(u,v)$ plane with pole at the vortex, so that 
\begin{equation}
	\label{eq:pseudosphere_polar_coordinates}
	\begin{cases}
		u=u_0+\rho(\vt)\cos\vt,\\
		v=v_0+\rho(\vt)\sin\vt,
	\end{cases}
\end{equation}
we write the unit tangent $\bm{\tau}$ to $\Line$ as in \eqref{eq:wall_tangent} and the bisection rule as in \eqref{eq:wall_jump_condition}, with $\nS$ replaced by $\nU$. Little labor shows that the corresponding wall equation then becomes
\begin{equation}
	\label{eq:pseuudosphere_wall_equation}
	\frac{\dd\rho}{\dd\vt}\cos\vt(r_0+\rho)=\rho[\sin\vt(r_0-\rho)-2v_0],
\end{equation}
where
\begin{equation}
	\label{eq:r_0_definition}
	r_0:=\sqrt{\rho^2+4v_0\rho\sin\vt+4v_0^2}
\end{equation}
is the distance between a point on $\Line$ and the point $(u_0,-v_0)$, which is the mirror image of the defect.

Equation \eqref{eq:pseuudosphere_wall_equation} has a conservation law,
\begin{equation}
	\label{eq:pseudosphere_conservation_law}
	r_0+\rho=c_0,
\end{equation}
where $c_0$ is an arbitrary constant. Simple algebraic manipulations reduce \eqref{eq:pseudosphere_conservation_law} to the following form
\begin{equation}
	\label{eq:pseudosphere_ellipse}
	\rho=\frac{\rho_0(1+e)}{1+e\sin\vt},
\end{equation}
where
\begin{equation}
	\label{eq:pseudosphere_constants}
	\rho_0:=\rho(\tfrac{\pi}{2})=\tfrac12(c_0-2v_0)\quad\text{and}\quad e:=\frac{v_0}{\rho_0+v_0}<1.
\end{equation}
Equation \eqref{eq:pseudosphere_ellipse} represents an \emph{ellipse} with eccentricity $e$ and major semiaxis $a=\rho_0+v_0$, so that its center lies at the point $(u_0,0)$ in the $(u,v)$ plane.

\begin{figure}
	\centering
	\begin{subfigure}{.45\linewidth}
		\centering
	\includegraphics[width=\linewidth]{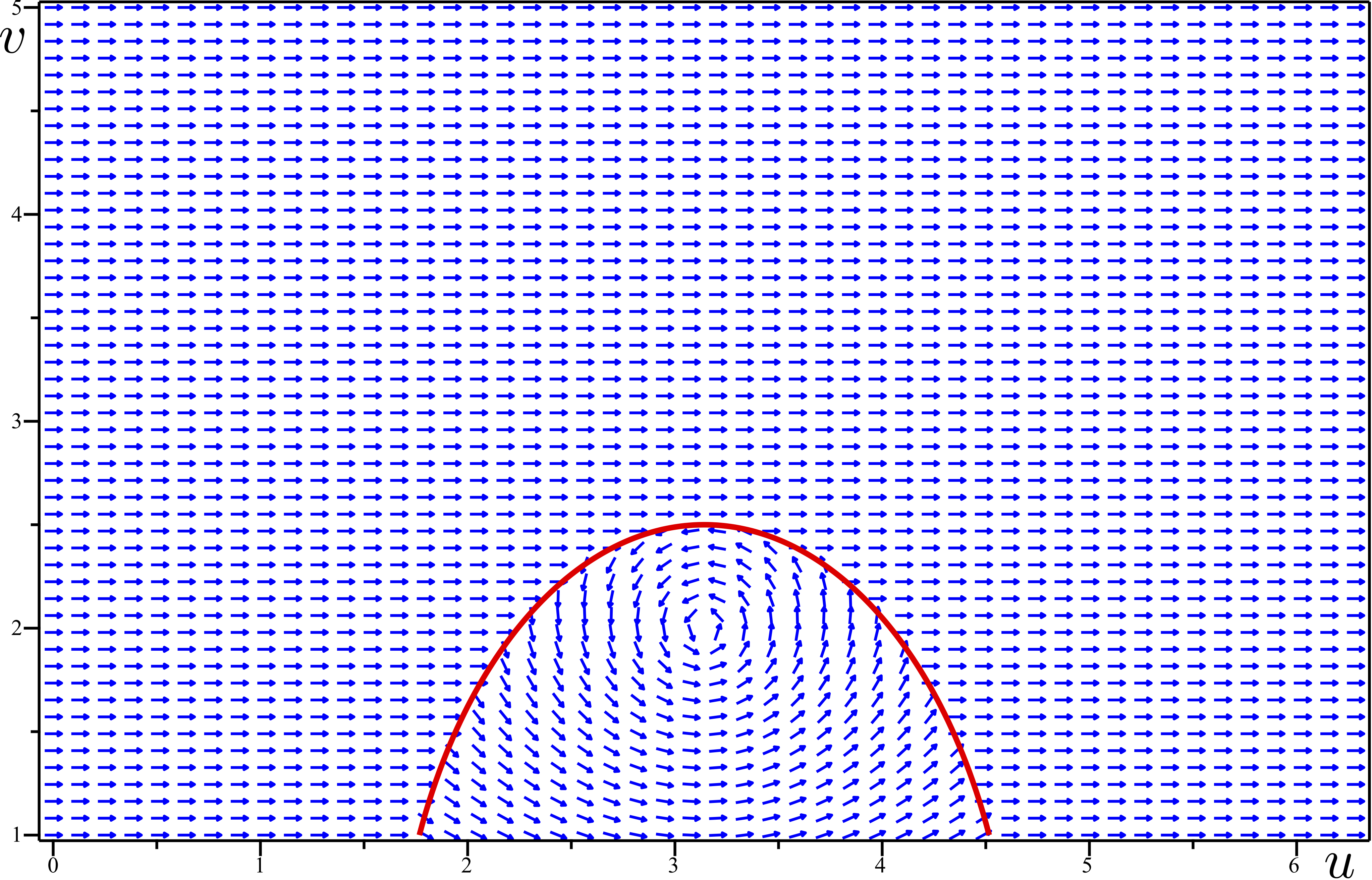}
	\caption{$\rho=\frac12$}
	\label{fig:pseudosphere_wall_rho_0_0p5}
	\end{subfigure}
	\qquad
	\begin{subfigure}{.45\linewidth}
			\centering
		\includegraphics[width=\linewidth]{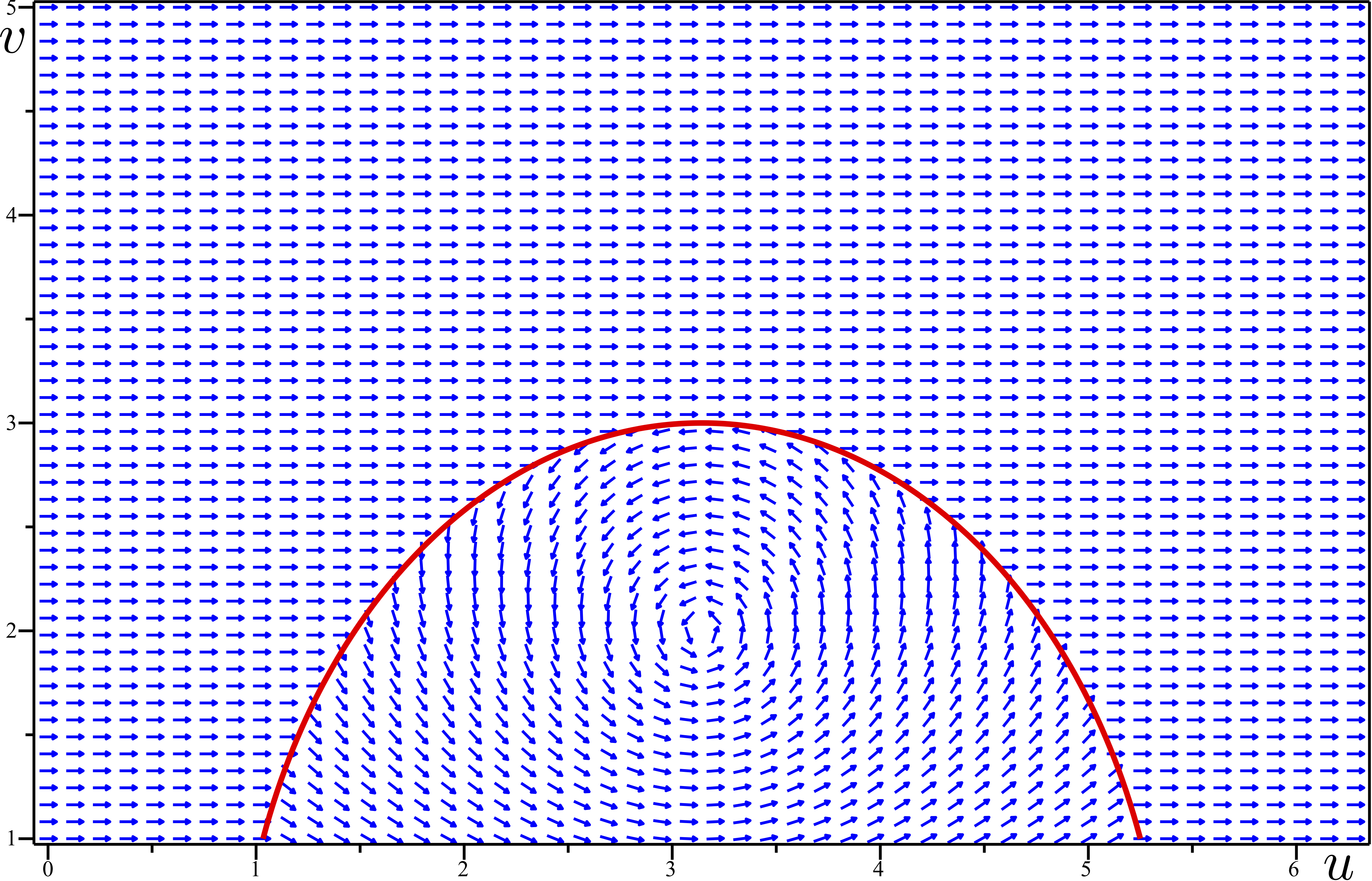}
		\caption{$\rho=1$}
		\label{fig:pseudosphere_wall_rho_0_1}
	\end{subfigure}
	\caption{The ellipse $\Line$ (red line) in \eqref{eq:pseudosphere_ellipse} for $v_0=2$ and two values of $\rho_0$. The field $\nA$ enclosed inside the wall and the field $\nU$ outside it  obey the bisection rule along $\Line$.}
	\label{fig:pseudosphere_wall}
\end{figure}

Since the solution \eqref{eq:pseudosphere_ellipse} is independent of $u_0$, by symmetry, we can take $u_0=\pi$. A typical wall $\Line$ is shown in Fig.~\ref{fig:pseudosphere_wall} for $v_0=2$ and two values of $\rho_0$.
For definiteness, we shall only consider the case where the field $\nA$ resides in the region inside $\Line$; as for the sphere, in the symmetric case, where the roles of the fields $\nA$ with $\nU$ are exchanged, the admissible  walls would be precisely the same.

\begin{figure}
	\centering
	\includegraphics[width=.5\linewidth]{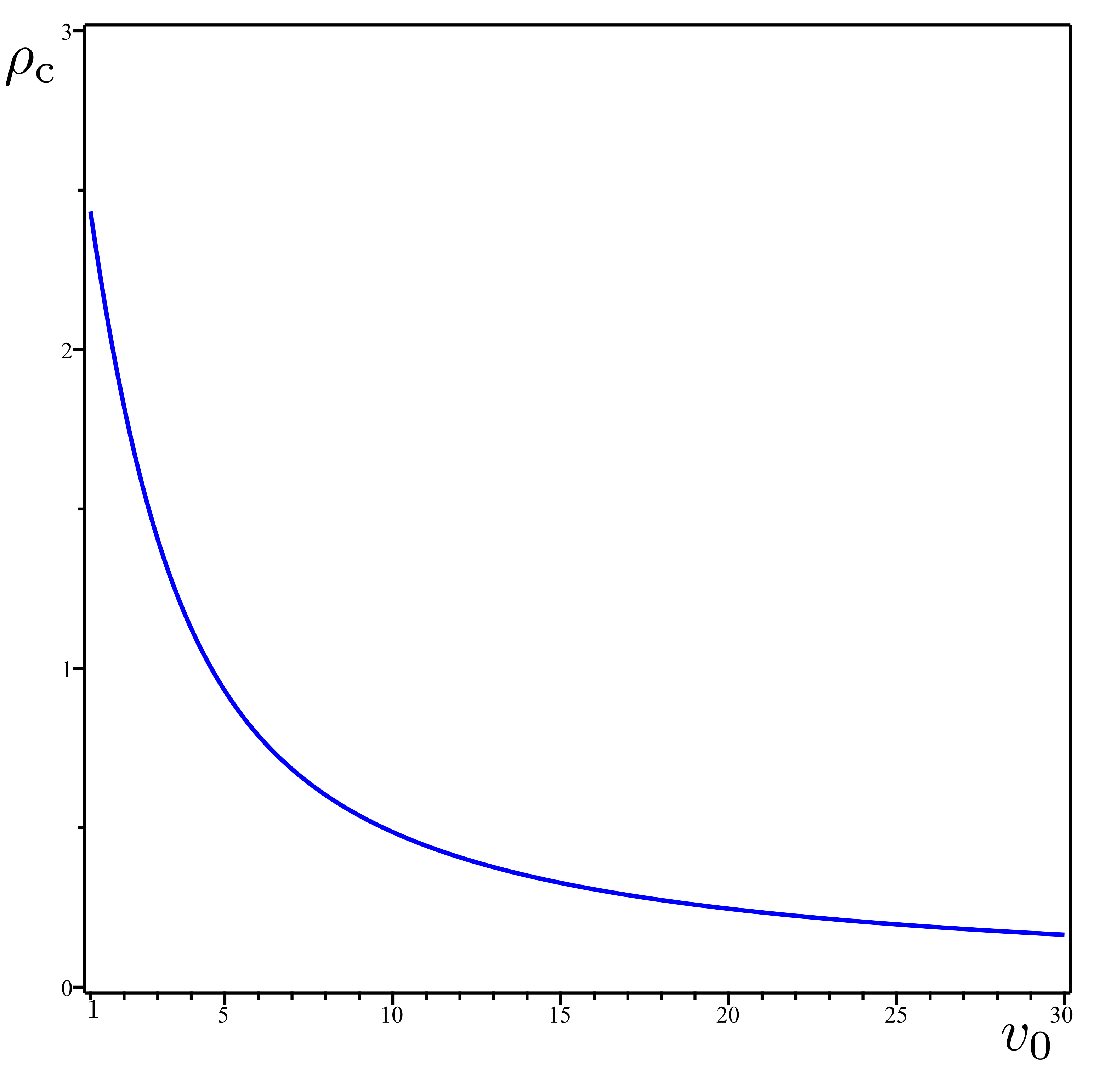}
	\caption{Graph of the function in \eqref{eq:pseudosphere_rho_c} that delivers the critical value $\rho_\mathrm{c}$ of $\rho_0$ below which the wall $\Line$ in \eqref{eq:pseudosphere_ellipse} does not hit the boundary of the strip $0\leqq u\leqq2\pi$ in the $(u,v)$ plane. For $v_0=1$, $\rho_\mathrm{c}\doteq2.43$.}
	\label{fig:rho_c}
\end{figure}

For given $v_0>1$, the ellipse represented by \eqref{eq:pseudosphere_ellipse} lies entirely within the strip $0\leqq u\leqq 2\pi$ whenever $\rho<\rho_\mathrm{c}$, where $\rho_\mathrm{c}$ is a critical value depending on $v_0$,
\begin{equation}
	\label{eq:pseudosphere_rho_c}
	\rho_\mathrm{c}:=\frac{1}{\sqrt{2}}\sqrt{v_0^2+\pi^2+1+\sqrt{[\pi^2+(v_0+1)^2][\pi^2+(v_0-1)^2]}}-v_0.
\end{equation}
This function decays like $\pi^2/2v_0$ as $v_0\to\infty$; its graph is plotted in Fig.~\ref{fig:rho_c}.

\begin{figure}
\centering
\includegraphics[width=.33\linewidth]{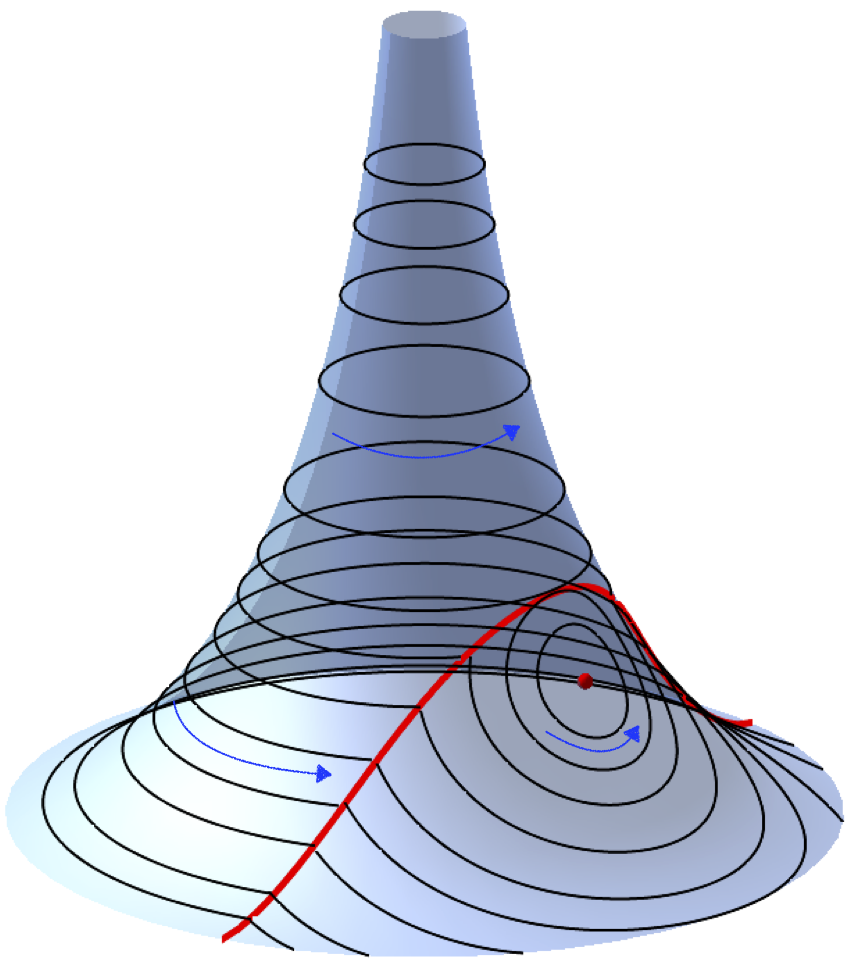}
\caption{Director field lines on the pseudosphere corresponding to the wall depicted in Fig.~\ref{fig:pseudosphere_wall_rho_0_0p5}, for $v_0=2$ and $\rho_0=\frac12$.}
\label{fig:pseudosphere_wall_rho_0_subcritical}
\end{figure}

Figure~\ref{fig:pseudosphere_wall_rho_0_subcritical} illustrates a typical director field with a vortex on the pseudosphere, enclosed by a wall with $\rho_0<\rho_\mathrm{c}$.

\begin{figure}
	\centering
\includegraphics[width=.45\linewidth]{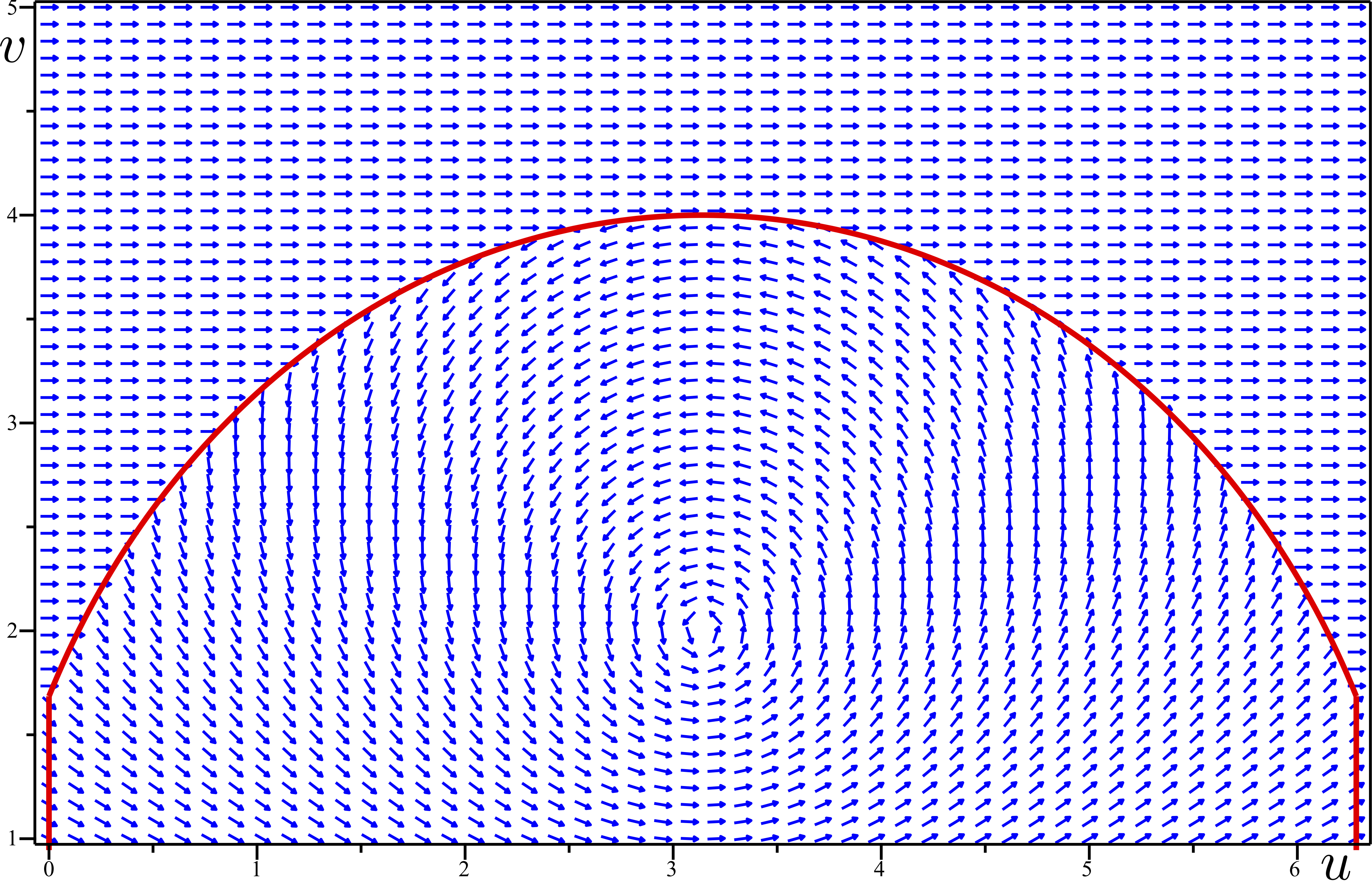}
\caption{Tailed wall for $v_0=2$ and $\rho_0=2$. According to  \eqref{eq:pseudosphere_rho_c}, $\rho_\mathrm{c}\doteq1.82.$}
\label{fig:pseudosphere_tailed_wall_plane}
\end{figure}

\begin{figure}
	\centering
	\begin{subfigure}{.3\linewidth}
		\centering
		\includegraphics[width=\linewidth]{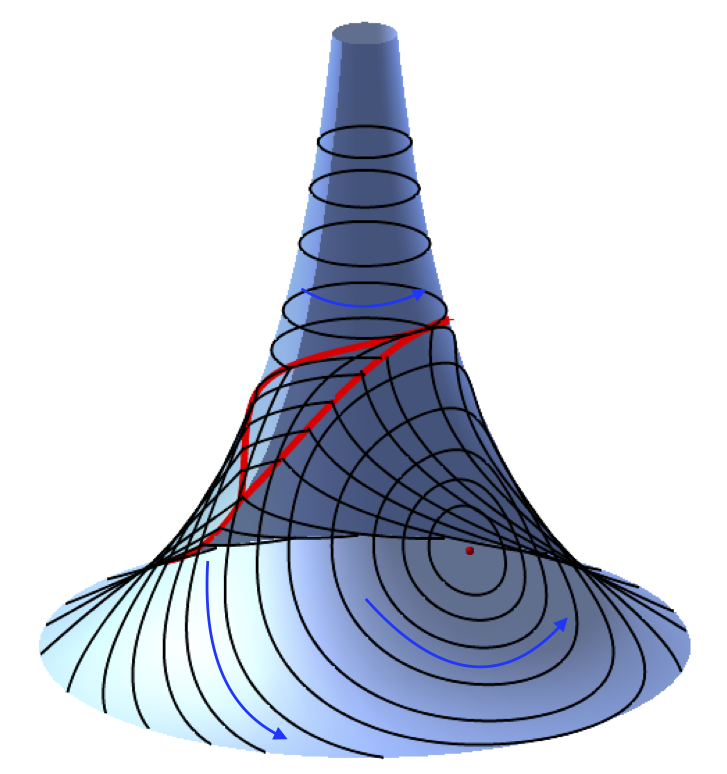}
		\label{fig:pseudosphere_tailed_wall_a}
	\end{subfigure}
	\quad
	\begin{subfigure}{.3\linewidth}
		\centering
	\includegraphics[width=\linewidth]{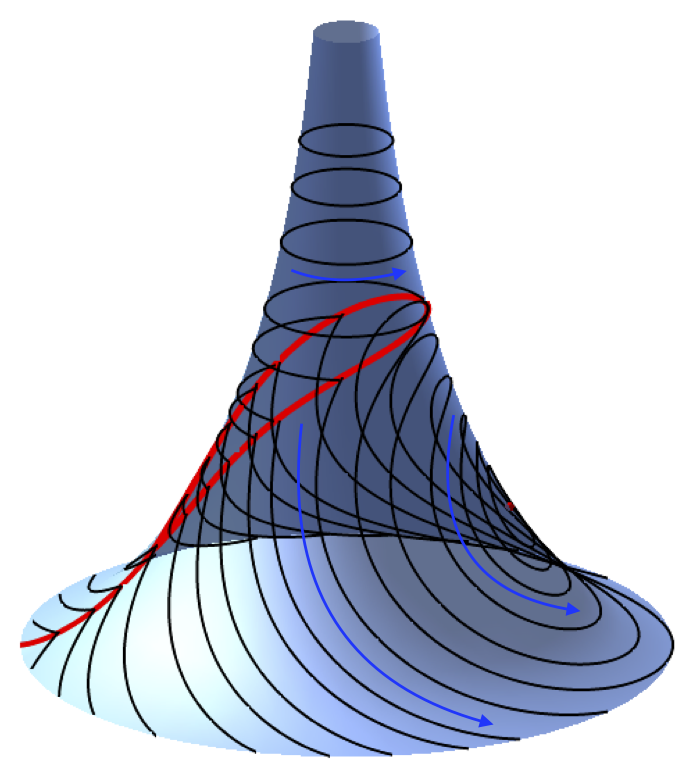}
		\label{fig:pseudosphere_tailed_wall_b}
	\end{subfigure}
	\quad
		\begin{subfigure}{.3\linewidth}
		\centering
		\includegraphics[width=\linewidth]{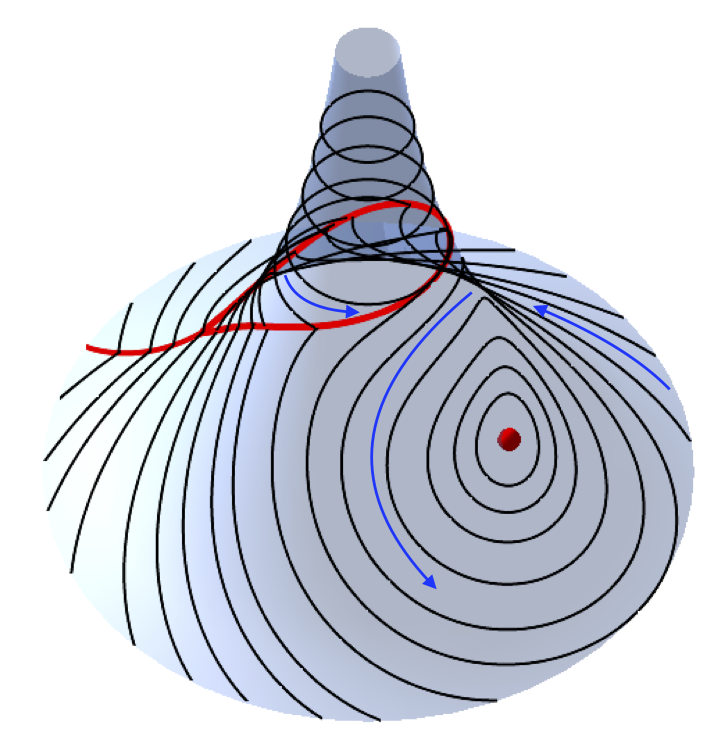}
		\label{fig:pseudosphere_tailed_wall_c}
	\end{subfigure}
	\caption{Three views of the director field lines on the pseudosphere corresponding to the tailed wall shown  in Fig.~\ref{fig:pseudosphere_tailed_wall_plane} for $v_0=2$ and $\rho_0=2$.}
	\label{fig:pseudosphere_tailed_wall}
\end{figure}

For $\rho_0>\rho_\mathrm{c}$, the wall $\Line$ hits the boundary of the strip $0\leqq u\leqq2\pi$ and develops a \emph{tail} along the single meridian corresponding to $u=0$ and $u=2\pi$.  By symmetry, the field $\nA$, which is discontinuous along this tail, satisfies there the bisection rule, and thus the added tail becomes part of the wall. Figure~\ref{fig:pseudosphere_tailed_wall_plane}
illustrates a tailed wall in the $(u,v)$ plane. The corresponding director field lines are shown in Fig.~\ref{fig:pseudosphere_tailed_wall}.

\begin{figure}
	\centering
	\includegraphics[width=.5\linewidth]{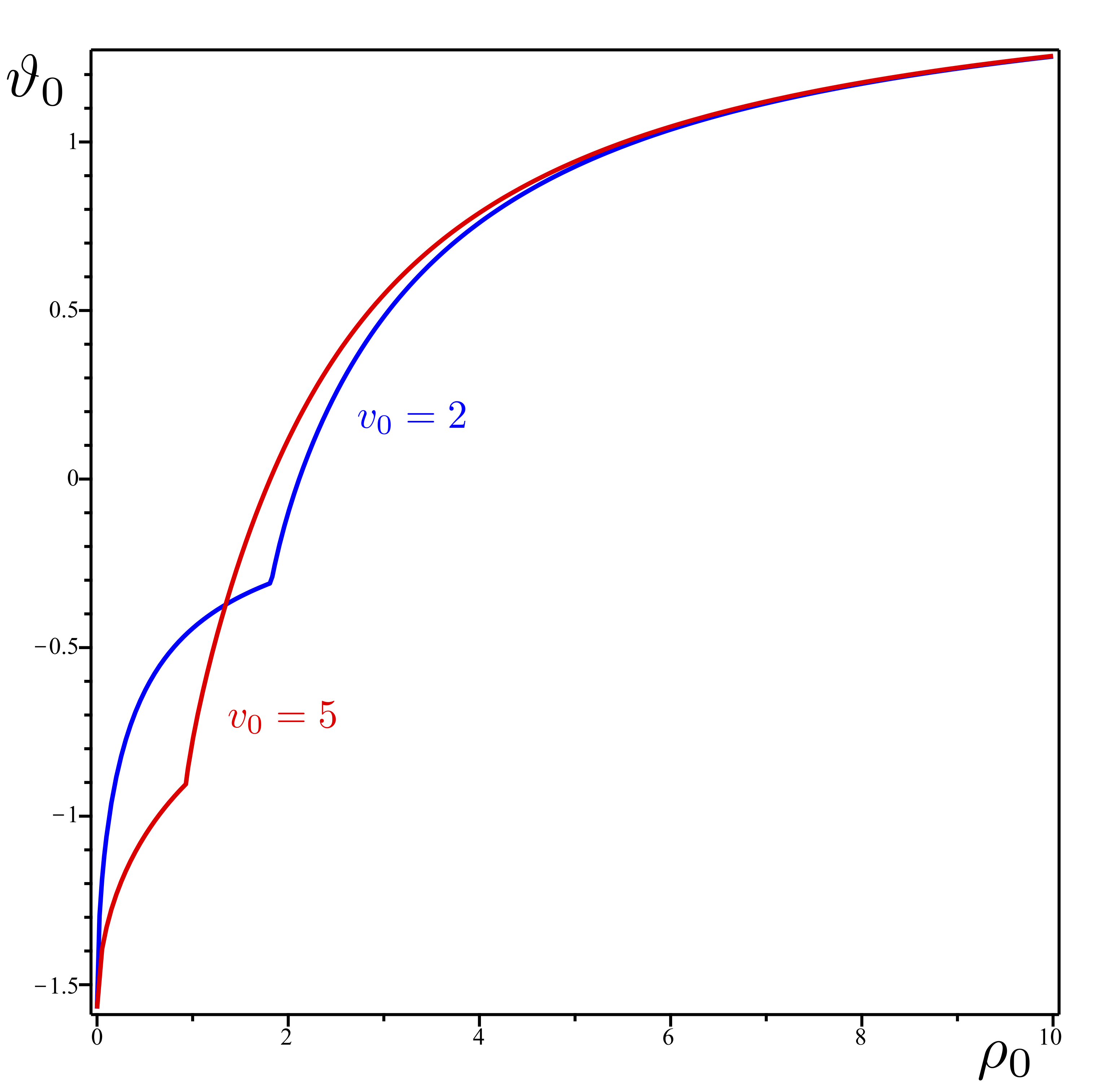}
	\caption{Graphs of $\vt_0$ against $\rho_0$ for $v_0=2$ (blue) and $v_0=5$ (red).}
	\label{fig:pseudosphere_theta_0}
\end{figure}

To compute the length $L(\Line)$ of the wall $\Line$ on the pseudosphere, we apply \eqref{eq:pseudosphere_line_element} and \eqref{eq:pseudosphere_polar_coordinates} to get
\begin{equation}
	\label{eq:pseudosphere_wall_length}
	L(\Line)=2\int_{\vt_0}^{\frac\pi2}\frac{\sqrt{\rho^2+\rho'^2}}{v_0+\rho\sin\vt}\dd\vt+H(\rho-\rho_\mathrm{c})\ln(v_0+\rho_0),
\end{equation}
where $H(\xi)$ is the standard Heaviside function, $\rho$ is as in \eqref{eq:pseudosphere_ellipse}, and $\vt_0=\vt(\rho_0,v_0)$ is a function that designates the polar angular fraction of the ellipse caught inside the coordinate strip $0\leqq u\leqq2\pi$, $v\geqq1$. Although $\vt_0$ can be given an explicit analytic expression, that is too cumbersome to be reproduced here; for any given $v_0>1$, $\vt(\rho_0,v_0)$ is a continuous, increasing function of $\rho_0$ ranging in $[-\frac\pi2,\frac\pi2]$ with a discontinuity in its derivative for $\rho_0=\rho_\mathrm{c}$. Figure~\ref{fig:pseudosphere_theta_0} shows $\vt_0$ as a function of $\rho_0$ for two selected values of $v_0$.

The second addendum in \eqref{eq:pseudosphere_wall_length} is the length of the tail of $\Line$; it is dominant in the limit as $\rho_0\to\infty$, as the first addendum converges to zero. For $\rho_0\to0$, $L(\Line)$ converges to $2\ln v_0$ and the vortex isolated within a shrinking $\Line$ simply evaporates in the limit.

\begin{figure}
	\centering
	\includegraphics[width=.5\linewidth]{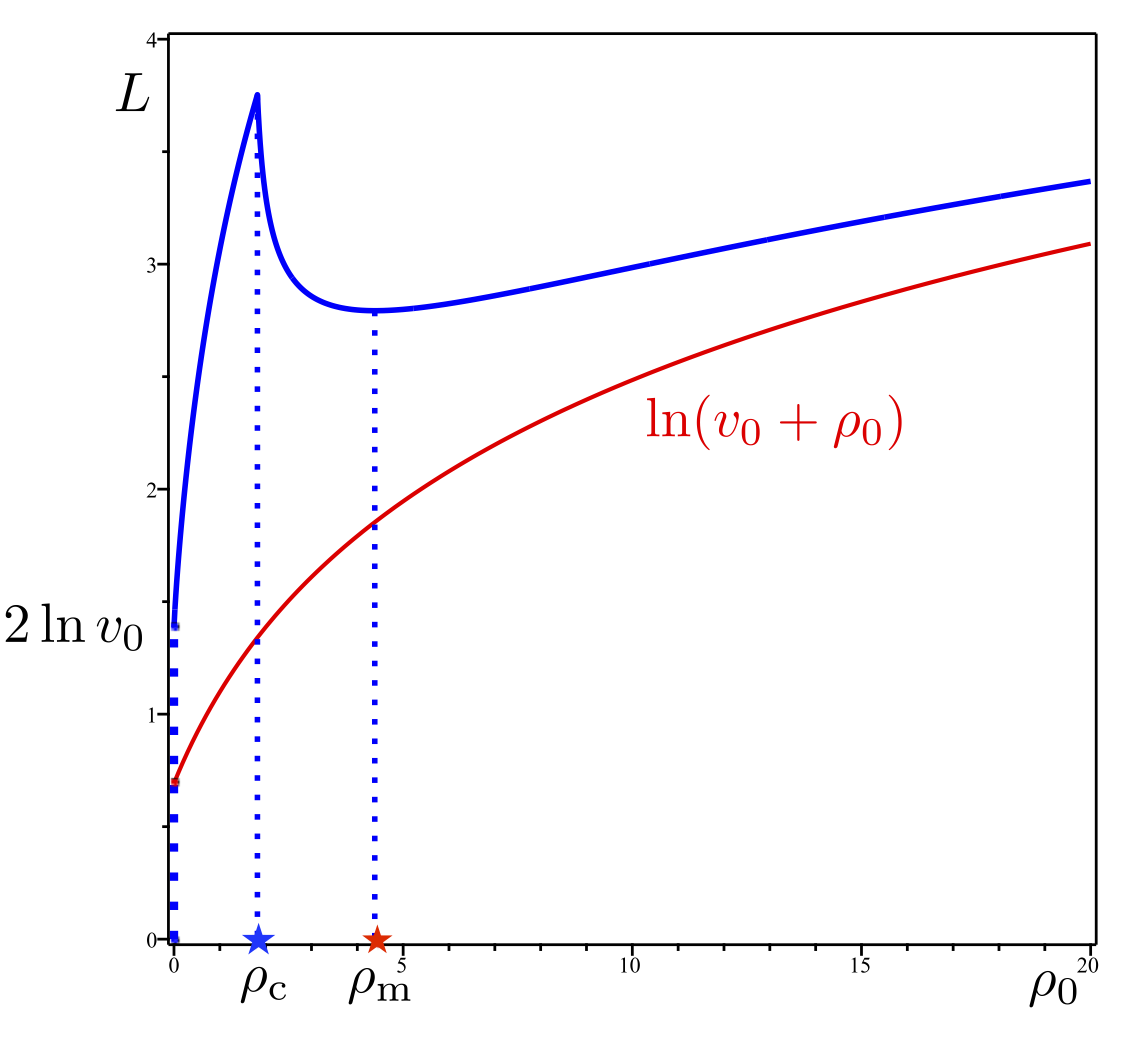}
	\caption{The graph of the length $L$ of walls $\Line$ corresponding to different values of $\rho_0$ (blue curve) is contrasted against the asymptotic behavior of the tail's length (red curve). Here $v_0=2$ and, according to \eqref{eq:pseudosphere_rho_c}, $\rho_\mathrm{c}\doteq1.82$, while the local minimum of $L$ falls at $\rho_\mathrm{m}\doteq4.38$ where $L=L_\mathrm{m}\doteq2.79$.}
	\label{fig:pseudosphere_wall_length}
\end{figure}

Figure~\ref{fig:pseudosphere_wall_length} illustrates $L(\Line)$ as a function of $\rho_0$ for $v_0=2$. 
The graph exhibits a relative minimum, which acts as an attractor for walls that happen to have $\rho_0>\rho_\mathrm{c}$: these walls are trapped, whereas all others manage to escape.

\begin{figure}
	\centering
	\begin{subfigure}[c]{0.25\linewidth}
		\centering
		\includegraphics[width=\linewidth]{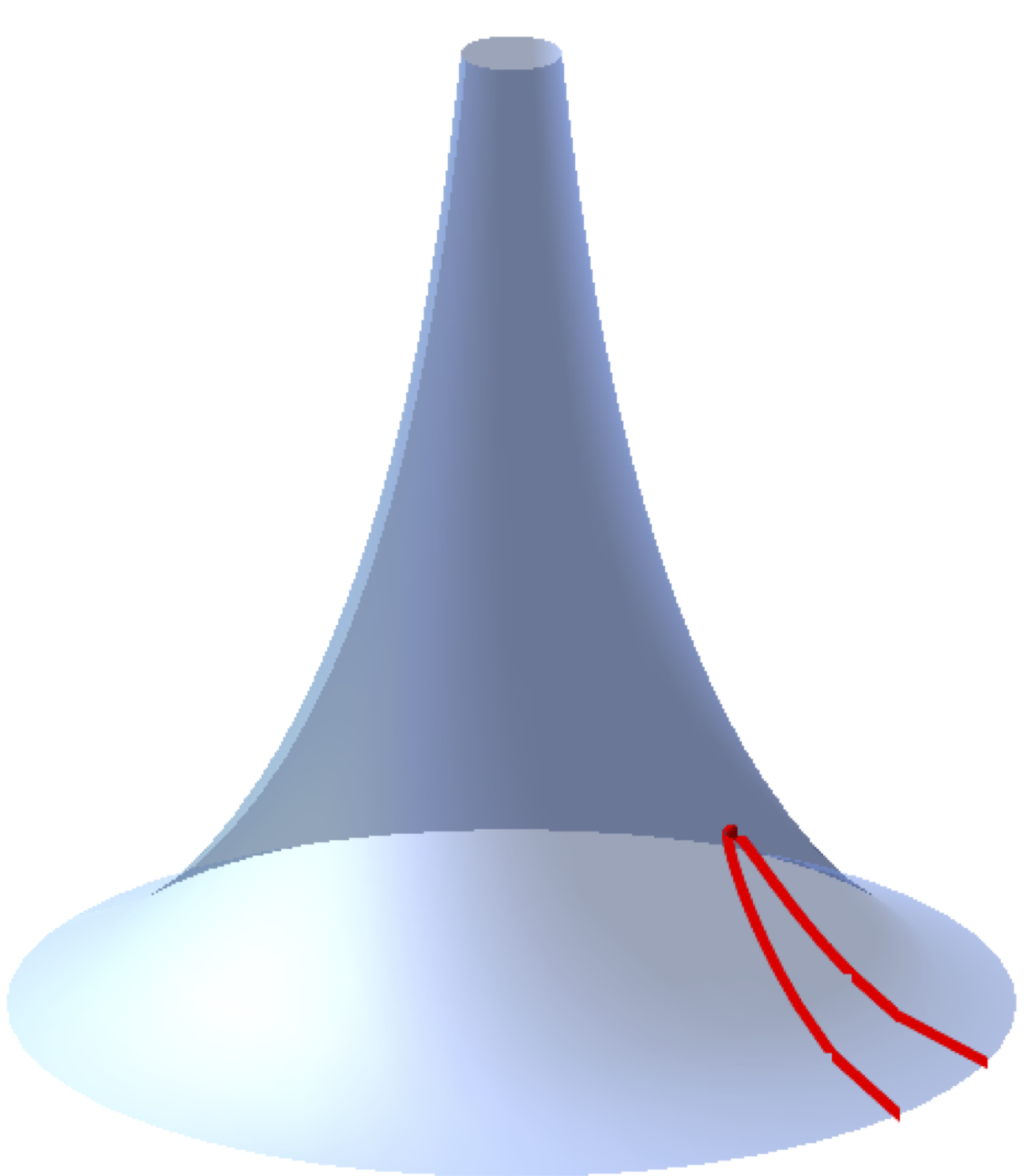}
		\caption{$\rho_0=0.01$}
		\label{fig:pseudosphere_gallery_a}
	\end{subfigure}
	\qquad
	\begin{subfigure}[c]{0.25\linewidth}
		\centering
		\includegraphics[width=\linewidth]{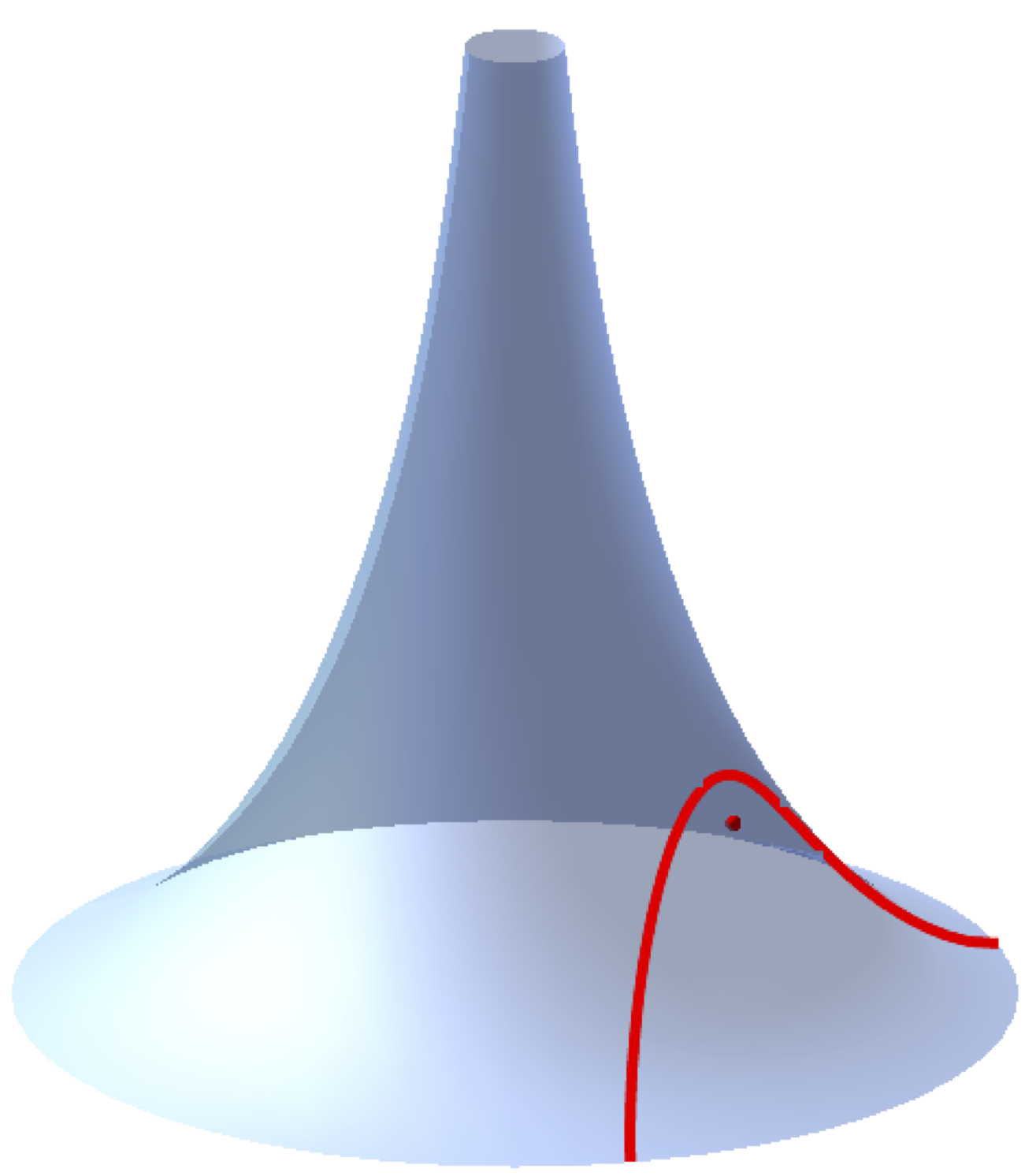}
		\caption{$\rho_0=0.2$}
		\label{fig:pseudosphere_gallery_b}
	\end{subfigure}
	\\
	\begin{subfigure}[c]{0.25\linewidth}
		\centering
	\includegraphics[width=\linewidth]{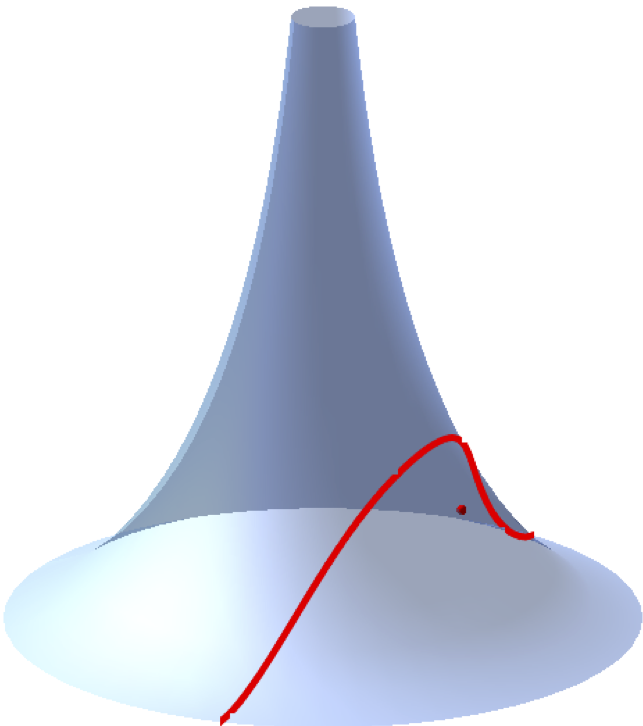}
		\caption{$\rho_0=0.5$}
		\label{fig:pseudosphere_gallery_c}
	\end{subfigure}
	\qquad
	\begin{subfigure}[c]{0.25\linewidth}
		\centering
		\includegraphics[width=\linewidth]{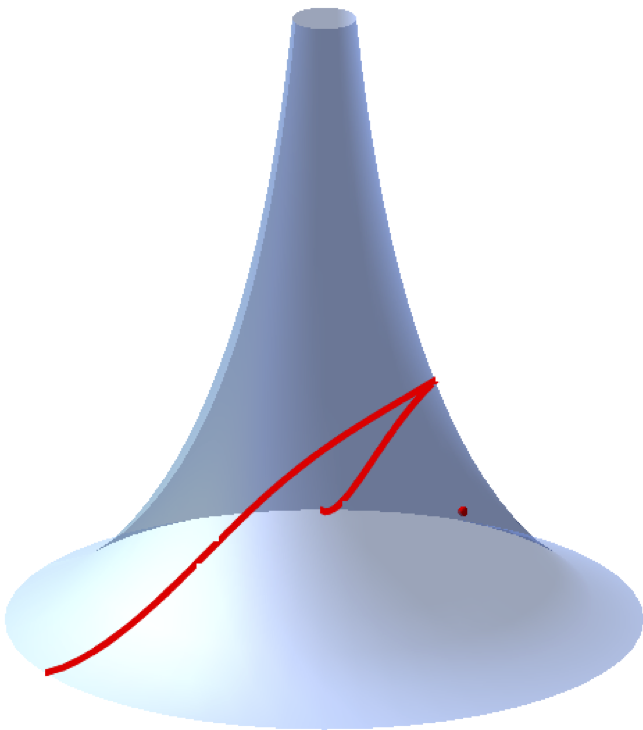}
		\caption{$\rho_0=1$}
		\label{fig:pseudosphere_gallery_d}
	\end{subfigure}
	\\
	\begin{subfigure}[c]{0.25\linewidth}
		\centering
		\includegraphics[width=\linewidth]{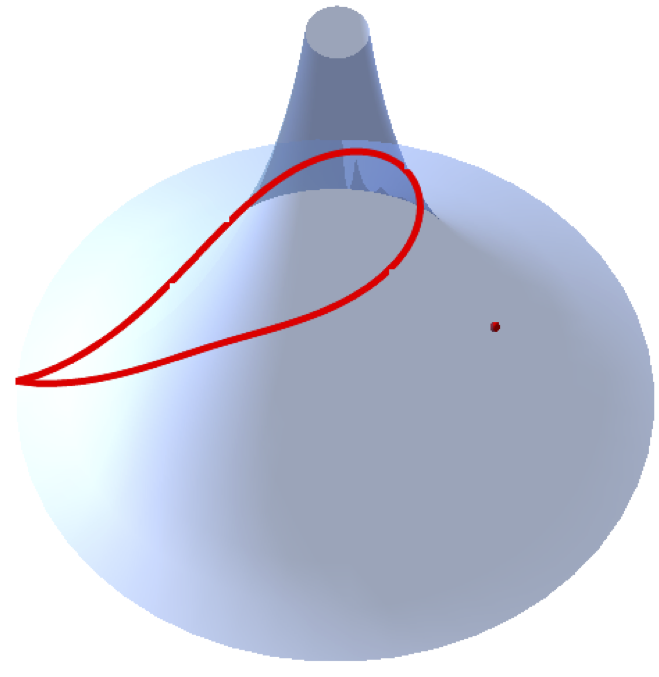}
		\caption{$\rho_0=\rho_\mathrm{c}$}
		\label{fig:pseudosphere_gallery_e}
	\end{subfigure}
	\\
	\begin{subfigure}[c]{0.25\linewidth}
		\centering
	\includegraphics[width=\linewidth]{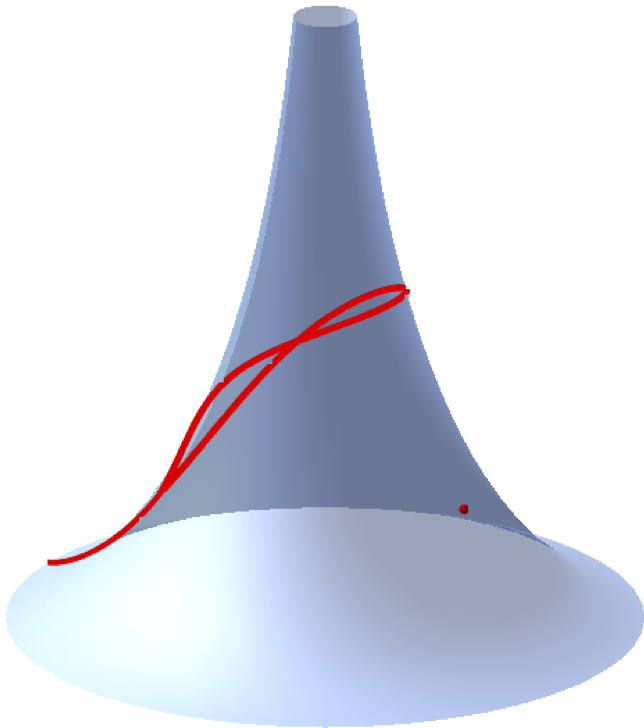}
		\caption{$\rho_0=2$}
		\label{fig:pseudosphere_gallery_f}
	\end{subfigure}
	\qquad
	\begin{subfigure}[c]{0.25\linewidth}
		\centering
		\includegraphics[width=\linewidth]{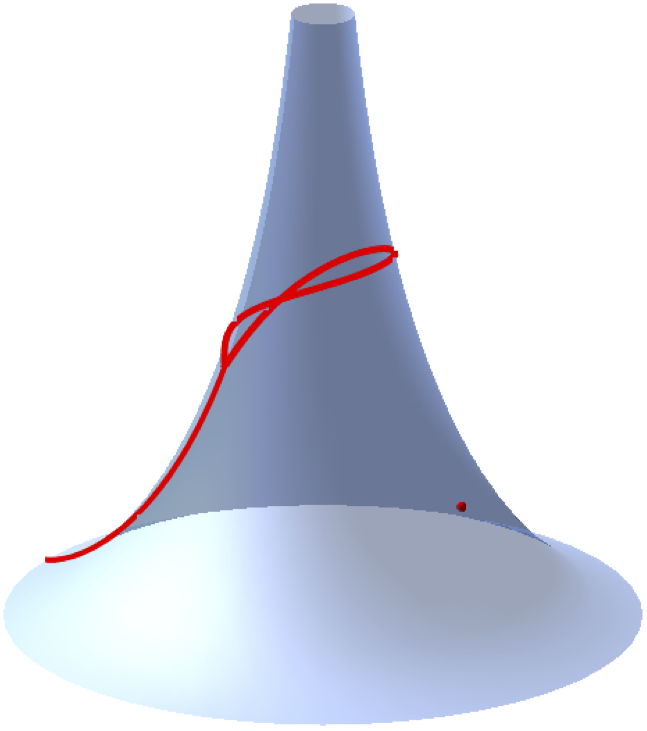}
		\caption{$\rho_0=2.5$}
		\label{fig:pseudosphere_gallery_g}
	\end{subfigure}
	\\
	\begin{subfigure}[c]{0.25\linewidth}
		\centering
		\includegraphics[width=\linewidth]{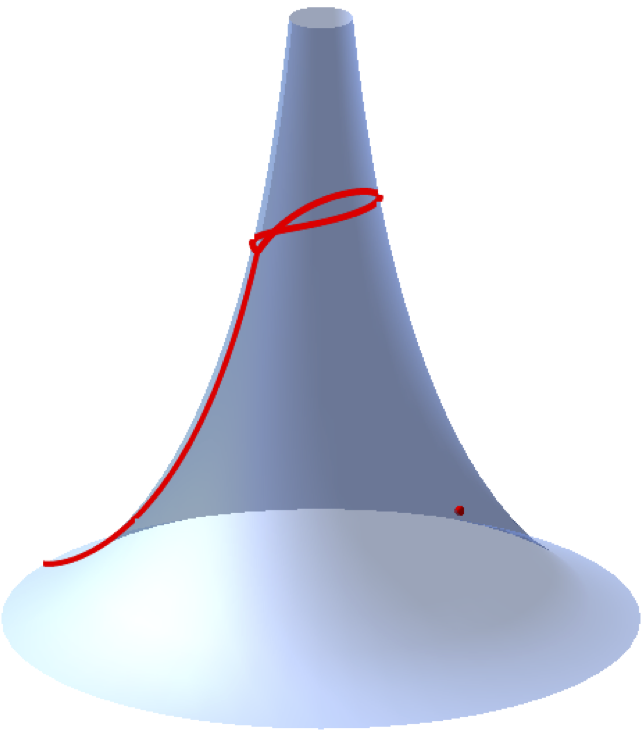}
		\caption{$\rho_0=3.5$}
		\label{fig:pseudosphere_gallery_h}
	\end{subfigure}
	\qquad
	\begin{subfigure}[c]{0.25\linewidth}
		\centering
		\includegraphics[width=\linewidth]{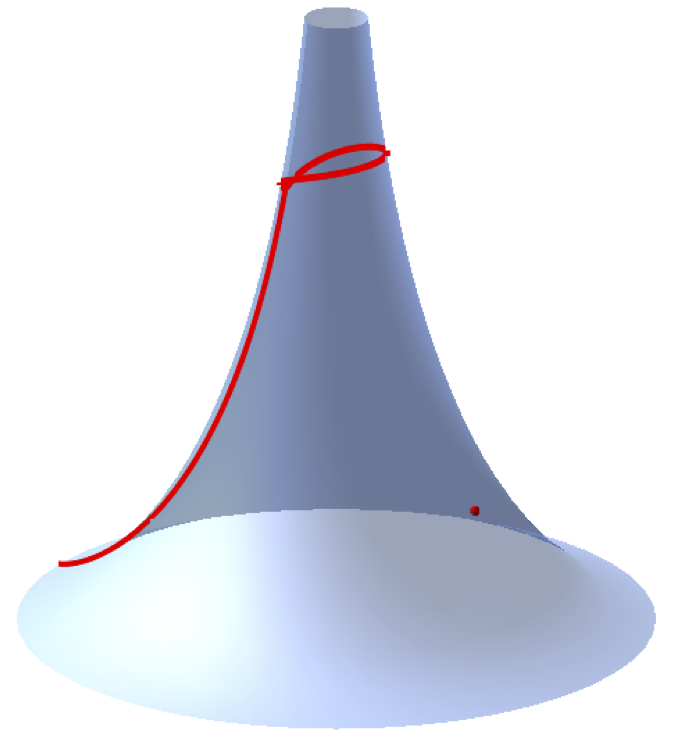}
		\caption{$\rho_0=\rho_\mathrm{m}$}
		\label{fig:pseudosphere_gallery_i}
	\end{subfigure}
	\caption{Walls on a pseudosphere. They all surround a vortex pinned at $u_0=\pi$ and $v_0=2$. The critical wall at $\rho_0=\rho_\mathrm{c}\doteq1.82$ separates escaping walls form trapped ones; the attracting wall falls at $\rho_0=\rho_\mathrm{m}\doteq4.38$.}
	\label{fig:pseudosphere_gallery}
\end{figure}

Figure~\ref{fig:pseudosphere_gallery} illustrates several walls surrounding a vortex; those in panels (e)--(d) are free to escape, whereas those in panels (f)--(i) are bound to be trapped. The critical wall is depicted in panel (e). The wall in panel (i) is the attracting one.

\section{Conclusions}\label{sec:conclusions}
The theoretical study of ferroelectric nematic liquid crystals could hardly avoid involving electrostatic interactions, which exhibit nonlocal features implying scenarios far more complicated than usually encountered for ordinary nematics. Approximations may help, provided they preserve the essence of the newly emerging phenomena.

An attempt was made here for thin films deposited on curved shells, under the assumption that the electrostatic energy strongly prevails over the elastic distortion energy, an assumption whose degree of confidence can be established quantitatively by comparing Debye screening length to a critical value emerging from Khachaturyan's classical instability.

Under this assumption, splay is completely banned from ferroelectric nematic shells, a phenomenon here called \emph{splay ostracism}, as it is reminiscent of the practice in use in ancient Greek cities to exile citizens whose influence was considered dangerous. A theorem reduced splay ostracism to a geometric condition, requiring the nematic director $\n$ (and so also the polarization vector $\pol$) to be orthogonal to geodesics on the surface $\surface$  representing the shell in three-dimensional space. When \emph{vortexes} of $\n$ are formed on $\surface$, they may happen to be surrounded by \emph{walls}, that is, lines $\Line$ on $\surface$ where $\n$ jumps, obeying a geodesic \emph{bisection} rule.

In this purely geometric setting, vortex configurations can only be distinguished on the basis of the wall's total length $L(\Line)$. A simple selection rule thus governs quasi static defect evolution: shortening $L(\Line)$.

By applying this rule, we saw that walls may either be trapped in local minima of $L(\Line)$ on $\surface$ or escape $\surface$ altogether. Both behaviors arguably reflect on defects, which may correspondingly be stabilized or expelled.

We worked out the details of this method for both $\surface$ a sphere and a pseudosphere to collect evidence on the possible role of the sign of the Gaussian curvature $K$ in favoring either trapped or escaping walls. Only escaping walls were found for a sphere, and both escaping and trapped ones for a pseudosphere. Clearly, we cannot draw any general conclusion from such limited information, apart from suggesting that the wall evolution is, as expected, sensitive to the sign of $K$.

In reality, splay ostracism may fail to be complete, in which case residual (nonlocal) electrostatic interactions may interfere with the simple geometric wall dynamics envisaged here. Similarly, distortion elasticity, which has been neglected here, may be another source of interference. These are important, but difficult issues that deserve to be addressed by future studies.

\appendix

\section{Proof of the splay ostracism theorem}\label{sec:proof}
The proof of Theorem~\ref{thm:splay} is split in four elementary steps.
\begin{enumerate}
	\item We first recall from \eqref{eq:distorion_measures} that the spin connector $\cv$ of the frame $\framen$ can in general be written as
	\begin{equation}
		\label{eq:spin_connector_components}
		\cv=-B\n+S\nperp.
	\end{equation}
	\item Assume that $S\equiv0$. We seek a curve $\Curve$ on $\surface$ passing through any selected point $\x\in\surface$ and such that $\n$ is \emph{parallel transported} (in the sense of Levi-Civita) along $\Curve$, see Fig.~\ref{fig:darboux}.\footnote{The notion of parallel transport of a tangential vector field on a surface  was introduced by Levi-Civita \cite{levi-civita:nozione} (see also \cite{persico:realizzazione} for a telling kinematic interpretation). The role played by this notion in our development is similar to the one played in \cite{pedrini:surface} to characterize all \emph{uniform} nematic fields, that is, fields with constant $S$ and $B$. The reader is also referred to \cite{pedrini:surface} for a general treatment of surfaces in three-dimensional space.}
	\begin{figure}
		\centering
		\includegraphics[width=.33\linewidth]{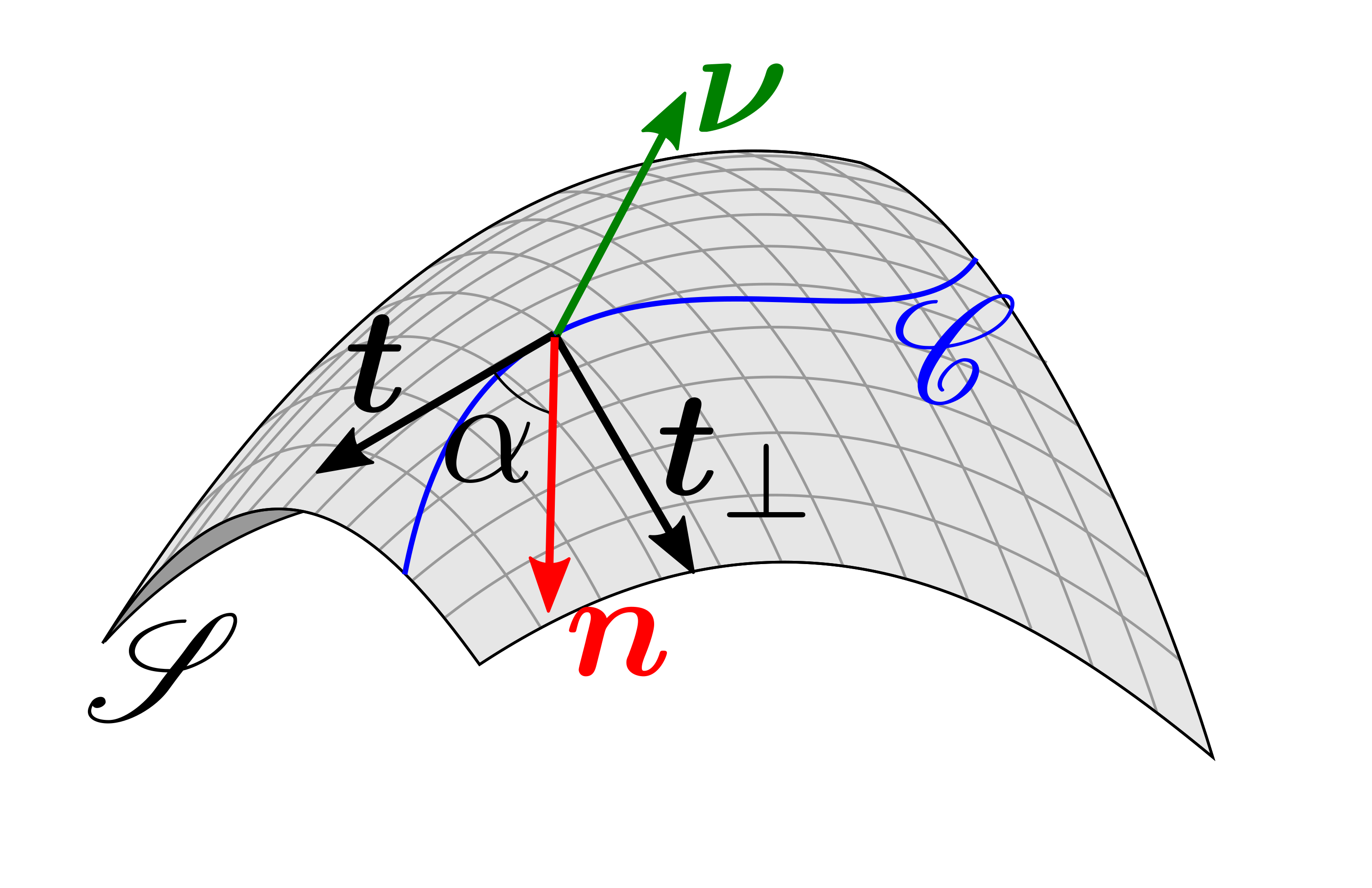}
		\caption{A curve $\Curve$ on $\surface$, along which $\n$ is parallel transported (in the sense of Leci-Civita); $\tangent$ is the unit tangent to $\Curve$ and $\tperp=\normal\times\tangent$.}
		\label{fig:darboux}
	\end{figure}
	Letting a prime $'$ denote differentiation with respect to the arc-length parameter along $\Curve$, and denoting by $\tangent$ the unit tangent to $\Curve$, we obtain from \eqref{eq:glidind_laws} that 
	\begin{equation}
		\label{eq:proof_step_2}
		\n'=(\nablas\n)\tangent=(\cv\cdot\tangent)\nperp+(\dv_1\cdot\tangent)\normal.
	\end{equation}
Since $\n$ is parallel transported along $\Curve$, $\n'$ must be parallel to  $\normal$, which requires that $\cv\cdot\tangent\equiv0$. Then \eqref{eq:spin_connector_components} implies that $\n\cdot\tangent\equiv0$, or equivalently
\begin{equation}
	\label{eq:tangent_nperp}
	\tangent=\pm\nperp.
\end{equation}
\item From \eqref{eq:tangent_nperp} and \eqref{eq:glidind_laws}, we compute
\begin{equation}
	\label{eq:proof_step_3}
	\tangent'=\pm(\nablas\nperp)\tangent=\pm(\dv_2\cdot\tangent)\normal,
\end{equation}
again because $\cv\cdot\tangent\equiv0$. This shows that all curves $\Curve$ that parallel transport a field $\n$ with $S\equiv0$ are geodesics of $\surface$.\footnote{Customarily, parallel transport is employed to generate  tangential fields on a surface. Here, conversely, we start from a certain tangential field and find that it is parallel transported by geodesics.} In particular, by \eqref{eq:tangent_nperp}, $\n$ is everywhere orthogonal to the system of geodesics that transport it.
\item To complete the proof, we only need to show that any tangential unit vector field orthogonal to a system of geodesics has no surface slay. To this end, we denote by $\tangent$ the unit vector field generated by the tangent to a system of geodesics on $\surface$ and set $\tperp:=\normal\times\tangent$. Letting $\n=\pm\tperp$, we compute
\begin{align}\label{eq:proof_step_4}
\divs\n&=\proj\cdot\nablas\n=\pm\proj\cdot\nablas\tangent_\perp\nonumber\\
	&=\pm(\tangent\otimes\tangent+\tangent_\perp\otimes\tangent_\perp)\cdot\nablas\tangent_\perp\\
	&=\pm[\tangent\cdot(\nablas\tangent_\perp)\tangent+\tangent_\perp\cdot(\nablas\tangent_\perp)\tangent_\perp].\nonumber
\end{align}
By the identity $\tangent_\perp\cdot(\nablas\tangent_\perp)\tangent_\perp=\tangent_\perp\cdot(\nablas\tangent_\perp)\trans\tangent_\perp$, and the fact that $(\nablas\tangent_\perp)\trans\tangent_\perp=\zero$, because $\tperp$ is a unit vector field,\footnote{Here $\trans$ denotes transposition of a second-rank tensor.} \eqref{eq:proof_step_4} implies that 
\begin{equation}
	\label{eq:proof_final_step}
	\divs\n=\pm\tangent\cdot\tperp'=\mp\tangent'\cdot\tperp=0.
\end{equation}
The last two equations follow from the identity $\tangent\cdot\tperp\equiv0$ and the fact that $\tangent'$ is parallel to $\normal$ because $\tangent$ is tangent to a geodesic.
\end{enumerate}

\section{Defect caustics}\label{sec:caustic}
We consider a general system of geodesics  in the plane.  Relative to Cartesian frame $(\e_x,\e_y)$ a pencil of straight lines is represented as
\begin{equation}
	\label{eq:caustic_pencil}
	\begin{cases}
		x=x_0(v)+u\cos\vt(v),\\
		y=u\sin\vt(v),
	\end{cases}
\end{equation}
where $x_0$ and $\vt$ are differentiable functions. The unit vectors $(\tangent,\tperp)$ associated with the pencil in \eqref{eq:caustic_pencil} are
\begin{equation}
	\label{eq:caustic_t_t_perp}
	\tangent=\cos\vt\e_x+\sin\vt\e_y\quad\text{and}\quad\tperp=-\sin\vt\e_x+\cos\vt\e_y.
\end{equation}
The Jacobian of the mapping $(u,v)\mapsto(x,y)$ represented by \eqref{eq:caustic_pencil} is easily found to be
\begin{equation}\label{eq:caustic_jacobian}
	J=u\vt'-x_0'\sin\vt,
\end{equation}	
where a prime now stands for differentiation with respect to $v$.

In a region of the plane where $J\neq0$, the pencil in \eqref{eq:caustic_pencil} is a system of geodesics with  coordinates $(u,v)$. $J$ does not vanish in the whole plane (except at most a point) only in two cases: 
\begin{inparaenum}[(a)]
\item for $u\in\real$, if $\vt'\equiv0$ and $x_0'\neq0$;
\item for $u\in\real^+$, if $x_0'\equiv0$ and $\vt'\neq0$.\footnote{The latter case could equivalently be phrased for $-u\in\real^+$.}
\end{inparaenum}
These cases correspond to a pencil of parallel straight lines and to a pencil of straight half lines emanating from a point, respectively. The corresponding field $\n=\tperp$ with $S\equiv0$ is either uniform or a vortex. 

These are precisely the instances referred to in Sec.~\ref{sec:theorem}. Other instances exist where $J=0$ on a larger set. These are associated with dense distributions of defects, which we call \emph{caustics}, an illustration of which is provided by the following example.

\begin{figure}
	\centering
	\includegraphics[width=.5\linewidth]{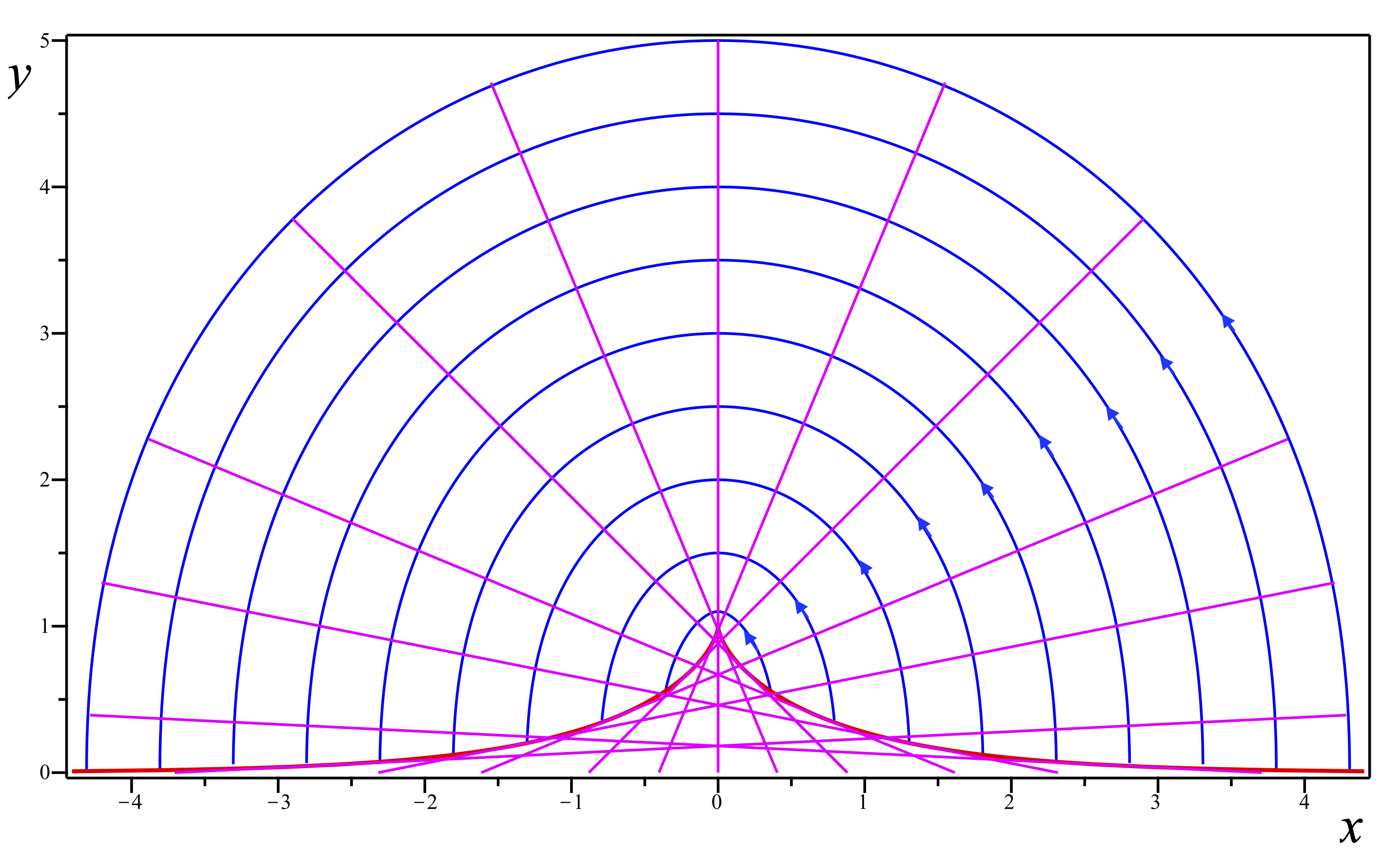}
	\caption{Example of a divergence free nematic field $\n$ in a region of the half plane $y\geqq0$ with a defect caustic. Geodesics represented by \eqref{eq:caustic_pencil_cartesian} in $y\geqq0$ are drawn in magenta. Their envelope is the red line represented by \eqref{eq:caustic_envelope}. The field lines of $\n$, oriented anticlockwise, are blue.}
	\label{fig:caustic}
\end{figure}

Letting $u=1$ in \eqref{eq:caustic_jacobian}, we set $\vt=v\in[0,\pi]$ and find that $J=0$ on the curve described parameterically by
\begin{equation}
	\label{eq:caustic_envelope}
	\begin{cases}
		x=\ln\dfrac{1-\cos v}{\sin v}+u\cos v,\\
		y=u\sin v.
	\end{cases}
\end{equation}
This curve (plotted in red in Fig.~\ref{fig:caustic}) is also the \emph{envelope} of the pencil \eqref{eq:caustic_pencil}. The latter can indeed be given the Cartesian form 
\begin{equation}
	\label{eq:caustic_pencil_cartesian}
	x-y\cot v-\ln\frac{1-\cos v}{\sin v}=0
\end{equation}
in the parameter $v$;
by applying to \eqref{eq:caustic_pencil_cartesian} the classical envelope costruction (see, for example, \cite[p.\,73]{struik:lectures}), we readily arrive at \eqref{eq:caustic_envelope}.

Figure~\ref{fig:caustic} shows geodesics (magenta), their envelope (red), and field lines of the director $\n=\tperp$ (blue). As could also be checked by direct inspection, the field $\n$ is divergence free in the region of the half plane $y\geqq0$ above the geodesic envelope, below which defects of $\n$ are densely distributed. A closer inspection of \eqref{eq:caustic_pencil_cartesian} shows that each point below the geodesic envelope is traversed by \emph{three} geodesics, and so  this region of the half-plane $y\geqq0$ is a defect \emph{caustic}.\footnote{Similar examples of defect caustics can also be found in \cite{pedrini:relieving}.} 


%

\end{document}